\pdfoutput=1
\documentclass[11pt,a4paper]{article}
\usepackage{jheppub}

\RequirePackage[utf8]{inputenc}

\usepackage{amsfonts}
\usepackage{amssymb}
\usepackage{comment}
\usepackage{graphicx}
\usepackage{amsmath}
\usepackage{tabu}
\usepackage{dsfont}
\usepackage{float}
\usepackage{amsthm}
\usepackage{slashed}
\usepackage{setspace}
\usepackage[labelformat=simple]{subcaption}

\usepackage{pgfplots}
\usepackage{tikz}
\usepackage{tikz-cd}
\usetikzlibrary{shapes.misc,shadows,fit,decorations.pathmorphing,positioning,trees,decorations.markings,decorations.pathreplacing,calc,shapes,patterns,arrows}
\usepackage{xcolor}
\usepackage[boxsize=0.5em,aligntableaux=center]{ytableau}
\usepackage{makecell}
\usepackage{environ}
\usepackage[utf8]{inputenc}
\usepackage{amsmath, amsthm, amssymb, amsfonts}
\usepackage[font=small, labelfont=bf]{caption}
\usepackage{amsmath, amsthm, amssymb, amsfonts}
\usepackage{multirow}
\usepackage{graphicx}
\usepackage{float}
\usepackage[numbers,sort&compress]{natbib}
\usepackage{jheppub}
\usepackage{enumerate}
\usepackage{hhline}
\usepackage{amsmath}
\usepackage{amsfonts}
\usepackage{amssymb}
\usepackage{graphicx}
\usepackage{cancel}
\usepackage{makecell}
\usepackage{tabularx}
\usepackage{cellspace}
\usepackage{longtable}
\usepackage{mathtools}
\usepackage{layouts}
\usepackage{empheq}
\usepackage{color}
\usepackage{hyperref}
\usepackage[capitalize,sort]{cleveref}

\crefname{figure}{Figure}{Figures}
\crefname{table}{Table}{Tables}

\usepackage{float}
\restylefloat{table}
\restylefloat{figure}

\newcommand{\nc}{\newcommand}
\nc{\beq}{\begin{equation}}
\nc{\eeq}{\end{equation}}
\nc{\be}{\begin{equation}}
\nc{\ee}{\end{equation}}
\nc{\bea}{\begin{eqnarray}}
\nc{\eea}{\end{eqnarray}}
\nc{\bi}{\begin{itemize}}
\nc{\ei}{\end{itemize}}
\nc{\ben}{\begin{enumerate}}
\nc{\een}{\end{enumerate}}

\def\ov{\overline}

\allowdisplaybreaks[2]
\numberwithin{equation}{section}

\allowdisplaybreaks[2]
\numberwithin{equation}{section}

\title{Classifying divisor topologies for string phenomenology}
\author[a,b]{Pramod Shukla}
\affiliation[a]{\small ICTP, Strada Costiera 11, Trieste 34151, Italy}
\affiliation[b]{\small Department of Physics, University of Allahabad, Prayagraj 211002, India.}
\emailAdd{pramodmaths@gmail.com}

\abstract{In this article we present a pheno-inspired classification for the divisor topologies of the {\it favorable} Calabi Yau (CY) threefolds with $1 \leq h^{1,1}(CY) \leq 5$ arising from the four-dimensional reflexive polytopes of the Kreuzer-Skarke database. Based on some empirical observations we conjecture that the topologies of the so-called coordinate divisors can be classified into two categories: (i). $\chi_{_h}(D) \geq 1$ with Hodge numbers given by $\{h^{0,0} = 1, \, h^{1,0} = 0, \, h^{2,0} = \chi_{_h}(D) -1, \, h^{1,1} = \chi(D) - 2 \chi_{_h}(D) \}$ and (ii). $\chi_{_h}(D) \leq 1$ with Hodge numbers given by $\{h^{0,0} = 1, \, h^{1,0} = 1 - \chi_{_h}(D), \, h^{2,0} = 0, \, h^{1,1} = \chi(D) + 2 - 4 \chi_{_h}(D)\}$, where $\chi_{_h}(D)$ denotes the Arithmetic genus while $\chi(D)$ denotes the Euler characteristic of the divisor $D$. We present the Hodge numbers of around 140000 coordinate divisors corresponding to all the CY threefolds with $1 \leq h^{1,1}(CY) \leq 5$ which corresponds to a total of nearly 16000 distinct CY geometries. Subsequently we argue that our conjecture can help in ``bypassing" the need of {\it cohomCalg} for computing Hodge numbers of coordinate divisors, and hence can be significantly useful for studying the divisor topologies of CY threefolds with higher $h^{1,1}$ for which {\it cohomCalg} gets too slow and sometimes even breaks as well. We also
demonstrate how these scanning results can be directly used for phenomenological model building, e.g. in estimating the $D3$-brane tadpole charge (under reflection involutions) which is a central ingredient for constructing explicit global models due to several different reasons/interests such as the de-Sitter uplifting through anti-$D3$ brane and (flat) flux vacua searches.
}

\begin{document}
\makeatletter
\let\old@fpheader\@fpheader
\renewcommand{\@fpheader}{\old@fpheader\hfill
arXiv:2205.05215}
\makeatother

\maketitle

\bigskip


\section{Introduction}
\label{sec_intro}

The study of Calabi Yau (CY) threefolds has been among the central tasks for constructing four-dimensional (semi)realistic models using superstring compactifications. In this context an enormous amount of effort has been made for constructing and classifying CY threefolds since more than three decades \cite{Green:1986ck, Candelas:1987kf, Green:1987cr, Candelas:1993dm, Batyrev:1993oya,Candelas:1994hw,Hosono:1994ax,Kreuzer:2000xy,Gray:2013mja}. The study of CY threefold can be broadly presented in two classes: 
\begin{itemize}
\item
Complete Intersection CY threefolds (CICYs): These are realized as multi-hypersurfaces in the product of projective spaces \cite{Green:1986ck}. A classification with complete list of 7890 such CICYs has been presented in  \cite{Candelas:1987kf} and subsequently their Hodge numbers $\{h^{1,1}, h^{2,1}\}$ have been computed in \cite{Green:1987cr}. This list of CICYs has been studied for further insights and classifications from time to time, e.g. the underlying fibration structures of CICYs have been presented in \cite{Gray:2014fla,Anderson:2017aux}. Such CICY threefolds have been heavily used for constructing local MSMS-like models (e.g. see \cite{Anderson:2011ns, Anderson:2012yf, Anderson:2013xka}), though other phenomenological issues such as moduli stabilization, inflation etc. have been mostly untouched using CICYs. However some recent works in this direction have been initiated, e.g. see \cite{Anderson:2010mh,Anderson:2011cza} for Heterotic moduli stabilization, and \cite{Bobkov:2010rf,Carta:2021sms,Carta:2021uwv, Carta:2022web,Carta:2022oex} for moduli stabilization in type IIB setups.  In addition, a complete list of explicit classification of CICY orientifolds with non-trivial $(1,1)$-cohomology has been also presented in \cite{Carta:2020ohw}. Moreover, a complete list of CICY fourfolds has been also presented in the meantime \cite{Gray:2013mja}.

\item
Toric Hypersurface CY threefolds (THCYs): Another class of CY threefolds consists of those which are realized as hypersurfaces in toric varieties \cite{Batyrev:1993oya} and we call them as Toric Hypersurface CY threefolds (THCYs) in our discussion. Such CY threefolds arising from the four-dimensional reflexive polytopes have been classified in \cite{Kreuzer:2000xy} which is popularly known as Kreuzer-Skarke (KS) database. This set of CY threefolds has been heavily utilized for a set of issues which have remained complementary to those in the CICYs studies. These issues include moduli stabilization, de Sitter realization, inflationary embedding etc. Although detailed computation of phenomenologically relevant topological properties of some of the THCYs have been initiated in early nineties \cite{Candelas:1993dm, Candelas:1994hw,Hosono:1994ax}, we have witnessed a huge surge in systematically exploring such CYs since the Kreuzer-Skarke classification has been presented in \cite{Kreuzer:2000xy}, followed by some efficient tools/packages such as ``Package for analyzing lattice polytopes" (PALP) \cite{Kreuzer:2002uu} and its new offspring version were launched in \cite{Braun:2011ik}. Similar to CICY case, the fibration structure of the THCY threefolds have been analyzed in \cite{Huang:2018gpl,Huang:2018esr}.

A more phenomenologist friendly dataset for THCY threefolds has been presented rather recently in \cite{Altman:2014bfa}. This so-called Altman-Gray-He-Jejjala-Nelson (AGHJN) dataset is a fantastic collection. In fact it is equipped with CY threefolds with $1\leq h^{1,1}(CY) \leq 6$ and has mostly all the necessary information about the CY threefold which one needs to begin with (while constructing explicit models) e.g. GLSM data for defining CY threefold, Hodge numbers of the CY threefold, Chern classes, Triple intersection numbers, Mori/K\"ahler cone etc. This dataset can be subsequently used for pheno model building or any other exhaustive classification one aims to look at, say depending on the topological properties of divisors etc. For example, see attempts in \cite{Altman:2017vzk, Cicoli:2018tcq}. Moreover this AGHJN-dataset has been recently updated with odd-orientifold constructions for CY threefolds in \cite{Altman:2021pyc} as an extension of the previous work in \cite{Gao:2013pra}. In addition, further extension of CY orientifold dataset of \cite{Altman:2014bfa, Altman:2021pyc} with the inclusion of THCYs with $h^{1,1}(CY) = 7$ has been initiated in \cite{Gao:2021xbs}.

\end{itemize}

\noindent
Let us also mention that the CICY database of \cite{Candelas:1987kf,Anderson:2017aux} has been also refereed as ``pCICY" database as there are additional so-called ``generalized" CICYs referred as ``gCICY" \cite{Anderson:2015iia} along with another possible dataset of the Toric CICY refereed as ``tCICY" ; a couple of examples of tCICYs have been presented in \cite{Cicoli:2021dhg}.

Which CY threefold is better or  more suitable for realistic string model building has been one of the prime questions that still remains (probably) too far from getting an answer up to a satisfactory extent. However one can always make a classification for the known lists of CY threefolds to suggest that ``some" class of CY threefold can be more useful for certain purposes as compared to the other ones. In this regard, the study of divisor topologies  plays a crucial role due to a series of reasons; the most central one being the fact that most of the (known) scalar potential contributions needed/used for moduli stabilization purpose are controlled/dictated by a set of divisor topologies. 

In the context of exploring the divisor topologies, the upgraded version of PALP \cite{Kreuzer:2002uu, Braun:2011ik} with \texttt{mori.x} module combined with software tools like SAGE \cite{sagemath} has turned out to be very helpful. Moreover, another very powerful tool \texttt{cohomCalg} has been launched during the same times \cite{Blumenhagen:2010pv,Blumenhagen:2011xn} which has made the computation of the Hodge numbers of various algebraic varieties quite efficient. These developments have enforced a significant amount of interest for ``global" model building in the type IIB CY orientifold compactifications leading to several interesting phenomenological models, e.g. see \cite{Blumenhagen:2008zz,Collinucci:2008sq,Cicoli:2011qg,Cicoli:2012vw,Blumenhagen:2012kz,Blumenhagen:2012ue,Cicoli:2013mpa,Cicoli:2013cha,Gao:2013rra,Cicoli:2016xae,Cicoli:2017axo,Cicoli:2017shd,AbdusSalam:2020ywo,Crino:2020qwk,Cicoli:2021dhg,Leontaris:2022rzj}. More recently, there has been a new tool, the so-called \href{https://cytools.liammcallistergroup.com/about/}{\texttt{CYTools}} which has been presented to compute the necessary topological data by directly using the reflexive polytopes of KS database \cite{Kreuzer:2002uu}. This package has been proven to be very efficient in triangulating the polytopes, including those corresponding to larger $h^{1,1}$, and it has been also demonstrated to perform divisor topology computations and its subsequent phenomenological implications have been initiated in \cite{Braun:2017nhi,Demirtas:2018akl,Demirtas:2020dbm}.

On these lines, knowing divisor topologies of the compactifying CY threefolds in a systematic way can be a useful step towards equipping one with ingredients needed for performing a classification for global model building. In fact, some of the direct points which motivates one for studying the divisor topologies can be argued via exploring suitable geometries which can help in inducing scalar potential terms, and hence can facilitate moduli stabilization process. These arguments can be elaborated along the following lines which can be taken as the main motivation for presenting a classification of the divisor topologies in the current work:

\begin{itemize}

\item
{\bf Non-perturbative superpotential:} In order to generate non-perturbative superpotential contributions which are central ingredients for K\"ahler moduli stabilization schemes, e.g. \cite{Kachru:2003aw,Balasubramanian:2005zx}, one needs suitable four-cycles with unit Arithmetic genus \cite{Witten:1996bn}, (see \cite{Blumenhagen:2010ja} also for zero-mode analysis). In this regard, rigid divisors (in particular the so-called del-Pezzo surfaces) are of crucial importance, and have received significant attention resulting in several classifications from time to time, which differ by the choice of input data collection, e.g. see \cite{Cicoli:2011it,Cicoli:2018tcq,Cicoli:2021dhg}.

In this context, there has been another divisor topology, namely the so-called ``Wilson divisor" which is necessary for realizing poly-instanton corrections \cite{Blumenhagen:2012kz}. These contributions appear as exponential corrections on top of the usual $E3$-instanton corrections leading to a schematic form of the superpotential given as $e^{-T_s + e^{-T_w}}$, where $T_s$ corresponds to the complexified four-cycle volume wrapping the $E3$-instanton while $T_w$ corresponds to the complexified four-cycle volume of the Wilson divisor. Inclusion of these effects generate sub-leading contributions for K\"ahler moduli stabilization which can help in driving inflation as well \cite{Blumenhagen:2012ue, Gao:2013hn}.

\item
{\bf Swiss-cheese structure:}
The CY threefolds with a particular divisors topology are found to be central in realizing the so-called LARGE Volume Scenario (LVS) scheme of moduli stabilization \cite{Balasubramanian:2005zx}. These divisor topologies are the so-called ``diagonal" del-Pezzo divisors in the sense that they can be shrinked to a point-like singularity by squeezing along a single direction \cite{Cicoli:2011it,Cicoli:2018tcq,Cicoli:2021dhg}.
 
\item
{\bf $K3$-fibred CY threefolds:}
The CY threefolds with $K3$-fibration exhibits some peculiar properties which have interesting phenomenological implications \cite{Cicoli:2008gp,Cicoli:2011it, Cicoli:2016xae, Cicoli:2017axo}.

\item
{\bf String-loop corrections:}
Although explicit results for string loop effects are known from the toroidal computations \cite{Berg:2005ja,Berg:2007wt}, their various insights can be extrapolated for the models based on (orientifolds of) CY threefolds as well. As argued in \cite{Berg:2007wt,Cicoli:2007xp}, the two classes of string-loop effects (known as KK-type and Winding-type) are induced with some very particular kind of brane settings \footnote{In fact a field theoretic argument for the existence of the so-called Winding loop corrections has been already proposed in \cite{vonGersdorff:2005bf}. The underlying argument for such corrections has been independent of the specific choices of brane-setting and can be argued to be extended to the CY orientifold based models beyond using the torus orientifolds.}. For example, KK-type correction needs the presence of non-intersecting stacks of $D7/O7$ and $O3$ planes while Winding-type effect needs intersecting stacks of $D7/O7$ configurations which intersect at some non-contractible two-cycles. However these requirements have been further revisited recently in \cite{Gao:2022uop} where it has been found that Winding-type corrections can appear more generically than what is expected from the very specific brane-setting arguments of \cite{Berg:2005ja,Berg:2007wt}, something which has been anticipated from \cite{vonGersdorff:2005bf}. 

In addition, there is a different class of string-loop effect; the so-called logarithmic loop-corrections \cite{Antoniadis:2018hqy,Antoniadis:2019rkh} for which three stacks of $D7$-branes with non-trivially specific intersection loci have been realized with the appropriate global construction in \cite{Leontaris:2022rzj}. 

\item
{\bf Higher derivative $F^4$-corrections:}
In the context of higher derivative effects, there are some terms which are beyond the two-derivative contributions via K\"ahler and superpotential, and appear directly to the scalar potential at $F^4$-order as proposed in \cite{Ciupke:2015msa}. It has been found that a topological quantity defined as $\Pi(D) = \int_{CY} c_2(CY) \wedge \hat{D}$ turns out to be of central importance. Here $c_2(CY)$ is the second Chern-class of the CY threefold and $\hat{D}$ denotes the dual $(1,1)$ class corresponding to the divisor $D$ of the CY threefold.

\item
{\bf Perturbatively flat flux vacua (PFFVs):} It has been recently proposed that an exponentially small value of the so-called Gukov-Vafa-Witten flux superpotential $|W_0|$ can be naturally attained in a dynamical way \cite{Demirtas:2019sip}; something which remains the main requirement of the KKLT scheme \cite{Kachru:2003aw} of the (K\"ahler) moduli stabilization. Applying this recipe for all the THCY geometries of $h^{1,1}(CY) = 2$ it has been observed in \cite{Carta:2021kpk} that the $K3$-fibred CY threefolds have significantly large number of (physical) PFFVs as compared to the so-called swiss-cheese CYs as well as those which are neither swiss-cheese nor $K3$-fibred.

In this context, let us also mention that the divisor topologies of all the pCICYs have been recently computed and classified in \cite{Carta:2022web} which have subsequently helped in classifying the PFFVs in \cite{Carta:2022oex} where it has been observed that pCICYs of $K3$-fibred type have a larger number of PFFV as compared to those which are not $K3$-fibred, and thus verifying the claims/observations of \cite{Carta:2021kpk}.

\item
{\bf Large tadpole charge and $\ov{D3}$ uplifting:}
Recently there has been a concrete global proposal about using $\ov{D3}$-brane uplifting to realize de-Sitter vacua within LVS framework \cite{Crino:2020qwk}. Although some strong phenomenological challenges have been posed to this class of models in \cite{Junghans:2022exo,Gao:2022fdi}, it is still very much desired to find the appropriate CY geometries with suitable orientifold constructions in terms of satisfying at least the topological ingredients needed in the proposal, while postponing the list of phenomenological constraints of \cite{Junghans:2022exo,Gao:2022fdi} to be addressed/settled in future. Having said that, the central ingredients of interest for such a de-Sitter uplifting proposal is the need of a CY orientifold which results in ``large enough" $D3$-tadpole charge along with at least two coincident $O3$-planes in the fixed point set of the chosen involution. Even though finding an exhaustive list of all possible involutions can be challenging, nevertheless focusing on the so-called reflection involutions (defined as $\sigma_i: x_i \to - x_i$ where $x_i$ denotes the toric coordinates defining the CY threefold), one can indeed test some of these topological conditions to reduce the huge set of suitable CY threefolds for this particular purpose.

\end{itemize}

\noindent
So far, we have briefly reviewed the so-called ``global" requirements which are at the core of those respective phenomenological models, and this is something well motivating for the study of divisor topologies of the CY threefolds which we plan to present in this work, with a broad classification depending on their properties. Let us note that although there have been some occasions where the computations/results of divisor topologies have been implicitly used, e.g. in attempts like constructing CY orientifolds with exchange involutions, one needs to identify the so-called ``non-trivially identical divisors" (NIDs) and subsequently divisors topologies have been inherently computed/used in such analysis, e.g. see \cite{Gao:2013pra} using $1\leq h^{1,1}(CY) \leq 4$, and \cite{Altman:2021pyc} using $1\leq h^{1,1}(CY) \leq 6$ from the database \cite{Altman:2014bfa}, however, a comprehensive/exhaustive analysis and classification of all the toric divisor topologies have not been reported or made available for readers so far, and we aim to (partially) fill this gap by taking a phenomenologists' eye on the subject. On these lines, we will also present a conjecture for the coordinate divisors of the favorable THCY threefolds which can help in computing the Hodge numbers for larger $h^{1,1}(CY)$. This can be of great importance in the sense that \texttt{cohomCalg} not only gets slow while computing the Hodge numbers of the divisors of CY threefolds with $h^{1,1} \geq 6$ but it breaks also (or shows no result) for some occasions, and therefore bypassing it by replacing with some alternative way could indeed be useful.

While stating the above let us also emphasize here the fact that the study of divisor topologies which we will be focusing on in this work is limited to the so-called `coordinate divisors' only. Given the fact that even for the case of a favorable CY threefold, generically there can be many more effective divisors on the CY as compared to those of its Ambient space, and therefore considering only the `coordinate divisors' can be a huge limitation for the purpose of an exhaustive classification of ``all" the interesting divisor topologies which could be relevant for phenomenological model building\footnote{We thank the referee for her/his useful comments and suggestions along these lines.}, however we also note that even such a limited subclass of divisors presents many interesting topologies useful for the generation of several sub-leading corrections to the effective scalar potential. For these reasons, {\it let us state that the mention of divisor topology throughout this work should be considered to be limited to the `coordinate divisors" only. }

The article is organized as follows: Section \ref{sec_div-topo} is devoted to the computation of divisor topologies for all the THCYs having $1\leq h^{1,1} \leq 5$ along with making some observations leading to a conjecture to bypass the need of \texttt{cohomCalg} for computing the Hodge number of divisors corresponding to CY threefolds with larger $h^{1,1}$. In section \ref{sec_classifications} we present a classification of all the distinct divisor topologies based on their phenomenological interests/relevance. Section \ref{sec_applications} presents the analysis about the Fixed-point set and some statistics about the possible orientifold constructions using reflection involutions, along with a couple of  global constructions for illustration purpose. Finally we present a summary with conclusions in section \ref{sec_conclusions}. In addition, given that every reader does not necessarily feel encouraged to download the huge dataset visiting an external website, we present a compact list of distinct 565 divisor topologies in appendix \ref{sec_appendix1} while attaching the complete list of 139740 divisor topologies in an ancillary. 

\vskip0.2cm
\noindent
{\bf Note:} While this work was in progress, \cite{Crino:2022zjk} appeared on arXiv which turns out to have overlapping interests with the current work. The model building aspects along with some detailed phenomenological applications of the current work are aimed to be presented in companion papers \cite{Cicoli:2022abc,AbdusSalam:2022krp}. For interested readers, some parts of these aspects can be found at the \href{https://www.youtube.com/watch?v=nDJy7QnNAZ8}{online forum} of the ``\href{https://sites.google.com/view/string-pheno-seminars/home}{Seminar series on string phenomenology}".


\section{Divisor topologies of CY threefolds}
\label{sec_div-topo}

In this section, we analyze of the divisor topologies corresponding to a subset of the CY threefold geometries arising from the four dimensional reflexive polytopes of the Kreuzer-Skarke database \cite{Kreuzer:2000xy}, which motivates for some pheno-inspired classification.

\subsection{Methodology}
Our current focus is limited to analyze the divisor topologies of the favorable CY geometries with $1 \leq h^{1,1}(CY) \leq 5$, arising from the triangulation of reflexive polytopes of \cite{Kreuzer:2000xy}, and for that purpose we use the topological data as available from the AGHJN database \cite{Altman:2014bfa}. In this regard, we present the estimates for possible number of CY geometries and their divisors in Table \ref{tab_number-of-space-and-divisors}, which we plan to study in our scan. 
\noindent
\begin{table}[h!]
\centering
\begin{tabular}{|c||c|c|c|c||c|c|} 
\hline
 & & & & & & \\
$h^{1,1}$ & polytope & Triang & Geom & fav-Geom & fav-Geom$^\ast$ & divisors of \\
 & & & & & & fav-Geom$^\ast$ \\
 \hline
 \hline
 1 & 5 & 5 & 5 & 5 & 4 & 20  \\
 2 & 36 & 48  & 39  & 39 & 37 & 222 \\
 3 & 244 & 569 & 306 & 305 & 300 & 2100 \\
 4 & 1197 & 5398 & 2014 & 2000 & 1994 & 15952 \\
 5 & 4990 & 57132 & 13635 & 13494 & 13494 & 121446 \\
\hline
Total \# & 6472 & 63152 & 15999 & 15843 & 15829 & 139740 \\ 
 \hline
\end{tabular}
\caption{Number of distinct CY geometries and their corresponding coordinate divisors for $1\leq h^{1,1}(CY) \leq 5$. Here fav-Geom$^\ast$ denotes those favourable CY geometries which have a trivial fundamental group, and the last column presents the total number of divisor topologies corresponding to a given $h^{1,1}(CY)$. These data can be read-off from \cite{Altman:2014bfa}.}
\label{tab_number-of-space-and-divisors}
\end{table}

Continuing with our previous methodology of classifications for del-Pezzo divisor topologies in \cite{Cicoli:2018tcq,Cicoli:2021dhg}, we take the following steps:
\begin{itemize}
\item 
As we have previously declared, we focus only on looking at the topology of the so-called `coordinate divisors' $D_i$ which are defined through setting the toric coordinates to zero, i.e. $x_i = 0$. Though we understand that such a choice can heavily limit the number of useful divisors, this can still produce a nice collection of interesting divisor topologies leading to a pheno-inspired classification as elaborated in the introduction.

Moreover it is also quite possible that the divisor combinations constructed out of the coordinate divisors may turn out to be non-smooth, e.g. it is quite frequent when a shrinkable del-Pezzo is involved in the combination, and hence becomes of less importance for phenomenological model building. Nevertheless, on technical grounds, it is anyway hard to make an exhaustive claim of any statistics for the generic set of divisors by considering all the various possible combinations of the coordinate divisors, and therefore our approach can be considered to be a pragmatic one. 

\item 
We consider the so-called ``favourable" triangulations and favourable geometries in the sense that all the toric divisors of the CY threefolds descends from the Ambient fourfold. 
We also exclude a couple of more examples which have non-trivial fundamental groups. We find this necessary to have a perfect match with the divisor Hodge diamonds obtained from the \texttt{cohomCalg} package \cite{Blumenhagen:2010pv,Blumenhagen:2011xn} and from what we call as ``direct" method using a couple of formulae 
as we will discuss later. 
However such spaces are very limited in number, in fact only 14 out of 15843 CY geometries as seen from Table \ref{tab_number-of-space-and-divisors}.

\end{itemize}

\noindent
Aiming for exploring their utilities for phenomenological purposes, we will study and classify the topologies of 139740 number of coordinate divisors corresponding to a total number of 15829  CY geometries as presented in Table \ref{tab_number-of-space-and-divisors}.

\subsection{Computation of Hodge numbers}
For a generic divisor ($D$) of the CY threefold $X$, there are only four independent Hodge numbers in the Hodge diamond, which are denoted as $h^{0,0}, h^{1,0}, h^{2,0}$ and $h^{1,1}$. Two of these can be computed from the Euler characteristics $\chi(D)$ and the Arithmetic genus $\chi_{_h}(D)$ of the divisor $D$ which can be computed via knowing the second Chern class of the $X$ along with the classical triple intersection numbers $\kappa_{ijk}$. The explicit formulae connecting these quantities can be given as below (e.g. see \cite{Blumenhagen:2008zz,Collinucci:2008sq, Bobkov:2010rf,Cicoli:2016xae}),
\begin{equation}
\begin{split}
\chi(D) &= 2 h^{0,0} - 4 h^{1,0} + 2 h^{2,0} + h^{1,1}= \int_{X} \left(\hat{D} \wedge \hat{D} \wedge \hat{D} + c_2(X) \wedge \hat{D} \right), \, \\
 \chi_{_h}(D) &= h^{0,0} - h^{1,0} + h^{2,0} = \frac{1}{12} \int_{X}\left(2\, \hat{D} \wedge \hat{D} \wedge \hat{D} + c_2(X) \wedge \hat{D} \right). \, \end{split}
\label{eq:chi-chih}
\end{equation}
Here we denote the second Chern class of the CY threefolds as $c_2(X)$ and, once again, $\hat{D}$ denotes the 2-forms dual to the divisor class. Thus, after knowing $\chi(D)$ and $\chi_{_h}(D)$ of a divisor using the second Chern class and the triple intersection numbers, one is practically left with computing only two out of the four Hodge numbers.

To begin with, we have used the \texttt{HodgeDiamond} module of the \texttt{cohomCalg} package \cite{Blumenhagen:2010pv,Blumenhagen:2011xn} to compute the divisor topologies for each of the so-called ``coordinate divisors" of all the favourable CY geometries with $1 \leq h^{1,1} \leq 4$ and partially for $h^{1,1} = 5$. However, we have subsequently observed that the program gets slower and sometimes does not give output for some examples of $h^{1,1} = 5$. Given that we effectively need to know only two Hodge numbers after having $\chi(D)$ and $\chi_{_h}(D)$, we subsequently used a different module \texttt{Lambda0CotangentBundle} of the \texttt{cohomCalg} package which computes the Hodge numbers $h^{0,0}, h^{1,0}$ and $h^{2,0}$ of a given divisor. Although this module is faster as compared to \texttt{HodgeDiamond}, it is also expected to be quite slow for larger $h^{1,1}$ of the CY threefolds. So we ask a question whether there is a way to somehow bypass the need of \texttt{cohomCalg} as we will argue with a conjecture we make based on the observations we make from our empirical results.

Basically, the input data from \cite{Altman:2014bfa} which we mainly use for our analysis consists of the followings:
\begin{itemize}
\item GLSM charges
\item Stanley-Reisner (SR) ideal
\item Second Chern-class $c_2(CY)$ of the CY threefold
\item Intersection tensor $\kappa_{ijk}$
\item Fundamental group
\end{itemize}

\noindent
Let us mention that while one needs GLSM charges and SR ideal for computing divisor topologies using  \texttt{cohomCalg}, one can compute two (out of the four) Hodge numbers using the second Chern class $c_2(CY)$ and the triple intersection numbers $\kappa_{ijk}$ in Eq.~(\ref{eq:chi-chih}). The need of including Fundamental group appears in order to remove the slight mismatch between the two sets of results. In fact, given that \texttt{cohomCalg} computes divisor topologies merely by using the GLSM charges and SR ideal, it does not distinguish among some geometries when they have the same GLSM/SR combinations but difference in their (non-trivial) fundamental groups. For example the first two examples in the KS database \cite{Kreuzer:2000xy} have the same set of GLSM charges and SR ideals given as WCP$^4[1,1,1,1,1]$ with SR:$\{x_1 x_2 x_3 x_4 x_5\}$ as seen from \cite{Altman:2014bfa}, and subsequently \texttt{cohomCalg} gives Hodge numbers which does not match with those computed using Eq.~(\ref{eq:chi-chih}) for the first example which has non-trivial fundamental group. However, this mismatch in two sets of results in very mild in the sense that there are not too many favorable CY geometries with non-trivial Fundamental group. In fact as seen from Table \ref{tab_number-of-space-and-divisors}, there are only 14 such cases for a total of 15843 CY geometries corresponding to $1 \leq h^{1,1}(CY) \leq 5$. 

\subsection{A conjecture to bypass the need of using cohomCalg}
Given that we are working with the favorable CY threefolds, it could be anticipated that divisors are smooth, in particular connected and hence $h^{0,0}(D) = 1$ is quite expected, and also turns out to be true for all the coordinate divisors in our explicit computation. Such observations have been made at multiple occasions, e.g. in the context of Kreuzer-Skarke database \cite{Cicoli:2018tcq,Cicoli:2021dhg} as well the recent classification of divisor topologies of the complete intersection CY threefolds \cite{Carta:2022web}. Further, from our explicit computations of divisor topologies of the CY geometries with $1\leq h^{1,1} \leq 4$ which correspond to scanning through around 18000 divisor topologies of more than 2300 CY spaces, we observe that all the divisors fall in one of the following three classes of the Hodge numbers:
\bea
\label{eq:conjecture}
& (i). & \quad h^{0,0}(D) = 1 , \quad h^{1,0}(D) = 0 , \quad h^{2,0}(D) = 0; \, \\
& (ii). & \quad h^{0,0}(D) = 1 , \quad h^{1,0}(D) = 0 , \quad h^{2,0}(D) \neq 0; \, \nonumber\\
& (iii). & \quad h^{0,0}(D) = 1 , \quad h^{1,0}(D) \neq 0 , \quad h^{2,0}(D) = 0. \, \nonumber
\eea
In other words, this observation means that we do not find $h^{1,0}(D)$ and $h^{2,0}(D)$ both non-zero for any of the divisors. Based on these findings, we conjecture that Eq. \eqref{eq:conjecture} should be true for all the coordinate divisors of all the favourable CY threefolds of the Kreuzer-Skarke database \cite{Kreuzer:2000xy}. Some arguments in support of this conjecture can be found in \cite{Braun:2016igl,Braun:2017nhi} where the Hodge numbers of a square-free divisor of a THCY threefold have been argued to be determined in terms of computing a set of combinatorial data.

Let us also note that both of these Hodge numbers are observed to be simultaneously  zero, and therefore one may have a classification of divisor topologies into three classes:

\begin{itemize}
\item{{\bf Rigid divisors:} Keeping in mind the conjecture (\ref{eq:conjecture}), these divisors which we will denote as $R$ can be described as four-cycles with unit Arithmetic genus, i.e. $\chi_{_h}(R) = 1$.
\bea
h^{0,0}(R) = 1 , \quad h^{1,0}(R) = 0 , \quad h^{2,0}(R) = 0, \quad h^{1,1}(R) = \chi(R)-2.
\eea
Some particular examples of such surfaces can be considered as ${\mathbb P}^2$ and the del-Pezzo surfaces dP$_n$ for $ 1 \leq n \leq 8$. Such surfaces arise by including eight generic blow-up points in ${\mathbb P}^2$, and the corresponding topological quantities are given as $\{\chi_{_h} (dP_n) =1, \, \chi(dP_n) = n+3\}$ where $n = 0$ corresponds to ${\mathbb P}^2$. Because of satisfying the Witten's unit Arithmetic genus condition \cite{Witten:1996bn} which is necessary for contributing to the non-perturbative superpotential, such  divisors have been of great interest in phenomenological model building and have attracted tremendous amount of interests in recent years, e.g. \cite{Blumenhagen:2010ja,Cicoli:2011it,Cicoli:2011qg,Gao:2013pra,Cicoli:2013cha} for initial attempts of concrete global model building.
}

\item{{\bf Non-rigid divisors:} These divisors which we will denote as $K$ can be described as divisors with non-zero deformations in the CY threefold, i.e. $h^{2,0}(K) \neq 0$ which according to our conjecture (\ref{eq:conjecture}) means that $h^{1,0}(K) = 0$ and hence leads to $\chi_{_h}(K) > 1$. In fact one has the following Hodge numbers to characterise such topologies,
\bea
h^{0,0}(K) = 1 , \quad h^{1,0}(K) = 0 , \quad h^{2,0}(K) = \chi_{_h}-1, \quad h^{1,1}(K) =\chi - 2 \chi_{_h}.
\eea
One particular example of such divisors can be considered as the $K3$  surfaces for which $\chi_{_h}(K3) = 2$ and $\chi(K3) = 24$. Another example one would like to consider can be the so-called `special deformation' divisors (${\rm SD}$) which appear very often in the overall scan as we will see later on. One class of such ${\rm SD}$ divisor is described by $\chi_{_h}({\rm SD}) = 3, \, \chi({\rm SD}) = 36$. Let us note that a prescription to ``rigidify" the non-rigid divisors so that they could contribute to the non-perturbative superpotential has been presented in \cite{Bianchi:2011qh, Louis:2012nb}.
}

\item{{\bf Wilson divisors:} These divisors which we will denote as $W$ can be described as rigid but non-simply connected divisors i.e. having zero deformations in the CY threefold, i.e. $h^{2,0}(W) = 0$ but $h^{1,0}(W) \neq 0$. Subsequently, according to our conjecture (\ref{eq:conjecture}) leads to $\chi_{_h}(W) \leq 0$. In fact one has the following Hodge number to characterize such topologies,
\bea
& & \hskip-0.75cm h^{0,0}(W) = 1 , \quad h^{1,0}(W) = 1 - \chi_{_h}, \quad h^{2,0}(W) = 0, \quad h^{1,1}(W) =\chi + 2 - 4 \chi_{_h}.
\eea
One particular example of such divisors can be considered as the so-called Wilson surfaces used for poly-instanton effects in \cite{Blumenhagen:2012kz} which are described as $\chi_{_h}(W) = 0$ and $\chi(W) = 0$. Having both $\chi$ and $\chi_{_h}$ zero makes this surface quite peculiar in many sense, adding to its utility for phenomenological model building \cite{Blumenhagen:2012kz}.
}
\end{itemize} 

\noindent
Let us recall that such a choice of considering three classes for divisor topologies was initiated in \cite{Gao:2013pra} in order to seek for the so-called ``non-trivially identical divisors" (NIDs) which are relevant for constructing CY orientifolds with non-trivial odd $(1,1)$-cohomology. However now we observe that it is generically true (in the sense the at least one of the two Hodge numbers $\{h^{1,0}, h^{2,0}\}$ always vanishes for the coordinate divisors exhausting all the possibilities, at least we are not aware of any contradiction through the known examples.

Our current detailed analysis shows that there are a total of 565 distinct divisor topologies corresponding to a total of 139740 coordinate divisors arising from a total of 15829 CY geometries with $ 1\leq h^{1,1}(CY) \leq 5$ as mentioned in Table \ref{tab_number-of-space-and-divisors}. These 565 distinct topologies are broadly classified in three categories as presented in Table \ref{tab_divisor-topologies}. 

\noindent
\begin{table}[H]
\centering
\begin{tabular}{|c||c|c|c|c|} 
\hline
Type & Divisor topology   & Distinct topology & Frequency & {$h^{1,1}$(CY)} \\
& $\{h^{0,0}, h^{1,0}, h^{2,0}, h^{1,1}\}$ & (out of 565) & (out of 139740) &  \\
\hline
$R_n$ & $\{1, 0, 0, n\}$  & 63 & 76839 & 2-5  \\
$K_n^m$ & $\{1, 0, m, n\}$  & 395 & 55972 & 1-5  \\
$W_n^m$ & $\{1, m, 0, n\}$  & 107 & 6929 & 2-5 \\
\hline
\end{tabular}
\caption{Distinct divisor topologies for favorable CY geometries and their frequencies of appearance. Also, $m$ and $n$ are positive integers. Further details for each of the three classes are collected in the full list of 565 topologies presented in Table \ref{tab_topo-list} of the Appendix \ref{sec_appendix1}.}
\label{tab_divisor-topologies}
\end{table}

\noindent
Let us also mention that the attached ancillary file can be read off in the following manner:
\bea
& & \hskip-1.5cm {\rm Data: = \{\{space \, \,\#, \, \, polytope \, \, Id, \, \, GLSM \, \, charge \, \, vectors, \, \, SR \,\, ideal, \, \, Divisor \, \, topology\}\}.}
\eea
For example, there are a total of 15829 CY threefolds mentioned as ``space \#" which are numbered from 1 to 15829 and respectively correspond to polytope Id 2 to 6472 of the KS database. To illustrate the data reading, let us mention the very first  example which corresponds to the Quintic threefold and the corresponding data for divisor topology is collected as below,
\bea
& & \{1, \{2\}, \{\{1\}, \{1\}, \{1\}, \{1\}, \{1\}\}, \{x_1 x_2 x_3 x_4 x_5\}, \\
& & \{\{1, \{1, 0, 4, 45\}\}, \{2, \{1, 0, 4, 45\}\}, \{3, \{1, 0, 4, 45\}\}, \{4, \{1, 0, 4, 45\}\}, \{5, \{1, 0, 4, 45\}\}\}\}.\nonumber
\eea
The first entry ``1" is the space \# in our collection, second entry $\{2\}$ corresponds to the polytope Id in the KS database, $\{\{1\}, \{1\}, \{1\}, \{1\}, \{1\}\}$ denotes the GLSM charge vectors corresponding to the five coordinate divisors while $\{x_1 x_2 x_3 x_4 x_5\}$ denotes the SR ideal. The second line shows that all the five coordinate divisors have the same Hodge numbers described by $\{h^{0,0}, h^{0,1}, h^{0,2}, h^{1,1}\} = \{1, 0, 4, 45\}$. 

Similarly, the topological data for the famous swiss-cheese CY threefold realized as a degree-18 hypersurface in WCP$^4[1,1,1,6,9]$ are collected in the following manner:
\bea
& & \hskip-1cm \{41, \{41\}, \{\{0, 1\}, \{0, 1\}, \{0, 1\}, \{2, 6\}, \{3, 9\}, \{1, 0\}\}, \{x_1x_2x_3,\, x_4x_5x_6\}, \\
& & \hskip-1cm \{\{1, \{1, 0, 2, 30\}\}, \{2, \{1, 0, 2, 30\}\}, \{3, \{1, 0, 2, 30\}\}, \{4, \{1, 0, 28, 218\}\}, \nonumber\\
& & \hskip-1cm \{5, \{1, 0, 65, 417\}\}, \{6, \{1, 0, 0, 1\}\}\}\},\nonumber
\eea
which shows that the sixth divisor having the Hodge numbers $\{h^{0,0}, h^{0,1}, h^{0,2}, h^{1,1}\} = \{1, 0, 0, 1\}$ corresponds to a ${\mathbb P}^2$ surface which appears as a result of resolving the singularity.

As a side remark, let us note that divisors with unit Arithmetic genus have a significant attraction due to being useful for inducing non-pertubative effects, and this has been the reason why we present them as a separate class, otherwise the three classes can be clubbed into two possibilities only, with the $\chi_{_h} =1$ case lying at the interface:
\bea
\label{eq:divisor-topology}
& & {\cal C}_1 \equiv
\begin{tabular}{ccccc}
    & & 1 & & \\
   & 0 & & 0 & \\
$(\chi_{_h}-1)$ \quad & & $(\chi - 2 \chi_{_h})$ \quad & & \quad $(\chi_{_h} -1)$ \\
   & 0 & & 0 & \\
    & & 1 & & \\
\end{tabular}, \qquad \chi_{_h} \geq 1;\\
& & \nonumber\\
& & {\cal C}_2 \equiv
\begin{tabular}{ccccc}
    & & 1 & & \\
   & $(1-\chi_{_h})$ & & $(1-\chi_{_h})$ & \\
0 \quad & & $(\chi + 2 - 4 \chi_{_h})$ \quad & & \quad 0 \\
   & $(1-\chi_{_h})$ & & $(1-\chi_{_h})$ & \\
    & & 1 & & \\
\end{tabular}, \qquad \chi_{_h} \leq 1. \nonumber
\eea
In the context of complete intersection CY threefolds (denoted as pCICYs which are) realised as hypersurfaces in the product of projective spaces, it has been observed in \cite{Carta:2022web} that the coordinate divisors of favourable pCICYs are always simply connected and subsequently all the divisors fall in the category ${\cal C}_1$ of (\ref{eq:divisor-topology}). Moreover, in the context of favourable pCICYs it has been further observed that there are only 11 distinct divisor topologies arising from a total number of 57885 coordinate divisors corresponding to 7820 pCICY threefolds \cite{Carta:2022web}. Unlike the case of very limited number of distinct divisor topologies for pCICYs, we find that there are several interesting divisor topologies for the CY threefolds of the KS database, which we plan to classify in a systematic way in the upcoming section.

\subsection{Diagonal divisors}
Using the relation in Eq. \eqref{eq:chi-chih}, it is easy to convince that divisors with vanishing self triple intersection $\int_{\rm CY} \hat{D} \wedge \hat{D} \wedge \hat{D} = 0$ satisfy the following relation,
\begin{equation}
\label{eq:Dcube=0}
h^{1,1}(D) = 10 h^{0,0}(D) - 8 h^{1,0}(D) +10 h^{2,0}(D). \,
\end{equation}
For smooth divisor $D$ which is also a connected divisor $h^{0,0}(D) = 1$, and assuming that $h^{1,1}(D) > 0$ one can classify the possibilities for divisors of vanishing cubic self-intersections. In the light of our conjecture (\ref{eq:conjecture}), we find that there are just a few possibilities for rigid coordinate divisors, i.e. those which have $h^{2,0}(D) = 0$, to satisfy the condition (\ref{eq:Dcube=0}). However for non-rigid cases, there can be a whole class of such divisor topologies. Some of the well known topologies which turns out to be satisfying the vanishing cubic condition in Eq.~(\ref{eq:Dcube=0}) are the followings:
\bea
\label{eq:Dcube=0-examples}
& & \hskip-1.5cm (i). \quad
\begin{tabular}{ccccc}
    & & 1 & & \\
   & $0$ & & $0$ & \\
0  & & $10$ & & 0 \\
   & $0$ & & $0$ & \\
    & & 1 & & \\
\end{tabular} \in R,  \, \qquad\qquad (ii). \quad 
\begin{tabular}{ccccc}
    & & 1 & & \\
   & $1$ & & $1$ & \\
0  & & $2$ & & 0 \\
   & $1$ & & $1$ & \\
    & & 1 & & \\
\end{tabular} \in W,\\
& &  \hskip-1.5cm (iii). \quad 
\begin{tabular}{ccccc}
    & & 1 & & \\
   & $0$ & & $0$ & \\
$n$ \quad & & $10(n+1)$ & & $n$ \\
   & $0$ & & $0$ & \\
    & & 1 & & \\
\end{tabular} \in {\rm K}, \qquad \qquad K3 : \{n=1\},   \, \quad  {\rm SD} :
\{n \geq 2\}. \nonumber
\eea
Moreover, we note that although we do not encounter any divisor topology corresponding to ${\mathbb T}^4$ surface given that is does not fall in the two classes we find in our scan as mentioned in Eq.~(\ref{eq:divisor-topology}), nevertheless it is worth noting that it also satisfies the condition (\ref{eq:Dcube=0}) as 
\bea
& & \hskip-1.5cm {\mathbb T}^4 = \begin{tabular}{ccccc}
    & & 1 & & \\
   & $2$ & & $2$ & \\
1  & & $4$ & & 1 \\
   & $2$ & & $2$ & \\
    & & 1 & & \\
\end{tabular}.
\eea
Now we mention another class of divisors $D_i$ (which we denote as $D_{\rm diag}$ and) that satisfies the following so-called `diagonality' condition \cite{Cicoli:2011it, Cicoli:2018tcq},
\bea
\label{eq:diag-divisor}
& & \hskip-1.5cm D_{\rm diag}: \qquad \kappa_{iii} \, \, \kappa_{i j k } = \kappa_{ii j} \, \, \kappa_{ii k}\,; \, \qquad \qquad i \,\,\, {\rm fixed} \quad {\rm and} \quad \forall \, \, \, j, k.
\eea
The reason why we call this a `diagonality' condition is the fact that if this condition is satisfied, then for divisors with non-vanishing cubic self-intersections i.e. $\kappa_{iii} \neq 0$, there exist a basis of coordinate divisors such that the volume of each of the four-cycles $D_i$ (denoted as $\tau_i$) is a complete-square quantity in terms of the two-cycle volume moduli (denoted as $t^i$). This can be argued with the following relation,
\bea
\label{eq:kappa0}
& & \hskip-1cm \tau_i \equiv \frac{1}{2}\, \kappa_{ijk} t^j \, t^k = \frac{1}{2 \, \kappa_0}\, \kappa_{ii j} \, \kappa_{ii k} t^j \, t^k = \frac{1}{2 \, \kappa_0}\, \left(\kappa_{iij} \,t^j \, \right)^2\,, \quad \kappa_0 = \kappa_{iii}.
\eea
Although the diagonality condition in Eq. (\ref{eq:diag-divisor}) is mainly used/invoked for searching the so-called diagonal del-Pezzo divisors needed to establish the swiss-cheese structure in the CY volume, we extend this nomenclature to all the possible divisors which satisfy the condition (\ref{eq:diag-divisor}) and we call them as generic ``diagonal" divisors, which may not be necessarily a del-Pezzo divisor. 

It is interesting to note that $K3$ and ${\mathbb T}^4$ surfaces not only satisfy the vanishing cubic relation (\ref{eq:Dcube=0}) but also the diagonality relation (\ref{eq:diag-divisor}), though it does so trivially and therefore may not always lead to complete square form of the four-cycle volume (\ref{eq:kappa0}). The later follows from a theorem by Oguiso \cite{OGUISO:1993} (see \cite{Douglas:2003um} also) which translates into the fact that only linear pieces of $K3/{\mathbb T}^4$ divisor are allowed in the intersection polynomial, i.e. $\kappa_{iii} = 0 = \kappa_{iij}$ for all $j$ whenever a divisor $D_i$ appears as a $K3/{\mathbb T}^4$ divisor of a $K3/{\mathbb T}^4$-fibred CY threefold. However given that the diagonality condition (\ref{eq:diag-divisor}) is trivially satisfied for these two surfaces, one may focus on only those divisors for which cubic self-intersections are non-vanishing, i.e. $ \kappa_{iii} \neq 0$ in order to utilize a complete square from of the four-cycle volume (\ref{eq:kappa0}). We present the occurrence of diagonal divisors arising from our scan in Table \ref{tab_diagonal-topologies}.

\noindent
\begin{table}[!htp]
\centering
\begin{tabular}{|c|c||c|c||c|c|c|c|}
\hline
$h^{1,1}$ & \#(CY) & \#(CY)  & \#($D_{\rm diag}$) & \#(CY) & \#($D_{\rm diag}$) & \#(CY) & \#($D_{\rm diag}$)  \\ 
& geom$^\ast$ & with $D_{\rm diag}$ &  & with $D_{\rm diag}$ &  & with $D_{\rm diag}$ & \\ 
& & & & ($\kappa_0 \neq 0$) &  ($\kappa_0 \neq 0$) & ($\kappa_0 = 0$) &  ($\kappa_0 = 0$) \\ 
\hline
1 & 4 & 4 & 20  & 4 & 20 & 0 & 0 \\ 
2 & 37 & 31 & 63 & 21 & 43 & 10 & 20 \\ 
3 & 300 & 227 & 501 & 134 & 213 & 136 & 288 \\ 
4 & 1994 & 1486 & 2884 & 780 & 1150 & 888 & 1734 \\ 
5 & 13494 & 9305 & 15696 & 4207 & 5426 & 6088 & 10270 \\ 
\hline
\end{tabular}
\caption{Topologies for diagonal divisors and their frequencies of appearance.}
\label{tab_diagonal-topologies}
\end{table}

\noindent
In order to understand the statistics presented in Table \ref{tab_diagonal-topologies} let us note the following points:
\begin{itemize}
\item
For $h^{1,1}(CY) =1$, all the four CY geometries have a total of 20 diagonal divisors as diagonality condition (\ref{eq:diag-divisor}) is trivially satisfied.

\item
For $h^{1,1}(CY) =2$, there are a total of 31 CY geometries which have a total of 63 diagonal divisors. As we will elaborate later on, this is because of the fact that there are 10 $K3$-fibred CY geometries and 21 swiss-cheese like CY geometries having diagonal del-Pezzo divisors. Being a special case in the sense that there are only two divisors in the basis, and hence if one is diagonal the remaining one has to be diagonal, and therefore these 31 spaces have at least 62 diagonal divisors. However, it is observed that one example, namely the famous one defined as degree 18 hypersurface in WCP$^4[1,1,1,6,9]$ has three diagonal divisors; in fact apart from the blown-up ${\mathbb P}^2$ divisor, the remaining two diagonal divisors corresponding to the charges ``6" and ``9" are dependent in the resolved toric GLSM representation.

\item
For $h^{1,1}(CY) =3$, there are 136 $K3$-fibred CYgeometries and $132$ swiss-cheese CY geometries (e.g.~see \cite{Cicoli:2018tcq}) along with 2 special cases with diagonal divisors\footnote{These include $K3$-like divisors (in the sense of having the same Hodge numbers) in some non $K3$-fibred CY threefolds which subsequently end up behaving like strong swiss-cheese examples.} which sums up to 270, and given that there are 43 common geometries which have $K3$-fibration along with swiss-cheese structure, and this justifies the total number of CY geometries with diagonal divisors to be 227.

\end{itemize}

\subsection{Some peculiar classes of divisors}

Knowing the Hodge numbers of the coordinate divisors determines the following four topological quantities which remain at the core of global model building and can be subsequently utilized at various stages,
\bea
\label{eq:topo-via-hodgenumbers}
& & \chi_{_h}(D) = h^{0,0}(D) - h^{1,0}(D) + h^{2,0}(D), \, \\
& & \chi(D) = 2 h^{0,0}(D) - 4 h^{1,0}(D) + 2 h^{2,0}(D) + h^{1,1}(D),\, \nonumber\\
& & \kappa_0(D) = \int_{X} \hat{D} \wedge \hat{D} \wedge \hat{D} = 10 h^{0,0}(D) - 8 h^{1,0}(D) + 10 h^{2,0}(D) - h^{1,1}(D),\, \nonumber\\
& & \Pi(D) = \int_{X}  c_2(X) \wedge \hat{D} = -8 h^{0,0}(D) + 4 h^{1,0}(D) - 8 h^{2,0}(D) + 2 h^{1,1}(D).\, \nonumber
\eea
Using these four relations, one can invoke various special cases, some of which could corresponds to interesting topologies. For example, one can consider the following four cases:
\begin{itemize}
\item
$D_{\chi_{h}}$: Divisors with vanishing Arithmetic genus $\chi_{_h}$.

\item
$D_\chi$: Divisors with vanishing Euler characteristics $\chi$.

\item
$D_\kappa$: Divisors with vanishing self cubic-intersections $\kappa_0$.

\item
$D_\Pi$: Divisors with vanishing $\Pi$.

\end{itemize}
\noindent
We find that there are 8 divisors topologies of $D_{\chi_{h}}$ type, 4 divisors topologies of $D_\chi$ type, 7 divisors topologies of $D_\kappa$ type, and only 2 divisors topologies of $D_\Pi$ type. In fact, the divisors of vanishing $\Pi = 0$ are those corresponding to a del-Pezzo surface of degree 6, i.e. a dP$_3$ and an exact Wilson divisor invoked for poly-instanton correction in \cite{Blumenhagen:2012kz}.

\noindent
\begin{table}[H]
\centering
\begin{tabular}{|c||c|c|c|c|c||c|c|c|c|c|c|} 
\hline
\# & $h^{p,q}$  & $\chi_{_h}$ & $\chi$ & $\kappa_0$ & $\Pi$ & $f_1$ & $f_2$ & $f_3$ & $f_4$ & $f_5$ & $f_{\rm all}$ \\
\hline
 459 & \{1,1,0,2\}& {0} & {0} & { 0} & { 0} & { 0} & { 0} & { 17} & { 214} & { 1940} & { 2171} \\
 460 & \{1,1,0,3\} & 0 & 1 & -1 & 2 & 0 & 0 & 0 & 1 & 66 & 67 \\
 461 & \{1,1,0,4\} & 0 & 2 & -2 & 4 & 0 & 0 & 0 & 8 & 177 & 185 \\
 462 & \{1,1,0,5\} & 0 & 3 & -3 & 6 & 0 & 0 & 0 & 5 & 153 & 158 \\
 463 & \{1,1,0,6\} & 0 & 4 & -4 & 8 & 0 & 0 & 0 & 4 & 55 & 59 \\
 464 & \{1,1,0,8\} & 0 & 6 & -6 & 12 & 0 & 0 & 0 & 0 & 42 & 42 \\
 465 & \{1,1,0,10\} & 0 & 8 & -8 & 16 & 0 & 0 & 0 & 1 & 3 & 4 \\
 466 & \{1,1,0,11\} & 0 & 9 & -9 & 18 & 0 & 0 & 0 & 0 & 1 & 1 \\
\hline
\end{tabular}
\caption{Divisor topologies $D_{\chi_{_h}}$ defined as those having $\chi_{_h}=0$, and their frequencies. }
\label{tab_chihZero-topo-list}
\end{table}

\noindent
\begin{table}[H]
\centering
\begin{tabular}{|c||c|c|c|c|c||c|c|c|c|c|c|} 
\hline
\# & $h^{p,q}$  & $\chi_{_h}$ & $\chi$ & $\kappa_0$ & $\Pi$ & $f_1$ & $f_2$ & $f_3$ & $f_4$ & $f_5$ & $f_{\rm all}$ \\
\hline
 459 & \{1,1,0,2\}& {0} & {0} & {0} & {0} & { 0} & { 0} & { 17} & { 214} & { 1940} & { 2171} \\
 471 & \{1,2,0,6\} & -1 & 0 & -12 & 12 & 0 & 0 & 0 & 8 & 48 & 56 \\
 484 & \{1,3,0,10\} & -2 & 0 & -24 & 24 & 0 & 0 & 0 & 6 & 55 & 61 \\
 495 & \{1,4,0,14\} & -3 & 0 & -36 & 36 & 0 & 0 & 0 & 0 & 3 & 3 \\
\hline
\end{tabular}
\caption{Divisor topologies $D_{\chi}$ defined as those having $\chi=0$, and their frequencies.}
\label{tab_chiZero-topo-list}
\end{table}

\noindent
\begin{table}[H]
\centering
\begin{tabular}{|c||c|c|c|c|c||c|c|c|c|c|c|} 
\hline
\# & $h^{p,q}$  & $\chi_{_h}$ & $\chi$ & $\kappa_0$ & $\Pi$ & $f_1$ & $f_2$ & $f_3$ & $f_4$ & $f_5$ & $f_{\rm all}$ \\
\hline
10 & \{1,0,0,10\} & 1 & 12 & 0 & 12 & 0 & 0 & 54 & 927 & 8983 & 9964 \\
65 &\{1,0,1,20\} & { 2} & { 24} & {0} & { 24} & {0} & { 24} & { 322} & { 1879} & { 10745} & { 12970} \\
84 & { \{1,0,2,30\}} & { 3} & {36} & {0} & { 36} & {0} & { 30} & { 235} & { 1145} & { 5851} & { 7261} \\
102 & \{1,0,3,40\} & 4 & 48 & 0 & 48 & 0 & 2 & 24 & 130 & 789 & 945 \\
120 & \{1,0,4,50\} & 5 & 60 & 0 & 60 & 0 & 0 & 1 & 5 & 103 & 109 \\
136 & \{1,0,5,60\} & 6 & 72 & 0 & 72 & 0 & 0 & 0 & 0 & 10 & 10 \\
459 & \{1,1,0,2\} & {0} & {0} & {0} & {0} & {0} & {0} & {17} & {214} & {1940} & {2171} \\
\hline
\end{tabular}
\caption{Divisor topologies $D_{\kappa}$ defined as those having $\kappa_0=0$, and their frequencies.}
\label{tab_kappaZero-topo-list}
\end{table}

\noindent
\begin{table}[H]
\centering
\begin{tabular}{|c||c|c|c|c|c||c|c|c|c|c|c|} 
\hline
\# & $h^{p,q}$  & $\chi_{_h}$ & $\chi$ & $\kappa_0$ & $\Pi$ & $f_1$ & $f_2$ & $f_3$ & $f_4$ & $f_5$ & $f_{\rm all}$ \\
\hline
4 & \{1,0,0,4\} & 1 & 6 & 6 & 0 & 0 & 0 & 4 & 152 & 2441 & 2597 \\
459 & \{1,1,0,2\} & {0} & {0} & {0} & {0} & {0} & {0} & {17} & {214} & {1940} & {2171} \\
\hline
\end{tabular}
\caption{Divisor topologies $D_{\Pi}$ defined as those having $\Pi=0$, and their frequencies.}
\label{tab_PiZero-topo-list}
\end{table}

\noindent
In addition, the four relations given in Eq. (\ref{eq:topo-via-hodgenumbers}) show that a divisor will have all the four quantities vanish if the following relation among the Hodge numbers hold,
\bea
\label{eq:hodge-for-all-zero}
& & h^{1,1}(D) = 2 \, h^{1,0}(D) = 2 \, h^{0,0}(D) + 2\,h^{2,0}(D).
\eea
As argued before, for a smooth divisor $D$ one can take $h^{0,0}(D) = 1$, and subsequently a generic divisor $D_{\rm zero}$ which satisfy all these relations can be given by a single parameter $n \in 2{\mathbb N}$ as below,
\begin{equation}
\label{eq:all-zero-topology}
D_{\rm zero} = \begin{tabular}{ccccc}
    & & 1 & & \\
   & $\frac{n}{2}$ & & $\frac{n}{2}$ & \\
$\frac{n}{2}-1$  & & $n$ & & $\frac{n}{2}-1$ \\
   & $\frac{n}{2}$ & & $\frac{n}{2}$ & \\
    & & 1 & & \\
\end{tabular}. \,
\end{equation}
Subsequently one may immediately recall a couple of examples in this class given as:
\bea
\label{eq:all-zero-topology-examples}
& & \hskip-1.5cm D_{\rm zero}^{n=2} = \begin{tabular}{ccccc}
    & & 1 & & \\
   & 1 & & 1 & \\
0  & & 2 & & 0 \\
   & 1 & & 1 & \\
    & & 1 & & \\
\end{tabular} \equiv W_{\rm poly}, \qquad D_{\rm zero}^{n=4} = \begin{tabular}{ccccc}
    & & 1 & & \\
   & 2 & & 2 & \\
1  & & 4 & & 1 \\
   & 2 & & 2 & \\
    & & 1 & & \\
\end{tabular} \equiv  {\mathbb T}^4 \,,
\eea
where the first one corresponds to $n=2$ while the 4-torus corresponds to $n =4$. These are well known complex 2D surfaces while the next one in this series corresponding to $n=6$ having the following Hodge diamond
\bea
D_{\rm zero}^{n=6} = \begin{tabular}{ccccc}
    & & 1 & & \\
   & 3 & & 3 & \\
2  & & 6 & & 2 \\
   & 3 & & 3 & \\
    & & 1 & & \\
\end{tabular},
\eea
is not a familiar face, at least to our knowledge. So we are able to present a class of divisor topologies for which the Euler characteristics $\chi$ as well the Arithmetic genus $\chi_{_h}$ vanish or equivalently self cubic-intersection $\kappa_0$ and the topological quantity $\Pi$ vanish. In this context, our detailed analysis shows that the topology $D_{\rm zero}$ which can appear as a coordinate divisor of a CY threefold of the Kreuzer-Skarke dataset is the only one corresponding to $n=2$. However, the other possibilities may appear from a combination of coordinate divisors which we do not aim to explore in the current work.


\section{Classification of divisor topologies}
\label{sec_classifications}

From Table \ref{tab_number-of-space-and-divisors} we recall that there is a total number of 139740 coordinate divisors from a total number of 15829 favourable CY threefolds with trivial fundamental group. In the detailed analysis we find that there are a total of 565 distinct topologies which arise from these CY threefold. In particular we find that there are 76839 divisors of $R$-type, 55972 divisors of $K$-type and 6929 divisors of $W$-type. Their detailed classification is presented in Table \ref{tab_topo-list} of the Appendix \ref{sec_appendix1}. However we discuss some insights of those results in this section under the three types of divisors we classified before.

\subsection{Rigid divisors with unit Arithmetic genus: $\chi_{_h}(D) = 1$}
The class of rigid divisors which are simply connected appears quite frequently in our scan. The Hodge diamond of such a divisor is given as:
\begin{equation}
\label{eq:R-topology}
R\equiv
\begin{tabular}{ccccc}
    & & 1 & & \\
   & 0 & & 0 & \\
0  & & $n$ & & 0 \\
   & 0 & & 0 & \\
    & & 1 & & \\
\end{tabular}. \,
\end{equation}
In our scan we find that there are 76839 divisors out of 139740 which fall in this category. However we find that there are only 63 distinct divisor topologies corresponding to a distinct value of $n$, where $1 \leq n \leq 109$. Although the full list is presented in the Table \ref{tab_topo-list} of the appendix \ref{sec_appendix1}, for illustration purpose at this stage we present first 10 rigid topologies along with additional five ones with highest $\chi$ in Table \ref{tab_rigid-topo-list}.

\noindent
\begin{table}[h!]
\centering
\begin{tabular}{|c|c||c|c|c|c|c||c|c|c|c|c|c|} 
\hline
&&&&&&&&&&&& \\
\# & \# in & $h^{p,q}$  & $\chi_{_h}$ & $\chi$ & $\kappa_0$ & $\Pi$ & $f_1$ & $f_2$ & $f_3$ & $f_4$ & $f_5$ & $f_{\rm all}$ \\
& Tab \ref{tab_topo-list}&&&&&&&&&&& \\
\hline
&&&&&&&&&&&& \\
1 & 1 & \{1,0,0,1\} & 1 & 3 & 9 & -6 & 0 & 8 & 59 & 372 & 2410 & 2849 \\
2 & 2 & \{1,0,0,2\} & 1 & 4 & 8 & -4 & 0 & 4 & 103 & 999 & 9224 & 10330 \\
3 & 3 & \{1,0,0,3\} & 1 & 5 & 7 & -2 & 0 & 0 & 4 & 160 & 2360 & 2524 \\
4& 4 & \{1,0,0,4\} & 1 & 6 & 6 & 0 & 0 & 0 & 4 & 152 & 2441 & 2597 \\
5& 5 & \{1,0,0,5\} & 1 & 7 & 5 & 2 & 0 & 0 & 6 & 103 & 1586 & 1695 \\
6& 6 & \{1,0,0,6\} & 1 & 8 & 4 & 4 & 0 & 0 & 9 & 198 & 2574 & 2781 \\
7& 7 & \{1,0,0,7\} & 1 & 9 & 3 & 6 & 0 & 2 & 21 & 250 & 2705 & 2978 \\
8& 8 & \{1,0,0,8\} & 1 & 10 & 2 & 8 & 0 & 4 & 67 & 714 & 5988 & 6773 \\
9&9 & \{1,0,0,9\} & 1 & 11 & 1 & 10 & 0 & 5 & 75 & 673 & 5772 & 6525 \\
10& 10 & \{1,0,0,10\} & 1 & 12 & 0 & 12 & 0 & 0 & 54 & 927 & 8983 & 9964 \\
& &&&&&&&&&&& \\
 \hline
& &&&&&&&&&&& \\
1& 59 & \{1,0,0,89\} & 1 & 91 & -79 & 170 & 0 & 0 & 0 & 2 & 1 & 3 \\
2& 60 & \{1,0,0,90\} & 1 & 92 & -80 & 172 & 0 & 0 & 0 & 0 & 2 & 2 \\
3& 61 & \{1,0,0,96\} & 1 & 98 & -86 & 184 & 0 & 0 & 0 & 0 & 1 & 1 \\
4& 62 & \{1,0,0,98\} & 1 & 100 & -88 & 188 & 0 & 0 & 0 & 4 & 0 & 4 \\
5& 63 & \{1,0,0,109\} & 1 & 111 & -99 & 210 & 0 & 0 & 1 & 0 & 0 & 1 \\
\hline
\end{tabular}
\caption{The first 10 $R$-type divisor topologies, along with five $R$-type divisor topologies with highest $|\chi|$. Here ``$f_n$" denotes the frequency with which a particular divisor appears for a given $n = h^{1,1}(CY)$.}
\label{tab_rigid-topo-list}
\end{table}

\noindent
Let us make some observations for this class of divisors:
\begin{itemize}
\item
The highest value of Euler characteristics for the $R$-type divisors is 111, and this belongs to a CY threefold with $h^{1,1}  = 3$. For more interested readers, let us mention that it corresponds to the polytope ID \# 277 of the AGHJN database \cite{Altman:2014bfa}.
\item
The highest value of $\Pi(D)$ for the $R$-type divisors is 210, and this corresponds to the same CY threefold with $h^{1,1}  = 3$. 
\item
One can observe that most of the topologies have negative self cubic-intersections. In fact 53 out of 63 which are numbered from 11-63 in Table \ref{tab_topo-list} have $\kappa_0 < 0$. The topology $R_{10}$ has vanishing $\kappa_0$ while the first 9 topologies have $\kappa = \{1, 2, .., 9\}$, and these are most likely to belong to the ${\mathbb P}^2$ surface along with the so-called del-Pezzo surfaces dP$_n$ obtained from blowing up eight generic points in ${\mathbb P}^2$.
\end{itemize}

\noindent
A particular subset of this class of divisors is the well known del-Pezzo surfaces defined as projective
algebraic surfaces with ample anti-canonical divisor class $-{\cal K}$ so that $-{\cal K} \cdot {\cal C} > 0$ for each curve ${\cal C}$.
Among the Fano surfaces we find also ${\mathbb P}^1 \times {\mathbb P}^1$ (sometimes also known as the Hirzebruch surface
${\mathbb F}_0$), with $h^{1,1} = 2$. Subsequently, in order to search the del-Pezzo divisors, first we scan for the divisors $D_s$ which satisfy the following two necessary conditions \cite{Cicoli:2011it},
\bea
\label{eq:dP}
& & \int_{X} \hat{D}_s \wedge \hat{D}_s \wedge \hat{D}_s = \kappa_{sss} > 0\, , \qquad \int_{X} \hat{D}_s \wedge \hat{D}_s \wedge \hat{D}_i \leq 0 \qquad \forall \, i \neq s \,.
\eea
Here the self-triple-intersection number $\kappa_{sss}$ corresponds to the degree of the del-Pezzo four-cycle dP$_n$ where $\kappa_{sss} = 9 - n$, which is always positive as $n \leq 8$ for del-Pezzo surfaces. Now the question is how may of the $R_m$-type divisors topologies are of del-Pezzo type. Naively one would say that all of the topologies with $1 \leq m \leq 9$ should be dP$_{m-1}$ setting $m = 1$ as ${\mathbb P}^2$. However merely looking at the Hodge numbers may not uniquely define the divisor topology in a genuine sense, as sometimes different geometries may have similar Hodge numbers. Just to give an example, one can take dP$_1$ and ${\mathbb P}^1\times{\mathbb P}^1$ which have identical Hodge diamonds. In fact comparing Table \ref{tab_rigid-topo-list} with Table \ref{tab_dPn-topo-list} shows that for a given $\chi$ with $3 \leq \chi \leq 11$, all the topologies of $R$-type are not necessarily a del-Pezzo surface despite having the same Hodge numbers. This can happen for various reasons, e.g. there could simply be an addition of isolated non-generic point which raises the $h^{1,1}(D)$ of the rigid divisor without satisfying the del-Pezzo conditions of anti-canonical bundle being ample as given in Eq.~(\ref{eq:dP}).

\noindent
\begin{table}[H]
\centering
\begin{tabular}{|c||c|c|c|c|c||c|c|c|c|c|c|} 
\hline
&&&&&&&&&&& \\
 dP$_n$ & $h^{p,q}$  & $\chi_{_h}$ & $\chi$ & $\kappa_0$ & $\Pi$ & $f_1$ & $f_2$ & $f_3$ & $f_4$ & $f_5$ & $f_{\rm all}$ \\
Type&&&&&&&&&&& \\
\hline
&&&&&&&&&&& \\
 dP$_0={\mathbb P}^2$ & \{1,0,0,1\} & 1 & 3 & 9 & -6 & 0 & 8 & 59 & 372 & 2410 & 2849 \\
 dP$_1 \, \& \, {\mathbb P}^1\times{\mathbb P}^1$ & \{1,0,0,2\} & 1 & 4 & 8 & -4 & 0 & 4 & 91 & 878 & 8038 & 9011 \\
 dP$_2$ & \{1,0,0,3\} & 1 & 5 & 7 & -2 & 0 & 0 & 4 & 155 & 2242 & 2401 \\
 dP$_3$ & \{1,0,0,4\} & 1 & 6 & 6 & {\bf 0} & 0 & 0 & 4 & 144 & 2271 & 2419 \\
 dP$_4$ & \{1,0,0,5\} & 1 & 7 & 5 & 2 & 0 & 0 & 2 & 55 & 947 & 1004 \\
 dP$_5$ & \{1,0,0,6\} & 1 & 8 & 4 & 4 & 0 & 0 & 9 & 184 & 2190 & 2383 \\
 dP$_6$ & \{1,0,0,7\} & 1 & 9 & 3 & 6 & 0 & 2 & 21 & 239 & 2459 & 2721 \\
 dP$_7$ & \{1,0,0,8\} & 1 & 10 & 2 & 8 & 0 & 4 & 67 & 689 & 5462 & 6222 \\
 dP$_8$ & \{1,0,0,9\} & 1 & 11 & 1 & 10 & 0 & 5 & 71 & 597 & 4692 & 5365 \\
\hline
Total &&&&&& 0 & 23 & 328 & 3313 & 30711 & 34375 \\
\hline
\end{tabular}
\caption{Collection of del-Pezzo divisor topologies satisfying Eq. (\ref{eq:dP}).}
\label{tab_dPn-topo-list}
\end{table}

\noindent
Comparing with frequencies for first 9 rigid divisors mentioned in Table \ref{tab_rigid-topo-list} with those of del-Pezzo divisors in Table \ref{tab_dPn-topo-list} we observed that for a given $\chi$, all the topologies of $R$-type are not necessarily a del-Pezzo surface despite having the same Hodge numbers. Although the frequency of ${\mathbb P}^2$ having the minimal Hodge numbers remains the same, there might be a couple of divisors containing some extra points without being necessarily according to the del-Pezzo blow-ups. For example, $h^{1,1}(CY) = 3$ has 20 such extra rigid surfaces which are not del-Pezzo divisors. However for usual model building purposes, this does not make much difference given that the topological quantities $\chi, \chi_{_h}$ etc. remains the same.

Given that $\kappa_0$ is always a non-zero positive integer for a dP$_n$ surface for $0\leq n \leq 8$, the Eq.~(\ref{eq:kappa0}) can be generically satisfied for a subset of del-Pezzo surfaces. Let us also emphasise here the fact that subsequently one can always shrink such a `diagonal' del-Pezzo ddP$_n$ to a point-like singularity by squeezing it along a single direction. For counting the CY threefolds which could support the LVS models, we perform the following two steps; first one corresponds to selecting the CY threefolds which have at least one del-Pezzo divisor, and subsequently in the second step we check for what is called as the `diagonality' of those del-Pezzo divisors as collected in Table \ref{tab_ddPn-topo-list}.

\noindent
\begin{table}[h!]
\centering
\begin{tabular}{|c||c|c|c|c|c||c|c|c|c|c|c|} 
\hline
&&&&&&&&&&& \\
 ddP$_n$ & $h^{p,q}$  & $\chi_{_h}$ & $\chi$ & $\kappa_0$ & $\Pi$ & $f_1$ & $f_2$ & $f_3$ & $f_4$ & $f_5$ & $f_{\rm all}$ \\
Type&&&&&&&&&&& \\
\hline
&&&&&&&&&&& \\
 dP$_0={\mathbb P}^2$ & \{1,0,0,1\} & 1 & 3 & 9 & -6 & 0 & 8 & 59 & 372 & 2410 & 2849 \\
 ${\mathbb P}^1 \times {\mathbb P}^1$ & \{1,0,0,2\} & 1 & 4 & 8 & -4 & 0 & 2 & 16 & 144 & 944 & 1106 \\
 ddP$_1$ & \{1,0,0,2\} & 1 & 4 & 8 & -4 & 0 & 0 & 0 & 0 & 0 & 0 \\
 ddP$_2$ & \{1,0,0,3\} & 1 & 5 & 7 & -2 & 0 & 0 & 0 & 0 & 0 & 0 \\
 ddP$_3$ & \{1,0,0,4\} & 1 & 6 & 6 & 0 & 0 & 0 & 0 & 0 & 0 & 0 \\
 ddP$_4$ & \{1,0,0,5\} & 1 & 7 & 5 & 2 & 0 & 0 & 0 & 0 & 0 & 0 \\
 ddP$_5$ & \{1,0,0,6\} & 1 & 8 & 4 & 4 & 0 & 0 & 0 & 0 & 0 & 0 \\
 ddP$_6$ & \{1,0,0,7\} & 1 & 9 & 3 & 6 & 0 & 2 & 17 & 109 & 624 & 752 \\
 ddP$_7$ & \{1,0,0,8\} & 1 & 10 & 2 & 8 & 0 & 4 & 40 & 277 & 827 & 1148 \\
 ddP$_8$ & \{1,0,0,9\} & 1 & 11 & 1 & 10 & 0 & 5 & 39 & 157 & 407 & 608 \\
\hline
Total &&&&&& 0 & 21 & 171 & 1059 & 5212 & 6463 \\
\hline
\end{tabular}
\caption{Collection of diagonal del-Pezzo (ddP$_n$) divisor topologies satisfying Eq. (\ref{eq:dP}) along with the diagonality condition (\ref{eq:diag-divisor}).}
\label{tab_ddPn-topo-list}
\end{table}

\subsection{Non-rigid divisors: $\chi_{_h}(D) > 1$}
The class of non-rigid divisors also appears quite frequently in our scan. The Hodge diamond of such a divisor is given as:
\begin{equation}
\label{eq:K-topology}
K \equiv
\begin{tabular}{ccccc}
    & & 1 & & \\
   & 0 & & 0 & \\
$m$  & & $n$ & & $m$ \\
   & 0 & & 0 & \\
    & & 1 & & \\
\end{tabular}. \,
\end{equation}
In our scan we find that there are 55972 divisors out of 139740 which fall in this category. Moreover we find that there are 395 distinct divisor topologies  (out of a total of 565) corresponding to a distinct set of values $\{m, n\}$, where the observed range for the values of $\{m, n\}$ is such that $1 \leq m \leq 65$ and $19 \leq n \leq 417$. As a side remark, let us mention that the divisor with highest $\chi = 549$ belong to this non-rigid class of divisors and correspond to the well know swiss-cheese CY threefold with $h^{1,1}(CY) =2$, which is realized as a degree-18 hypersurface in WCP$^4[1,1,1,6,9]$.

Two of the most familiar topologies of this category are the $K3$ surfaces and the so-called `special deformation" (SD) divisors which are respectively characterised by $\{m, n\} = \{1, 20\}$ and $\{m, n\} = \{2, 30\}$. Both of these satisfy another interesting condition (\ref{eq:Dcube=0}) guaranteeing the vanishing of self-cubic intersections. Some selected topologies of this class are presented in Table \ref{tab_non-rigid-topo-list}. In the context of non-rigid divisors, it is appropriate to discuss the $K3$-surfaces appearing as divisors of a CY threefold. The condition for a CY threefold to be a $K3/{\mathbb T}^4$-fibred space can be correlated with the fact that the corresponding $K3/{\mathbb T}^4$ divisor should appear only linearly in the intersection polynomial \cite{OGUISO:1993, Douglas:2003um}. In this regard, we present the CY threefolds geometries which are $K3$-fibred in Table \ref{tab_K3fibred-topologies}.

\noindent
\begin{table}[h!]
\centering
\begin{tabular}{|c|c||c|c|c|c|c||c|c|c|c|c|c|} 
\hline
&&&&&&&&&&&& \\
\# & \# in & $h^{p,q}$  & $\chi_{_h}$ & $\chi$ & $\kappa_0$ & $\Pi$ & $f_1$ & $f_2$ & $f_3$ & $f_4$ & $f_5$ & $f_{\rm all}$ \\
& Tab \ref{tab_topo-list}&&&&&&&&&&& \\
\hline
&&&&&&&&&&&& \\
1& 65 & \{1,0,1,20\} & 2 & 24 & 0 & 24 & 0 & 24 & 322 & 1879 & 10745 & 12970 \\
2& 66 & \{1,0,1,21\} & 2 & 25 & -1 & 26 & 0 & 8 & 124 & 965 & 6601 & 7698 \\
3& 84 & \{1,0,2,30\} & 3 & 36 & 0 & 36 & 0 & 30 & 235 & 1145 & 5851 & 7261 \\
4& 67 & \{1,0,1,22\} & 2 & 26 & -2 & 28 & 0 & 4 & 53 & 391 & 2976 & 3424 \\
5& 100 & \{1,0,3,38\} & 4 & 46 & 2 & 44 & 4 & 26 & 88 & 423 & 2063 & 2604 \\
6& 68 & \{1,0,1,23\} & 2 & 27 & -3 & 30 & 0 & 0 & 20 & 213 & 1376 & 1609 \\
7& 83 & \{1,0,2,29\} & 3 & 35 & 1 & 34 & 3 & 16 & 63 & 250 & 1136 & 1468 \\
8& 69 & \{1,0,1,24\} & 2 & 28 & -4 & 32 & 0 & 4 & 25 & 125 & 927 & 1081 \\
9& 99 & \{1,0,3,37\} & 4 & 45 & 3 & 42 & 4 & 10 & 45 & 166 & 809 & 1034 \\
10& 102 & \{1,0,3,40\} & 4 & 48 & 0 & 48 & 0 & 2 & 24 & 130 & 789 & 945 \\
& &&&&&&&&&&& \\
 \hline
& &&&&&&&&&&& \\
1 & 458 & \{1,0,65,417\} & 66 & 549 & 243 & 306 & 0 & 1 & 0 & 0 & 0 & 1 \\
2 & 457 & \{1,0,59,384\} & 60 & 504 & 216 & 288 & 0 & 0 & 0 & 0 & 3 & 3 \\
3 & 456 & \{1,0,58,374\} & 59 & 492 & 216 & 276 & 0 & 0 & 4 & 0 & 0 & 4 \\
4 & 455 & \{1,0,57,368\} & 58 & 484 & 212 & 272 & 0 & 0 & 0 & 1 & 0 & 1 \\
5 & 454 & \{1,0,55,357\} & 56 & 469 & 203 & 266 & 0 & 0 & 2 & 0 & 0 & 2 \\
\hline
\end{tabular}
\caption{List of 10 $K$-type divisor topologies with highest frequency, along with five $K$ type divisor topologies with highest $|\chi|$. Here ``$f_n$" denotes the frequency with which a particular divisor appears for a given $n = h^{1,1}(CY)$.}
\label{tab_non-rigid-topo-list}
\end{table}

\noindent
\noindent
\begin{table}[H]
\centering
\begin{tabular}{|c|c||c|c|}
\hline
$h^{1,1}$ & \# of CY & \# of $K3$-fibred & Total \# of  \\ 
& geom$^\ast$ &  CY  & $K3$-divisors \\ 
\hline
1 & 4 & 0 & 0  \\ 
2 & 37 & 10 & 20 \\ 
3 & 300 & 136 & 285 \\ 
4 & 1994 & 865 & 1700 \\ 
5 & 13494 & 5970 & 10128 \\ 
\hline
\end{tabular}
\caption{Number of $K3$-fibred CY threefolds with $K3$ divisors and their frequencies.}
\label{tab_K3fibred-topologies}
\end{table}

\noindent
Let us note a slight difference in the values presented in the first row of the Table \ref{tab_non-rigid-topo-list}  as compared to those presented in the fourth column of the Table  \ref{tab_K3fibred-topologies}. This is because of the fact that not all $K3$-topologies as anticipated from the Hodge number computations are genuine in the sense of satisfying the linearity arguments of \cite{OGUISO:1993, Douglas:2003um}. This argument/observation is similar to the fact that not all del-Pezzo topologies as anticipated from the Hodge number computations are necessarily del-Pezzo and it will be dictated by the additional condition of anti-canonical bundle being ample as given in Eq.~(\ref{eq:dP}).

\subsection{Wilson divisors: $\chi_{_h}(D) < 1$}
This class of rigid divisors consists of those which are non-simply connected and appears quite frequently in our scan. The Hodge diamond of such a divisor is given as:
\begin{equation}
\label{eq:W-topology}
W \equiv
\begin{tabular}{ccccc}
    & & 1 & & \\
   & $m$ & & $m$ & \\
0  & & $n$ & & 0 \\
   & $m$ & & $m$ & \\
    & & 1 & & \\
\end{tabular}. \,
\end{equation}
In our scan we find that there are 6929 divisors out of 139740 which fall in this category. Moreover we find that there are 107 distinct divisor topologies (out of a total of 565) corresponding to a distinct set of values $\{m, n\}$, where the observed range for the values of $\{m, n\}$ is such that $1 \leq m \leq 55$ and $2 \leq n \leq 18$.

The simplest of these 107 topologies is the so-called ``Exact" Wilson divisor characterized by $h^{1,0} = 1$ \cite{Blumenhagen:2012kz} and due to our conjecture (\ref{eq:conjecture}) this means those divisor which satisfy vanishing of holomorphic Euler characteristics i.e. $\chi_{_h} = 0$. As presented in Table \ref{tab_chihZero-topo-list} there are eight topologies of this kind and the most special one corresponds to $\{m ,n\} = \{1, 2\}$ which means that $\chi$ and $\chi_{_h}$ both vanish for these surface. In fact this is quite a peculiar topology which corresponds to a ${\mathbb P}^1$ fibration over ${\mathbb T}^2$'s \cite{Blumenhagen:2012kz}. Moreover, these are used for generating poly-instanton effects on top of the usual non-perturbative corrections to the superpotential which are induced through the usual ${\cal O}(1)$ type $E3$-instanton or gaugino condensation effects on stacks of $D7$-brane \cite{Blumenhagen:2012kz}. In Table \ref{tab_wilson-topo-list} we present 10 most frequent topologies of Wilson-type arising from our scan, and as said above, the so-called an ``Exact Wilson"  divisor numbered as first example in Table \ref{tab_wilson-topo-list} turns out to be the most often appearing topology of this class. In addition we have presented 5 more topologies in Table \ref{tab_wilson-topo-list} which have the largest value of $\chi_{_h}$ and $|\chi|$ as well as $\Pi$. 

\noindent
\begin{table}[h!]
\centering
\begin{tabular}{|c|c||c|c|c|c|c||c|c|c|c|c|c|} 
\hline
&&&&&&&&&&&& \\
\# & \# in & $h^{p,q}$  & $\chi_{_h}$ & $\chi$ & $\kappa_0$ & $\Pi$ & $f_1$ & $f_2$ & $f_3$ & $f_4$ & $f_5$ & $f_{\rm all}$ \\
& Tab \ref{tab_topo-list}&&&&&&&&&&& \\
\hline
&&&&&&&&&&&& \\
1& 459 & \{1,1,0,2\} & 0 & 0 & 0 & 0 & 0 & 0 & 17 & 214 & 1940 & 2171 \\
2& 467 & \{1,2,0,2\} & -1 & -4 & -8 & 4 & 0 & 1 & 16 & 145 & 925 & 1087 \\
3& 477 & \{1,3,0,2\} & -2 & -8 & -16 & 8 & 0 & 1 & 17 & 127 & 773 & 918 \\
4& 498 & \{1,6,0,2\} & -5 & -20 & -40 & 20 & 0 & 0 & 8 & 35 & 273 & 316 \\
5& 486 & \{1,4,0,2\} & -3 & -12 & -24 & 12 & 0 & 1 & 11 & 26 & 275 & 313 \\
6& 461 & \{1,1,0,4\} & 0 & 2 & -2 & 4 & 0 & 0 & 0 & 8 & 177 & 185 \\
7& 513 & \{1,9,0,2\} & -8 & -32 & -64 & 32 & 0 & 0 & 1 & 23 & 160 & 184 \\
8& 481 & \{1,3,0,6\} & -2 & -4 & -20 & 16 & 0 & 0 & 1 & 17 & 149 & 167 \\
9& 462 & \{1,1,0,5\} & 0 & 3 & -3 & 6 & 0 & 0 & 0 & 5 & 153 & 158 \\
10& 469 & \{1,2,0,4\} & -1 & -2 & -10 & 8 & 0 & 0 & 1 & 19 & 136 & 156 \\
& &&&&&&&&&&& \\
 \hline
& &&&&&&&&&&& \\
1& 561 & \{1,43,0,2\} & -42 & -168 & -336 & 168 & 0 & 0 & 0 & 0 & 9 & 9 \\
2& 562 & \{1,45,0,2\} & -44 & -176 & -352 & 176 & 0 & 0 & 0 & 2 & 2 & 4 \\
3& 563 & \{1,48,0,2\} & -47 & -188 & -376 & 188 & 0 & 0 & 0 & 0 & 1 & 1 \\
4& 564 & \{1,49,0,2\} & -48 & -192 & -384 & 192 & 0 & 0 & 0 & 4 & 0 & 4 \\
5& 565 & \{1,55,0,2\} & -54 & -216 & -432 & 216 & 0 & 0 & 1 & 0 & 0 & 1 \\
\hline
\end{tabular}
\caption{List of 10 Wilson divisor topologies with highest frequency, along with five Wilson divisor topologies with highest $|\chi|$. Here ``$f_n$" denotes the frequency with which a particular divisor appears for a given $n = h^{1,1}(CY)$.}
\label{tab_wilson-topo-list}
\end{table}





\section{Fixed point set and dS uplifting via $\ov{D3}$-brane}
\label{sec_applications}

After knowing the divisor topologies of all the coordinate divisors in terms of their Hodge numbers one can use them for several phenomenological purposes, e.g. in computing the tadpole charges after constructing explicit CY orientifolds, which one needs for 4D ${\cal N} =1$ type IIB based model building, For that purpose one needs to consider an appropriate involution and the most commonly used ones are the so-called reflection involution corresponding to flipping one (or more) of the coordinates defining the toric ambient space via $\sigma: x_i \to - x_i$. Given that the $D3$-tadpole charge can be estimated by the Euler characteristics of the four-cycles wrapped by the $D7/O7$ configurations along with the $O3$-planes present in the global brane setting, one can subsequently perform a systematic study of these aspects for all the CY threefolds, in particular for THCYs up to $1 \leq h^{1,1}(CY) \leq 5$ for our current interest.

There can be numerous ways by which our results can be useful for various phenomenological purposes. One such application can be to systematise the proposal of \cite{Crino:2020qwk} which presents an elegant recipe for realizing de-Sitter vauca via $\ov{D3}$ uplifting within LVS framework. Although all the details are not relevant for the plan/goal of the current work, let us mention that this uplifting recipe demands the following two central ingredients on the global model building side, apart from the usual need of at least a diagonal del-Pezzo divisor to support LVS:

\begin{itemize}
\item
A significantly large $D3$-brane charge consistent with the tadpole cancellation requirements in needed.

\item
The choice of orientifold involution should be such that it allows for (at least) two coincident $O3$-planes in the fixed point loci.

\end{itemize} 

\noindent
Naively speaking these two requirements seem to be not that strong, however it turns out that this is not precisely the case, at least with CY orientifolds with smaller $h^{1,1}(CY)$. In this regard, our dataset and available scanning results are directly applicable to explore this possibility. All one needs to do is to compute the fixed point set for each of the possible reflection involutions $\sigma_i : x_i \to - x_i$ which flips the sign of the toric coordinate $x_i$ corresponding to the coordinate divisor $D_i$ which can be defined as $x_i = 0$.

\subsection{Estimating the $D3$-tadpole charge}

The $S$-dual pair of fluxes $(F_3, H_3)$ carry a non-trivial $D3$ charge given as,
\bea
& & N_{\rm flux} =  \int_{CY} F_3 \wedge H_3 = H^\Lambda\, F_\Lambda - H_\Lambda \, F^\Lambda > 0,
\eea
where the positivity condition $N_{\rm flux} > 0$ follows from a self-duality relation between the RR flux $F_3$ and Hodge dual of the NS-NS three-form flux $\ast H_3$. Moreover, $N_{\rm flux}$ has to satisfy the following tadpole cancellation condition \cite{Blumenhagen:2008zz},
\bea
\label{eq:tadpole1}
& & \hskip-1cm N_{D3} + \frac{N_{\rm flux}}{2} +  N_{\rm gauge} = \frac{N_{O3}}{4} + \sum_{O7_i} \frac{\chi(\Gamma_{O7_i})}{12} + \sum_{D7_i} N_{D7_i} \frac{\chi(\Gamma_{D7_i})}{24} , 
\eea
where 
\bea
& & N_{\rm gauge} = -  \frac{1}{2} \sum_{D7_i}  \int_{D7_i} tr[{\cal F}^2_{D7_i}],
\eea
and the number of $D7$-branes in the $i^{\rm th}$ stack wrapping a four-cycle $\Gamma_{D7_i}$ is denoted as $N_{D7_i}$ while the total number of $D3$-branes and $O3$-planes are denoted as $N_{D3}$ and $N_{O3}$ respectively. For the case, when four $D7$-branes and their image branes are placed on top of the $O7$-planes, and the ${\cal F}$ fluxes are absent, the above condition simplifies to the following form,
\bea
\label{eq:tadpole1-ontop}
& & N_{D3} + \frac{N_{\rm flux}}{2}  =  \frac{N_{O3}}{4} + \sum_{O7_i} \frac{\chi(\Gamma_{O7_i})}{4} \equiv {\cal Q}_{D3}^{\rm simp},
\eea
where we have introduced ${\cal Q}_{D3}^{\rm simp}$ which denotes a value half of the total $D3$ charge induced by $D7/O7/O3$ configurations for this special case. Moreover, the special condition (\ref{eq:tadpole1-ontop}) can be of crucial importance not only for the brane settings in which $D7$-brane are placed on top of the $O7$-planes but also as a test for ensuring that the involution is well-behaved. It can be done by checking if RHS of Eq.~(\ref{eq:tadpole1-ontop}) is divisible by 4 or not. Say if it is not the case and one gets some fractional value, then there can be a set of possible reasons; e.g. it could be possible that the involution is introducing some singularity which means that the CY orientifold is not smooth or there could be a possibility of some missing non-generic $O3/O7$-planes which are sometimes not easy to capture in a generic fashion\footnote{For example, see the orientifold of ``Example C" in \cite{Blumenhagen:2012kz}.}. In order to avoid singularities affecting the pheno (inspired) model, it could be safe to discard all those involutions for which ${\cal Q}_{D3}^{\rm simp}$ turn out to be fractional. In this way, the condition (\ref{eq:tadpole1-ontop}) is not only useful for constructing brane setting with configurations of $D7$-branes placed on top of $O7$-planes but it can also be helpful for checking that a given involution $\sigma: x_i \to - x_i$ is genuine/safe or not, even if one wants to use less simple brane setting in their model building. However a shortcoming of this approach could be the risk of missing some involutions which might be interesting as well, after giving a bit of more attention.

In addition one can have non-local placing of $D7$-branes in the form of a Whitney-brane configuration  \cite{Collinucci:2008pf} which has been found to be useful in designing large $D3$ tadpole charge \cite{Crino:2020qwk}. It has two kinds of contributions, one is purely geometric while other one is flux dependent and hence model specific. It has been suggested in \cite{Crino:2022zjk} to consider the geometric piece of Whitney brane configurations to have some estimates on the allowed $D3$ charge corresponding to a given involution $\sigma_i$. Following that line, one can consider two $D7$-brane configurations described as local and non-local classes such that

\begin{itemize}

\item
{\bf Local class:} The placing of four $D7$-branes and their image branes on top of the $O7$-planes corresponds to the so-called local-class and in this case for the $i$'th involution $\sigma_i$ one has the following expression for the $D3$ charge,
\bea
\label{eq:D3-local}
& & {\cal Q}_{D3}^{\rm local} = 2\, {\cal Q}_{D3}^{\rm simp},
\eea

\item
{\bf Non-local class:} For a given fixed involution $\sigma_i$, in this case the ``geometric part" of the $D3$ charge can be given as \cite{Crino:2022zjk}
\bea
\label{eq:D3-nonlocal}
& & {\cal Q}_{D3_i}^{\rm nonlocal} = \frac{N_{O3_i}}{2} + \frac{\Pi(D_i)}{3} + \frac{43}{3} \kappa_0 + \frac{\chi(O7_i)}{6} + \sum_{j \neq i}  \frac{\chi(O7_j)}{6},
\eea
where the last piece corresponds to the cases where there are multiple components of the $O7$-planes.

\end{itemize}

\noindent
For our current purpose, we consider the involution $\sigma: x_i \to - x_i$ for all the 139740 toric coordinates corresponding to 15829 CY geometries as listed in Table \ref{tab_number-of-space-and-divisors}. These are the same coordinates defining the so-called ``coordinate divisors" as $D_i: x_i = 0$ for which we have analysed the divisor topologies in the current work. We also limit our involution to reflecting only one coordinate at a time, and so do not study any involution with multiple reflecting coordinates. Within these assumptions the details on some statistics about the Fixed-point analysis is presented in Table \ref{tab_space-Fixed-point-set} and Table \ref{tab_involution-Fixed-point-set}.

\noindent
\begin{table}[h!]
\centering
\begin{tabular}{|c||c|c|c|c|c|} 
\hline
$h^{1,1}$ & 1 & 2 & 3 & 4 & 5 \\ 
\hline
\# of fav-CY  geom$^\ast$ ($X$) & 4 & 37 & 300 & 1994 & 13494 \\ 
\#($X$) with $\sigma_i$ satisfying ${\cal Q}_{D3}^{\rm simp} \in {\mathbb Z}$& 4 & 30 & 247 & 1559 & 9742 \\ 
 &  &  &  &  &  \\ 
\#($X$) without any $O3$ for any of the $\sigma_i$ & 0 & 5 & 31 & 94 & 219 \\ 

 &  &  &  &  &  \\ 
\#($X$) having at least one $\sigma_i$ with $\#(O7_i) \geq1$ & 4 & 30 & 247 & 1559 & 9742 \\ 
\#($X$) having at least one  $\sigma_i$ with $\#(O7_i) \geq2$ & 0 & 17 & 192 & 1383 & 9166 \\ 
\#($X$) having at least one  $\sigma_i$ with $\#(O7_i) \geq3$ & 0 & 0 & 2 & 61 & 887 \\ 
\#($X$) having at least one  $\sigma_i$ with $\#(O7_i) \geq4$ & 0 & 0 & 1 & 11 & 78 \\ 
\#($X$) having at least one  $\sigma_i$ with $\#(O7_i) \geq5$ & 0 & 0 & 0 & 6 & 57 \\ 
\#($X$) having at least one  $\sigma_i$ with $\#(O7_i) \geq6$ & 0 & 0 & 0 & 0 & 0 \\ 
 &  &  &  &  &  \\ 
 \#($X$) having at least one  $\sigma_i$ with $\#(O3_i) \geq1$ & 4 & 25 & 216 & 1465 & 9523 \\ 
 \#($X$) having at least one  $\sigma_i$ with $\#(O3_i) \geq2$ & 4 & 25 & 214 & 1461 & 9446 \\ 
 \#($X$) having at least one  $\sigma_i$ with $\#(O3_i) \geq3$ & 3 & 22 & 181 & 1312 & 8466 \\ 
 \#($X$) having at least one  $\sigma_i$ with $\#(O3_i) \geq4$ & 2 & 19 & 148 & 1060 & 6608 \\ 
 \#($X$) having at least one  $\sigma_i$ with $\#(O3_i) \geq5$ & 2 & 13 & 98 & 612 & 3554 \\ 
 \#($X$) having at least one  $\sigma_i$ with $\#(O3_i) \geq15$ & 0 & 0 & 1 & 10 & 44 \\ 
 \#($X$) having at least one  $\sigma_i$ with $\#(O3_i) \geq20$ & 0 & 0 & 0 & 3 & 8 \\ 
  \#($X$) having at least one  $\sigma_i$ with $\#(O3_i) \geq25$ & 0 & 0 & 0 & 0 & 0 \\ 
\hline
\end{tabular}
\caption{Number of CY geometries with $O3/O7$-planes for a given involution $\sigma_i : x_i \to - x_i$. Here we see that almost all the CY geometries have at least one involution which has $O3$-planes along with $O7$-planes which are always present, at least with a single component.}
\label{tab_space-Fixed-point-set}
\end{table}

\noindent
\begin{table}[h!]
\centering
\begin{tabular}{|c||c|c|c|c|c|} 
\hline
$h^{1,1}$ & 1 & 2 & 3 & 4 & 5 \\ 
\hline
\# of fav-CY  geom$^\ast$ ($X$) & 4 & 37 & 300 & 1994 & 13494 \\ 
\# of div. topologies of $X$  & 20 & 222 & 2100 & 15952 & 121446 \\ 
(i.e. \# max. possible $\sigma_i$) &  &  &  &  &  \\ 
 &  &  &  &  &  \\ 
\#($X$) with $\sigma_i$ satisfying ${\cal Q}_{D3}^{\rm simp} \in {\mathbb Z}$& 4 & 30 & 247 & 1559 & 9742 \\ 
\# of all possible $\sigma_i$ satisfying ${\cal Q}_{D3}^{\rm simp} \in {\mathbb Z}$& 20 & 180 & 1729 & 12472 & 87678 \\ 
 &  &  &  &  &  \\ 
\#($\sigma_i$) without any $O3$ & 0 & 26 & 191 & 413 & 801 \\ 
 &  &  &  &  &  \\ 
\#($\sigma_i$) with $\#(O7_i) \geq1$ & 20 & 180 & 1729 & 12472 & 87678 \\ 
\#($\sigma_i$) with $\#(O7_i) \geq2$ & 0 & 34 & 513 & 4546 & 36640 \\ 
\#($\sigma_i$) with $\#(O7_i) \geq3$ & 0 & 0 & 7 & 200 & 2784 \\ 
\#($\sigma_i$) with $\#(O7_i) \geq4$ & 0 & 0 & 1 & 17 & 135 \\ 
\#($\sigma_i$) with $\#(O7_i) \geq5$ & 0 & 0 & 0 & 12 & 110 \\ 
\#($\sigma_i$) with $\#(O7_i) \geq6$ & 0 & 0 & 0 & 0 & 0 \\ 
 &  &  &  &  &  \\ 
\#($\sigma_i$) with $\#(O3_i) \geq1$ & 17 & 105 & 987 & 8491 & 64915 \\ 
\#($\sigma_i$) with $\#(O3_i) \geq2$ & 16 & 90 & 806 & 6960 & 51962 \\ 
\#($\sigma_i$) with $\#(O3_i) \geq3$ & 12 & 68 & 528 & 4320 & 29358 \\ 
\#($\sigma_i$) with $\#(O3_i) \geq4$ & 8 & 51 & 383 & 2810 & 17671 \\ 
\#($\sigma_i$) with $\#(O3_i) \geq5$ & 8 & 28 & 196 & 1102 & 6609 \\ 
\#($\sigma_i$) with $\#(O3_i) \geq15$ & 0 & 0 & 1 & 13 & 54 \\ 
\#($\sigma_i$) with $\#(O3_i) \geq20$ & 0 & 0 & 0 & 4 & 9 \\ 
\#($\sigma_i$) with $\#(O3_i) \geq25$ & 0 & 0 & 0 & 0 & 0 \\ 
\hline
\end{tabular}
\caption{Statistics with respect to the number of distinct involutions $\sigma_i : x_i \to - x_i$ corresponding to divisor topology $D_i$.}
\label{tab_involution-Fixed-point-set}
\end{table}

\subsection{A couple of lessons from CY geometries with $ h^{1,1} = 2$}
There are 17 CY geometries with  $h^{1,1} = 2$ which have at least one involution resulting in two coincident $O3$-planes. However if we impose that there is a diagonal del-Pezzo which is also part of the $O7$-planes, then there are only two possibilities. One Example corresponds to the degree-14 CY threefold realized in WCP$^4[1,1,2,3,7]$ which was studied in \cite{Crino:2020qwk}, while the other CY geometry corresponds to a degree-10 hypersurface embedded in WCP$^4[1,1,1,2,5]$. These are the only possibilities one can have with reflection involution $\sigma_i : x_i \to - x_i$. In this subsection we will make a couple of observations which may help in understanding some features of the de-Sitter uplifting proposal \cite{Crino:2020qwk} specially those requirements which are needed in the global model building context.

\subsubsection{Example 1: CY as a degree-18 hypersurafce in WCP$^4[1,1,1,6,9]$}
The very first example which comes to our mind for swiss-cheese CY is the famous degree-18 hypersurafce embedded in WCP$^4[1,1,1,6,9]$. This corresponds to the polytope-id 41 of AGHJN database \cite{Altman:2014bfa}, and it's toric data is given as:
\begin{center}
\begin{tabular}{|c||cccccc|}
\hline
CY Hypersurface & $x_1$  & $x_2$  & $x_3$  & $x_4$  & $x_5$ & $x_6$     \\
\hline
18 & 1 & 1 & 1 & 6 & 9 & 0  \\
6 & 0 & 0 & 0 & 2 & 3 & 1  \\
\hline
Topology \# as  & 84  & 84 & 84 &  400 & 458 & 1   \\
listed in Table \ref{tab_topo-list} & &  &  &    &  & \\
\hline
\end{tabular}
\end{center}
The Hodge numbers for this swiss-cheese are $(h^{2,1}, h^{1,1}) = (272, 2)$, the Euler number is $\chi(X)=-540$ and the SR ideal is given as:
\bea
& & {\rm SR} =  \{x_1 x_2 x_3, \,x_4 x_5 x_6\}. \nonumber
\eea
However this CY geometry is not suitable for realizing the $\ov{D3}$-brane uplifting of de-Sitter proposal of \cite{Crino:2020qwk}. The main reason for this argument is the fact that the involution reflecting the coordinate with respect to the divisor with highest $|\chi|$ (as needed for large $D3$-tadpole charge) does not have any $O3$-plane present which is one of the central requirement of \cite{Crino:2020qwk}. 

To have some estimates, let us quickly recall some orientifold constructions; in the case of orientifold involutions where four D$7$-brane and its image are put on top of the O$7$ components, the divisor with highest GLSM charges can usually lead to larger value of the $D3$-tadpole charge. Therefore one may like to test the involution $\sigma: x_5 \to - x_5$ which results in the following two components of the O$7$-planes
\begin{equation}
O7_1 = D_5, \qquad O7_2 = D_6 , 
\end{equation}
while there are no $O3$-planes present in the fixed point set. This leads to the following condition for cancellation of the D$3$-brane tadpole \cite{Blumenhagen:2008zz},
\begin{equation}
{\cal Q}_{D3}^{\rm simp}= \frac{\chi(O7)}{4} =\frac{\chi(O7_1) + \chi(O7_2)}{4} = \frac{549+3}{4} = 138. 
\end{equation}
This involution was considered in \cite{Louis:2012nb}. Moreover if we consider the non-local scenario for this involution, then using Eq.~(\ref{eq:D3-nonlocal}) we find 
\bea
& & {\cal Q}_{D3}^{\rm nonlocal} = \frac{N_{O3_i}}{2} + \frac{\Pi(D_i)}{3} + \frac{43\,\kappa_0}{3}  + \frac{\chi(O7_i)}{6} + \sum_{j \neq i}  \frac{\chi(O7_j)}{6}, \\
& & \hskip1.5cm = 0 +  \frac{306}{3} + \frac{43 \times 243}{3} +  \frac{549}{6} + \frac{3}{2} = 3678,\nonumber
\eea
where we have used the $\kappa_0 = 243$ and $\Pi = 306$ for the divisor $D_5$ from the Table \ref{tab_topo-list}. This value of $D3$ charge corresponds to the highest value of ${\cal Q}_{D3}^{\rm nonlocal}$ for all the CY geometries with $1 \leq h^{1,1}(CY) \leq 5$ in our scan, and it was claimed in \cite{Crino:2022zjk} as well. However, notice that this involution does not result in $O3$-planes in the fixed point set which we will elaborate more later on.

Similarly if we consider the involution to be $\sigma: x_4 \to - x_4$ which reflects the divisor with lower GLSM charge than that of the highest one, then the fixed point set has only one O$7$-plane denoted as O$7 = D_4$, and there are no O$3$-points present. Subsequently, one has the following estimates for the $N_{\rm flux}$ value corresponding to the D$7$-brane being on top of O$7$ scenario,
\begin{equation}
{\cal Q}_{D3}^{\rm simp} =\frac{\chi(O7)}{4} =\frac{\chi(O7_4)}{4} = \frac{276}{4} = 69. 
\end{equation}
We note that this value is relatively lower than the previous value. Moreover, due the absence of coincident $O3$-planes, this CY orientifold is also not suitable for uplifting proposal of \cite{Crino:2020qwk}. 

Unlike these two involutions, the remaining three equivalent involutions corresponding to $\sigma_i : x_i \to - x_i$ for $i \in \{1, 2, 3\}$. In this case there are $O3$-planes present along with the a single $O7_i = D_i$ for each $i \in \{1, 2, 3\}$. These are located as the following points of the CY threefold:
\bea
& (i). \quad & \sigma_1: O7= \{D_1\}, \qquad O3: \,\, D_2 D_3 D_5 = 3, \quad D_2 D_3 D_6 = 1;\\
& (ii). \quad & \sigma_2: O7= \{D_2\}, \qquad O3: \,\, D_1 D_3 D_5 = 3, \quad D_1 D_3 D_6 = 1;\nonumber\\
& (iii). \quad & \sigma_3: O7= \{D_3\}, \qquad O3: \,\, D_1 D_2 D_5 = 3, \quad D_1 D_2 D_6 = 1.\nonumber
\eea
Subsequently one gets the following $D3$-tadpole charges to be canceled,
\begin{equation}
{\cal Q}_{D3}^{\rm simp} =\frac{\chi(O7)+\chi(O3)}{4} =\frac{36+3+1}{4} = 10. 
\end{equation}
So this time we do have coincident $O3$-planes for these three involutions but the $\chi$ of respective divisors being not so large, it does not help in getting a large $D3$-tadpole charge. The reason for presenting this example is not only the fact that it is well known (probably the most studied one due to LVS phenomenology) but the following additional lessons one can take from this example:
\begin{itemize}

\item Though our scan is performed with CY geometries in the range $1 \leq h^{1,1} \leq 5$, this CY still enjoys to possess the (coordinate) divisor of highest $\chi$ and $\chi_{_h}$ with their values being given as 549 and 66 respectively.

\item For the involutions in which four $D7$-brane and their image branes are placed on top of the $O7$-planes, this CY geometry produces the largest value ${\cal Q}_{D3}^{\rm simp}$ throughout our scan. 


\item Given that the first three involutions (which are equivalent) can indeed give coincident $O3$-planes and so merely having coincident $O3$-planes is not a difficult task but having it on top of a divisor in the $O7$-plane with large $\chi$ (say more than 400) is indeed a challenging condition to fulfil.

\item A possible way out to realize large $D3$-brane tadpole charge can be through the realization of Whitney-brane configurations \cite{Crino:2020qwk} rather than placing all the $D7$-branes along with their images on top of the $O7$-planes. However as we will see in the next examples, that can be possible with reflecting the coordinates with a bit less-simple GLSM charges as compared to what we currently have, namely $\{1,0\}$, corresponding to first three coordinates of this CY geometry which can lead to $O3$-planes as well. The two coordinates which have such GLSM charges, and could have served for the purpose, namely $x_4$ and $x_5$ does not lead to involutions with $O3$-planes as we discussed before.

\end{itemize}

\subsubsection{Example 2: CY as a degree-14 hypersurafce in WCP$^4[1,1,2,3,7]$}
This CY corresponds to the polytope-id 39 of AGHJN database \cite{Altman:2014bfa}. The toric data for this swiss-cheese CY geometry is given as:
\begin{center}
\begin{tabular}{|c||cccccc|}
\hline
CY Hypersurface & $x_1$  & $x_2$  & $x_3$  & $x_4$  & $x_5$ & $x_6$     \\
\hline
14  & 0 & 1 & 1 & 2 & 3 & 7 \\
4  & 1 & 0 & 0 & 0 & 1 & 2  \\
\hline
Topology \# as  & 1& 65  & 65 & 102 &  145 & 410    \\
listed in Table \ref{tab_topo-list} & &  &  &    &  & \\
\hline
\end{tabular}
\end{center}
The Hodge numbers for this swiss-cheese are $(h^{2,1}, h^{1,1}) = (132, 2)$, the Euler number is $\chi(X)=-260$ and the SR ideal is given as:
\bea
& & {\rm SR} =  \{x_2 x_3 x_4, \,x_1 x_5 x_6\}. \nonumber
\eea
The fixed-point sets corresponding to reflecting each of the six coordinates are given as below,
\bea
& (i). \quad & \sigma_1: O7= \{D_1, D_5\}, \qquad O3: \,\, D_2 D_3 D_6 = 2;\\
& (ii). \quad & \sigma_2: O7= \{D_2\}, \qquad O3: \,\, D_1 D_3 D_6 = 1, \quad D_3 D_5 D_6 = 7;\nonumber\\
& (iii). \quad & \sigma_3: O7= \{D_3\}, \qquad O3: \,\, D_1 D_3 D_6 = 1, \quad D_3 D_5 D_6 = 7;\nonumber\\
& (iv). \quad & \sigma_4: O7= \{D_4\}, \qquad {\rm Trivial} \, \, O3 \, \, {\rm only};\nonumber\\
& (v). \quad & \sigma_5: O7= \{D_1, D_5\}, \qquad O3: \,\, D_2 D_3 D_6 = 2;\nonumber\\
& (vi). \quad & \sigma_6: O7= \{D_6\}, \qquad O3: \,\, D_1 D_2 D_3 = 1, \quad D_2 D_3 D_5 = 1.\nonumber
\eea
In this example, one can see that the involution $\sigma_6$ corresponding to the largest GLSM can have two $O3$-planes but they are not coincidental in nature. For $\sigma_1: x_1 \to - x_1$ or equivalently $\sigma_5: x_5 \to - x_5$ we have the following relation,
\begin{equation}
{\cal Q}_{D3}^{\rm simp}=\frac{\chi(O7_1)+\chi(O7_2)+\chi(O3)}{4} =\frac{3+75+2}{4} = 20. 
\end{equation}
This again shows that the simplistic case of placing $D7$-branes on top of $O7$-planes does not help in getting large $D3$-tadpole charge. However given that GLSM charge corresponding to the coordinate $x_5$ is quite non-trivial, for involution $\sigma_5$ one can construct a Whitney-brane configuration to have a non-local scenario for this involution. Subsequently using Eq.~(\ref{eq:D3-nonlocal}) we find 
\bea
& & {\cal Q}_{D3}^{\rm nonlocal} = \frac{N_{O3_i}}{2} + \frac{\Pi(D_i)}{3} + \frac{43\,\kappa_0}{3}  + \frac{\chi(O7_i)}{6} + \sum_{j \neq i}  \frac{\chi(O7_j)}{6}, \\
& & \hskip1.5cm = \frac{2}{2} +  \frac{66}{3} + \frac{43 \times 9}{3} +  \frac{75}{6} + \frac{3}{2} = 166,\nonumber
\eea
where we have used the $\kappa_0 = 9$ and $\Pi = 66$ for the divisor $D_5$ from the Table \ref{tab_topo-list}. This actually shows that inclusion of coincident $O3$-planes requirement reduces the maximum value of ${\cal Q}_{D3}^{\rm nonlocal}$ from 3678 corresponding to CY defined in WCP$^4[1,1,1,6,9]$ to 166 corresponding to this CY defined in WCP$^4[1,1,2,3,7]$.

\subsubsection{Example 3: CY as a degree-10 hypersurafce in WCP$^4[1,1,1,2,5]$}
There is another swiss-cheese CY threefold which has two $O7$-plane components along with coincident $O3$-planes. The toric data for this CY is given as:
\begin{center}
\begin{tabular}{|c||cccccc|}
\hline
CY Hypersurface & $x_1$  & $x_2$  & $x_3$  & $x_4$  & $x_5$ & $x_6$     \\
\hline
4 & 0 & 0 & 1 & 0 & 1 & 2  \\
10 & 1 & 1 & 0 & 2 & 1 & 5  \\
\hline
Topology \# as  & 65  & 65 & 9 & 102  & 83  &  388  \\
listed in Table \ref{tab_topo-list} & &  &  &    &  & \\
\hline
\end{tabular}
\end{center}
The Hodge numbers for this swiss-cheese are $(h^{2,1}, h^{1,1}) = (116, 2)$, the Euler number is $\chi(X)=- 228$ and the SR ideal is given as:
\bea
& & {\rm SR} =  \{x_1 x_2 x_4, \,x_3 x_5 x_6\}. \nonumber
\eea
The fixed-point sets corresponding to reflecting each of the six coordinates are given as below,
\bea
& (i). \quad & \sigma_1: O7= \{D_1\}, \qquad O3: \,\, D_2 D_3 D_6 = 3, \quad D_2 D_5 D_6 = 5 ;\\
& (ii). \quad & \sigma_2: O7= \{D_2\}, \qquad O3: \,\, D_1 D_3 D_6 = 3, \quad D_1 D_5 D_6 = 5;\nonumber\\
& (iii). \quad & \sigma_3: O7= \{D_3, D_5\}, \qquad O3: \,\, D_1 D_2 D_6 = 2;\nonumber\\
& (iv). \quad & \sigma_4: O7= \{D_4\}, \qquad {\rm Trivial} \, \, O3 \, \, {\rm only};\nonumber\\
& (v). \quad & \sigma_5: O7= \{D_3, D_5\}, \qquad O3: \,\, D_1 D_2 D_6 = 2;\nonumber\\
& (vi). \quad & \sigma_6: O7= \{D_6\}, \qquad O3: \,\, D_1 D_2 D_3 = 1, \quad D_1 D_2 D_5 = 1.\nonumber
\eea
In this example, one can see that the involution $\sigma_6$ corresponding to the largest GLSM can have two $O3$-planes but they are of 1+1 type and not coincidental in nature. For $\sigma_3: x_3 \to - x_3$ or equivalently $\sigma_5: x_5 \to - x_5$ we have the following relation,
\begin{equation}
{\cal Q}_{D3}^{\rm simp}=\frac{\chi(O7_1)+\chi(O7_2)+\chi(O3)}{4} =\frac{35+11+2}{4} = 12. 
\end{equation}
This again shows that the simplistic case of placing $D7$-branes on top of $O7$-planes does not help in getting large $D3$-tadpole charge. However given that GLSM charge corresponding to the coordinate $x_5$ is quite non-trivial, for involution $\sigma_5$ one can construct a Whitney-brane configuration to have a non-local scenario for this involution. Subsequently using Eq.~(\ref{eq:D3-nonlocal}) we find 
\bea
& & {\cal Q}_{D3}^{\rm nonlocal} = \frac{N_{O3_i}}{2} + \frac{\Pi(D_i)}{3} + \frac{43\,\kappa_0}{3}  + \frac{\chi(O7_i)}{6} + \sum_{j \neq i}  \frac{\chi(O7_j)}{6}, \\
& & \hskip1.5cm = \frac{2}{2} +  \frac{34}{3} + \frac{43 \times 1}{3} +  \frac{35}{6} + \frac{11}{2} = 38,\nonumber
\eea
where we have used the $\kappa_0 = 1$ and $\Pi = 34$ for the divisor $D_5$ from the Table \ref{tab_topo-list}. Such a low value is due to $\kappa_0 = 1$ being the minimal value and note that with a factor ot $43/3$ this cubic self-intersection number matters most for having large values of ${\cal Q}_{D3}^{\rm nonlocal}$. Therefore, this example is worse than the Example-2 we discussed before.

\subsection{Global model candidates for $\ov{D3}$ uplifting}

As we mentioned before, the uplifting proposal of \cite{Crino:2020qwk} demands to have coincident $O3$-planes in the fixed point set along with a large value of the $D3$ tadpole charge. Keeping this in mind we have analyzed all the possibilities including the so-called local and non-local cases, and we subsequently find that the value of $D3$ charge significantly decreases after imposing the requirement of the coincident $O3$-planes as seen from Table \ref{tab_D3-tadpole-charge}. We find that most of the involutions resulting in very high value of ${\cal Q}_{D3}$ via ${\cal Q}_{D3}^{\rm local}$ or ${\cal Q}_{D3}^{\rm nonlocal}$ cases do not have (coincident) $O3$-planes. Moreover, the estimated in Table \ref{tab_D3-tadpole-charge} are presented without taking into account LVS requirements, i.e. all the shortlisted CY geometries in Table \ref{tab_D3-tadpole-charge} need not be necessarily suitable for LVS. The possibility of meeting these topological requirements in LVS framework has been presented in Table \ref{tab_LVS-D3-tadpole-charge} where we also distinguish LVS models into two categories, one being $K3$-fibred LVS and other being non-$K3$-fibred LVS. From Table \ref{tab_LVS-D3-tadpole-charge} we also observe that a good percent of CY geometries have involutions resulting in coincident $O3$-planes, however this number significantly reduces after demanding that $D3$ charge is large, say ${\cal Q}_{D3} \geq 100$ or so.

\noindent
\begin{table}[h!]
\centering
\begin{tabular}{|c||c|c|c|c|} 
\hline
  &  &   &  &  \\ 
$h^{1,1}$  & 2 & 3 & 4 & 5 \\ 
  &  &   &  &  \\ 
\hline
  &  &   &  &  \\ 
\# of fav-CY  geom$^\ast$ ($X$)  & 37 & 300 & 1994 & 13494 \\ 
\# of divisor topologies of $X$  & 222 & 2100 & 15952 & 121446 \\ 
(i.e. \# max. possible $\sigma_i$) &   &  &  &  \\ 
 &  &   &  &  \\ 
\#($X$) with $\sigma_i$ satisfying ${\cal Q}_{D3}^{\rm simp} \in {\mathbb Z}$ & 30 & 247 & 1559 & 9742 \\ 
\# of all possible $\sigma_i$ satisfying ${\cal Q}_{D3}^{\rm simp} \in {\mathbb Z}$  & 180 & 1729 & 12472 & 87678 \\
 &  &  &  &  \\  
 \hline
 &  &  &  &  \\  
max(${\cal Q}_{D3}^{\rm local}$)   & 276 & 248 & 244 & 256 \\ 
max(${\cal Q}_{D3}^{\rm local}$) with  ``coincident" $\#(O3_i) \geq2$ & 42 & 110 & 112 & 128 \\ 
 &  &  &  &  \\ 
 \hline
  &  &  &  &   \\  
max(${\cal Q}_{D3}^{\rm nonlocal}$) & 3678 & 3272 & 3212 & 3280 \\ 
max(${\cal Q}_{D3}^{\rm nonlocal}$) with  ``coincident" $\#(O3_i) \geq2$ & 238 & 810 & 868 & 1374 \\ 
 &  &  &  &  \\ 
\hline
\end{tabular}
\caption{Maximum value of ${\cal Q}_{D3}^{\rm local}$ and ${\cal Q}_{D3}^{\rm nonlocal}$ with and without imposing the two (or more) coincident $O3$-planes condition, corresponding to an involution $\sigma_i$ for $2\leq h^{1,1}(CY) \leq 5$.}
\label{tab_D3-tadpole-charge}
\end{table}

\noindent
\begin{table}[h!]
\centering
\begin{tabular}{|c||c|c|c|c|} 
\hline
  &  &   &  &  \\ 
$h^{1,1}$  & 2 & 3 & 4 & 5 \\ 
  &  &   &  &  \\ 
\hline
  &  &   &  &  \\ 
\# of fav-CY  geom$^\ast$ ($X$)  & 37 & 300 & 1994 & 13494 \\ 
\# of divisor topologies of $X$  & 222 & 2100 & 15952 & 121446 \\ 
(i.e. \# max. possible $\sigma_i$) &   &  &  &  \\ 
 &  &   &  &  \\ 
\#($X$) with $\sigma_i$ satisfying ${\cal Q}_{D3}^{\rm simp} \in {\mathbb Z}$ & 30 & 247 & 1559 & 9742 \\ 
\# of all possible $\sigma_i$ satisfying ${\cal Q}_{D3}^{\rm simp} \in {\mathbb Z}$  & 180 & 1729 & 12472 & 87678 \\
 &  &  &  &  \\  
 \hline
   &  &  &  &  \\ 
{\bf (A)}: \#($X$) having at least one $\sigma_i$ with $\#(O7_i) \geq1$  & 25 & 214 & 1461 & 9446 \\ 
along with at least two ``coincident" $O3$-planes &   &  &  &  \\ 
  &  &  &  &  \\ 
\#($X$) with {\bf (A)} and max(${\cal Q}_{D3}$) $\geq 100$  & 3 & 55 & 451 & 4412 \\ 
  &  &  &  &  \\ 
\#($X$) with {\bf (A)} and ddP (LVS)  & 14 & 105 & 566 & 3025 \\ 
\#($X$) with {\bf (A)} which are $K3$-fibred  & 10 & 98 & 662 & 4386 \\ 
\#($X$) with {\bf (A)} and $K3$-fibred LVS  & 0 & 37 & 135 & 739 \\ 
  &  &  &    &  \\ 
\hline
  &  &  &  &  \\ 
{\bf (B)}: \#($X$) having at least one $\sigma_i$ with $\#(O7_i) \geq2$  & 2 & 49 & 653 & 5692 \\ 
along with at least two ``coincident" $O3$-planes &   &  &  &  \\ 
  &  &  &  &  \\ 
\#($X$) with {\bf (B)} and max(${\cal Q}_{D3}$) $\geq 100$  & 1 & 15 & 228 & 2546 \\ 
  &  &  &    &  \\ 
\#($X$) with {\bf (B)} and ddP (LVS)  & 2 & 35 & 298 & 2042 \\ 
\#($X$) with {\bf (B)} which are $K3$-fibred  & 0 & 17 & 249 & 2394 \\ 
\#($X$) with {\bf (B)} and $K3$-fibred LVS  & 0 & 13 & 72 & 503 \\ 
  &  &  &    &  \\ 
\hline
\end{tabular}
\caption{Number of LVS models with and without imposing the two (or more) coincident $O3$-planes condition, corresponding to an involution $\sigma_i$ for $2\leq h^{1,1}(CY) \leq 5$.}
\label{tab_LVS-D3-tadpole-charge}
\end{table}

\noindent
Let us close the discussion by presenting a swiss-cheese CY threefold which has multiple $O7$-plane components along with coincident $O3$-planes resulting in the largest ${\cal Q}_{D3} = 1374$ corresponding to $h^{1,1}(CY) = 5$. The toric data for this CY is given in Table \ref{tab_toric-data-h11=5} and this CY corresponds to the polytop-id: 6369. The Hodge numbers for this swiss-cheese are $(h^{2,1}, h^{1,1}) = (143, 5)$, the Euler number is $\chi(X)=- 276$ and the SR ideal is given as:
\bea
& & {\rm SR} =  \{{x_1 x_2, x_1 x_6, x_1 x_9, x_2 x_6, x_2 x_8, x_5 x_8, x_6 x_7, x_2 x_3 x_4, x_3 x_4 x_7, x_3 x_4 x_9, x_5 x_7 x_9}\}. \nonumber
\eea

\begin{table}[h!]
\centering
\begin{tabular}{|c||ccccccccc|}
\hline
CY Hypersurface & $x_1$  & $x_2$  & $x_3$  & $x_4$  & $x_5$ & $x_6$ & $x_7$ & $x_8$ & $x_9$     \\
\hline
3 & 0 & 0 & 0 & 0 & 1 & 1 & 0 & 1 & 0 \\
11 & 0 & 0 & 1 & 1 & 3 & 5 & 0 & 0 & 1 \\
14 & 0 & 0 & 1 & 1 & 4 & 7 & 1 & 0 & 0 \\
22& 0 & 1 & 2 & 2 & 6 & 11 & 0 & 0 & 0 \\
6 & 1 & 0 & 0 & 0 & 2 & 3 & 0 & 0 & 0 \\
\hline
\hline
Topology \# as  & 1  & 2 & 84 & 84  & 192  &  394 & 4 & 4 & 527 \\
listed in Table \ref{tab_topo-list} & &  &  &    &  & & & &  \\
\hline
$\chi_{_h}$ & 1&  1&  3&  3&  11&  28&  1&  1&  -10 \\
$\chi$ & 3&  4&  36&  36&  114&  247&  6&  6&  -38 \\
$\kappa_0$ & 9&  8&  0&  0&  18&  89&  6&  6&  -82 \\
$\Pi$ & -6&  -4&  36&  36&  96&  158&  0&  0 & 44 \\
\hline
\end{tabular}
\caption{Toric data for the swiss-cheese CY with largest $D3$ charge for $h^{1,1}(CY) \leq 5$.}
\label{tab_toric-data-h11=5}
\end{table}

\noindent
The fixed-point sets corresponding to reflecting each of the nine coordinates are given as below,
\bea
& (i). \quad & \sigma_1: O7= \{D_1, D_2, D_6\}, \qquad O3: \,\, D_7 D_8 D_9 = 2 ;\\
& (ii). \quad & \sigma_2: O7= \{D_1, D_2, D_6\}, \qquad O3: \,\, D_7 D_8 D_9 = 2 ;\nonumber\\
& (iii). \quad & \sigma_3: O7= \{D_3\}, \qquad O3: \,\, D_1 D_4 D_7 = 1, \quad D_2 D_4 D_7 = 3;\nonumber\\
& (iv). \quad & \sigma_4: O7= \{D_4\}, \qquad O3: \,\, D_1 D_3 D_7 = 1, \quad D_2 D_3 D_7 = 1;\nonumber\\
& (v). \quad & \sigma_5: O7= \{D_5\}, \qquad {\rm Trivial} \, \, O3 \, \, {\rm only};\nonumber\\
& (vi). \quad & \sigma_6: O7= \{D_1, D_2, D_6\}, \qquad O3: \,\, D_7 D_8 D_9 = 2 ;\nonumber\\
& (vii). \quad & \sigma_7: O7= \{D_7\}, \qquad O3: \,\, D_1 D_3 D_4 = 1, \quad D_3 D_4 D_6 = 3, \quad D_6 D_8 D_9 = 2;\nonumber\\
& (viii). \quad & \sigma_8: O7= \{D_8\}, \qquad O3: \,\, D_2 D_7 D_9 = 2;\nonumber\\
& (ix). \quad & \sigma_9: O7= \{D_9\}, \qquad O3: \,\, D_1 D_7 D_8 = 2.\nonumber
\eea
In this example, the involution $\sigma_6: x_6 \to - x_6$ we have the following relation,
\begin{equation}
{\cal Q}_{D3}^{\rm simp}=\frac{\chi(O7_1)+\chi(O7_2)+\chi(O7_3)+\chi(O3)}{4} =\frac{3+4+247+2}{4} = 64.
\end{equation}
Having ${\cal Q}_{D3}^{\rm simp}$ divisible by 4 means that this involution should be a priori well defined, and leads to ${\cal Q}_{D3}^{\rm local} = 2\,{\cal Q}_{D3}^{\rm simp} = 128$ as mentioned in Table \ref{tab_D3-tadpole-charge}. 
This again shows that the simplistic case of placing $D7$-branes on top of $O7$-planes may not help in getting large enough $D3$-tadpole charge. 

However given that GLSM charge corresponding to the coordinate $x_6$ is quite non-trivial, for involution $\sigma_6$ one can construct a Whitney-brane configuration to have a non-local scenario for this involution. Subsequently using Eq.~(\ref{eq:D3-nonlocal}) we find 
\bea
& & {\cal Q}_{D3}^{\rm nonlocal} = \frac{N_{O3_i}}{2} + \frac{\Pi(D_i)}{3} + \frac{43\,\kappa_0}{3}  + \frac{\chi(O7_i)}{6} + \sum_{j \neq i}  \frac{\chi(O7_j)}{6}, \\
& & \hskip1.5cm = \frac{2}{2} +  \frac{158}{3} + \frac{43 \times 89}{3} +  \frac{247}{6} + \frac{3+4}{2} = 1374,\nonumber
\eea
which has been mentioned in the Table \ref{tab_D3-tadpole-charge}. In the last line we have used the $\kappa_0 = 89$ and $\Pi = 158$ for the divisor $D_6$ from the Table \ref{tab_toric-data-h11=5}. Notice that such a large $D3$ tadpole estimate is mainly due to a large value of the self cubic intersection being $\kappa_0 = 89$.


\section{Conclusions}
\label{sec_conclusions}

There have been several databases of CY threefolds which have got tremendous amount of attention in recent years, and the two main ones are the so-called the Kreuzer-Skarke (KS) dataset of four-dimensional reflexive polytopes \cite{Kreuzer:2000xy} and the complete intersection Calabi Yau (CICY) database \cite{Candelas:1987kf}. 

In this article we present a pheno-inspired classification for the divisor topologies of the {\it favorable} Calabi Yau (CY) threefolds with $1 \leq h^{1,1}(CY) \leq 5$ arising from the four-dimensional reflexive polytopes of the Kreuzer-Skarke database. Our main focus is to study the topologies corresponding to the so-called `coordinate divisors' which descend from the Ambient space. In this regard, firstly we have computed the Hodge numbers of all such coordinate divisors for CY threefolds up to $h^{1,1}(CY) \leq 4$ using {\it cohomCalg} and subsequently based on some peculiar observations and patterns we conjecture that there can be only two classes of divisor topologies; either having $\chi_h(D) \geq 1$ or $\chi_h(D) \leq 1$ such that their corresponding Hodge numbers are respectively given by $\{h^{0,0} = 1, \, h^{1,0} = 0, \, h^{2,0} = \chi_{_h}(D) -1, \, h^{1,1} = \chi(D) - 2 \chi_{_h}(D) \}$ and $\{h^{0,0} = 1, \, h^{1,0} = 1 - \chi_{_h}, \, h^{2,0} = 0, \, h^{1,1} = \chi(D) + 2 - 4 \chi_{_h}(D)\}$, where $\chi_{_h}(D)$ denotes the Arithmetic genus while $\chi(D)$ denotes the Euler characteristic of the divisor $D$. Given that the two topological quantities $\chi_{_h}(D)$ and $\chi(D)$ can be directly computed from the classical triple intersection numbers and the second Chern class of the CY threefolds, we present the Hodge numbers of 139740 coordinate divisors corresponding to a total of nearly 15829 CY geometries with $1 \leq h^{1,1}(CY) \leq 5$ as listed in Table \ref{tab_number-of-space-and-divisors}.  In the detailed analysis we find that there are a total of 565 distinct topologies which arise from these CY threefold. In particular we find that there are 76839 divisors of what we call as $R$-type, 55972 divisors of $K$-type and 6929 divisors of $W$-type. Given that every reader is not always equally encouraged to read the data file by downloading the same as an external attachment, for their convenience a brief classification has been presented in Table \ref{tab_topo-list} of the Appendix \ref{sec_appendix1}. In addition we have discussed some of the salient features and the insights of the classification results in separate sections.

Subsequently we have argued that our conjecture can help in ``bypassing" the need of {\it cohomCalg} for computing Hodge numbers of coordinate divisors, and hence can be significantly useful for studying the divisor topologies of CY threefolds with higher $h^{1,1}$ for which {\it cohomCalg} gets too slow and sometimes even breaks as well. We also demonstrate how these scanning results can be directly used for phenomenological model building, e.g. in estimating the $D3$-brane tadpole charge (under reflection involutions) which is a central ingredient for constructing explicit global models based on several different reasons/interests such as (flat) flux vacua searches and the de-Sitter uplifting through anti-$D3$ brane. 

\subsection*{Some insights with observations}
Based on the classification of the distinct topologies, we make the following interesting observations:
\begin{itemize}

\item From our divisor topology analysis which has led us to the conjecture in Eq.~(\ref{eq:conjecture}), we classify all the coordinate divisor topologies in three classes. In fact these can be effectively clubbed into two as given in Eq.~(\ref{eq:divisor-topology}) by placing the third one at the boundary of the remaining two classes as per being $\chi_{_h}(D) \leq 1$ or $\chi_{_h}(D) \geq 1$.

\item We observe that there are a huge number of rigid surfaces present in the divisor topologies. In particular there are several rigid surface which satisfy unit Arithmetic genus condition and hence can be a priori suitable for generating non-perturbative superpotential effects. This properties of the divisors of THCYs is unlike the case with the other database consisting of pCICYs defined by some multi-hypersurface constraints in the product of $ {\mathbb P}^n$'s for which no coordinate divisor was found to be rigid \cite{Carta:2022web}.

\item From Table \ref{tab_non-rigid-topo-list} we can see that the $K3$-surface is the most frequent divisor topology which appears 12970 times. We find that the \% of $K3$-fibred CY threefolds increases with increasing $h^{1,1}(CY)$ as seen from the Table \ref{tab_K3fibred-topologies}.

\item We also observe that there are no ${\mathbb T}^4$ surfaces present as a coordinate divisor of the favorable CY threefolds of the Kreuzer-Skarke database. In fact, to the best of our knowledge the only CY threefolds for which ${\mathbb T}^4$ surfaces appears to be a coordinate divisor are among the Shoen manifolds which are non-favorable CICY threefolds \cite{Carta:2022web}.

\item We observe that the self cubic-intersection number $\int_{_{\rm CY}} \hat{D} \wedge {\hat D} \wedge {\hat D}$ vanishes for 7 topologies as listed in Table \ref{tab_kappaZero-topo-list} which means their Hodge numbers satisfy Eq.~(\ref{eq:Dcube=0}). In fact, the most famous ones in this class can be thought of to be given as $dP_9$, $K3$ and the so-called special deformation divisor\footnote{Actually there has been a trend of calling the divisors with $h^{2,0}(D) > 1$ as `special deformation' divisors from \cite{Gao:2013pra} because it was observed to be frequently appearing in our scan of the so-called NIDs, and hence worthy of some `special' attention. Otherwise one may like to name divisors with $h^{2,0}(D) = 1$ as `special' or $K3$ or $K3$-like, while those with $h^{2,0}(D) > 1$ as `generic' deformation rather than `special' deformation. The simplest one of this kind corresponds to the Hodge number $\{1, 0, 2, 30\}$ which, for example, also appears as the first three divisors of the famous WCP$^4[1,1,1,6,9]$ model used for realizing LVS.} SD numbered as 84 in our collection of 565 distinct divisor topologies in Table \ref{tab_topo-list}.

\item
We observe that $\Pi$ is always an even number taking the values in the range $-6 \leq \Pi \leq 306$. Moreover there are two divisor topologies of vanishing $\Pi$ as collected in Table \ref{tab_PiZero-topo-list}. These are the so-called dP$_3$ and the ``Exact Wilson" divisor with $h^{1,1} = 2$. 

\item We classify peculiar divisors which have $\chi_{_h}$ and $\chi$ both vanishing, and equivalently the cubic self-intersections $\kappa_0$ as well as $\Pi$ also vanish for these surfaces. From our scan, we find that there is just a unique topology of this kind which corresponds to the ``Exact Wilson" divisor with $h^{1,1} = 2$ as encountered in \cite{Blumenhagen:2012kz}.

\item
We find that the divisor with maximum $\chi$ with a value $\chi = 549$ corresponds to the well known degree-18 CY threefold realized in WCP$^4[1,1,1,6,9]$ corresponding to $h^{1,1}(CY) = 2$. in fact, this Cy has a divisor which has the highest value for all the four topological quantities we focused in the current study. These are $\chi_{_h} = 66, \chi = 549, \kappa_0 = 243$ and $\Pi = 306$. So this vanilla CY threefold extensively considered in LVS model building and related studies continues to amaze us with interesting properties.

\item
From Table \ref{tab_topo-list} we observe that for all the rigid divisors of $R$-types as well as $W$-type, the self triple-intersections $\kappa_0$ are non-positive, except for $R_n$ topologies with $1\leq n \leq 9$. This corresponds to ${\mathbb P}^2$ and the eight del-Pezzo divisors obtained by blowing up generic points in it. For all other divisors with $h^{2,0}(D) = 0$, the cubic self-intersection $\kappa_0$ vanishes. This also means that all divisors of $W$-type  always have non-positive $\kappa_0$, and in total, there are only 9 divisor topologies out of 170 of $R+W$-type, which can have positive $\kappa_0$.

\end{itemize}

\noindent
Recalling the days of early nineties when it used to take a tremendous amount of effort for working out the necessary global model building ingredients for a few spaces \cite{Candelas:1993dm,Candelas:1994hw,Hosono:1994ax}, as a model builder we have certainly been much more equipped in current times with extra ordinary tools for studying the CY properties, and the recent serge in this direction is very encouraging. Some of the inital phenomenological applications of the current work in the context of inflation and de-Sitter realization are to be discussed in companion work \cite{Cicoli:2022abc,AbdusSalam:2022krp}.


\section*{Acknowledgments}
I am very thankful to Ralph Blumenhagen, Andreas Braun, Federico Carta, Michele Cicoli, Chiara Crin\`o, Xin Gao,  I\~naki Garc\'\i{}a-Etxebarria, Arthur Hebecker, Christoph Mayrhofer, Alessandro Mininno, Fernando Quevedo, Thorsten Rahn, Andreas Schachner and Roberto Valandro for useful discussions at various different stages. I would also like to thank Roberto Valandro for sharing the information about their work \cite{Crino:2022zjk} being in progress which turned out to have some significantly overlapping interests with the current work. In addition, I would like to thank the SISSA/ICTP HPC Cluster for allowing the access, and in particular Benvenuto Bazzo, Ivan Girotto and Johannes Grassberger at ICTP for their crucial technical help. Moreover, I am thankful to the organizers of the ``\href{https://sites.google.com/view/string-pheno-seminars/home}{Seminar series on string phenomenology}" (especially Viraf Mehta, Fabian Ruehle, and Pablo Soler) for the opportunity to present (a part of) this work at their \href{https://www.youtube.com/watch?v=nDJy7QnNAZ8}{online forum}. I am also grateful to Paolo Creminelli, Atish Dabholkar and Fernando Quevedo for their continued support.


\appendix
\setcounter{equation}{0}


\section{A complete list of distinct topologies of all the coordinate divisors}
\label{sec_appendix1}
In Table \ref{tab_topo-list}, we list the topological quantities corresponding to all the coordinate divisors of the favourable CY geometries with $1 \leq h^{1,1}(CY) \leq 5$ arising from the Kreuzer-Skarke database \cite{Kreuzer:2000xy}. In the collection of Table \ref{tab_topo-list}, $h^{p,q}$ denotes the set of four Hodge numbers of the divisors collected as: $h^{p,q} = \{h^{0,0}, h^{1,0}, h^{2,0},h^{1,1}\}$. Further, $\chi_{_h}$ denotes the Arithmetic genus, $\chi$ denotes the Euler characteristics, $\kappa_0$ denotes the self-triple intersection number and $\Pi$ denotes the quantity $\int c_2(CY)\wedge \hat{D}$.
\vskip-0.5cm
\noindent
\begin{center}
\renewcommand\arraystretch{1}
\begin{longtable}{|c||c|c|c|c|c||c|c|c|c|c|c|} 
\caption{List of all distinct divisor topologies for CY threefolds with $1 \leq n \leq 5$, where $n = h^{1,1}(CY)$ and ``$f_n$" denotes the frequency of a divisor for a given $n$.}\\
\hline
&&&&&&&&&&& \\
\# & $h^{p,q}$  & $\chi_{_h}$ & $\chi$ & $\kappa_0$ & $\Pi$ & $f_1$ & $f_2$ & $f_3$ & $f_4$ & $f_5$ & $f_{\rm all}$ \\
&&&&&&&&&&& \\
\hhline{|=|=|=|=|=|=|=|=|=|=|=|=|}
\endhead
\label{tab_topo-list}
&&&&&&&&&&& \\
 1 & \{1,0,0,1\} & 1 & 3 & 9 & -6 & 0 & 8 & 59 & 372 & 2410 & 2849 \\
 2 & \{1,0,0,2\} & 1 & 4 & 8 & -4 & 0 & 4 & 103 & 999 & 9224 & 10330 \\
 3 & \{1,0,0,3\} & 1 & 5 & 7 & -2 & 0 & 0 & 4 & 160 & 2360 & 2524 \\
 4 & \{1,0,0,4\} & 1 & 6 & 6 & 0 & 0 & 0 & 4 & 152 & 2441 & 2597 \\
 5 & \{1,0,0,5\} & 1 & 7 & 5 & 2 & 0 & 0 & 6 & 103 & 1586 & 1695 \\
 6 & \{1,0,0,6\} & 1 & 8 & 4 & 4 & 0 & 0 & 9 & 198 & 2574 & 2781 \\
 7 & \{1,0,0,7\} & 1 & 9 & 3 & 6 & 0 & 2 & 21 & 250 & 2705 & 2978 \\
 8 & \{1,0,0,8\} & 1 & 10 & 2 & 8 & 0 & 4 & 67 & 714 & 5988 & 6773 \\
 9 & \{1,0,0,9\} & 1 & 11 & 1 & 10 & 0 & 5 & 75 & 673 & 5772 & 6525 \\
 10 & \{1,0,0,10\} & 1 & 12 & 0 & 12 & 0 & 0 & 54 & 927 & 8983 & 9964 \\
 11 & \{1,0,0,11\} & 1 & 13 & -1 & 14 & 0 & 1 & 35 & 516 & 5364 & 5916 \\
 12 & \{1,0,0,12\} & 1 & 14 & -2 & 16 & 0 & 1 & 34 & 605 & 5508 & 6148 \\
 13 & \{1,0,0,13\} & 1 & 15 & -3 & 18 & 0 & 0 & 29 & 383 & 3758 & 4170 \\
 14 & \{1,0,0,14\} & 1 & 16 & -4 & 20 & 0 & 1 & 30 & 354 & 3190 & 3575 \\
 15 & \{1,0,0,15\} & 1 & 17 & -5 & 22 & 0 & 2 & 9 & 103 & 1054 & 1168 \\
 16 & \{1,0,0,16\} & 1 & 18 & -6 & 24 & 0 & 1 & 22 & 258 & 1737 & 2018 \\
 17 & \{1,0,0,17\} & 1 & 19 & -7 & 26 & 0 & 1 & 11 & 103 & 970 & 1085 \\
 18 & \{1,0,0,18\} & 1 & 20 & -8 & 28 & 0 & 0 & 3 & 64 & 614 & 681 \\
 19 & \{1,0,0,19\} & 1 & 21 & -9 & 30 & 0 & 0 & 2 & 37 & 520 & 559 \\
 20 & \{1,0,0,20\} & 1 & 22 & -10 & 32 & 0 & 0 & 7 & 121 & 517 & 645 \\
 21 & \{1,0,0,21\} & 1 & 23 & -11 & 34 & 0 & 1 & 8 & 51 & 317 & 377 \\
 22 & \{1,0,0,22\} & 1 & 24 & -12 & 36 & 0 & 0 & 0 & 29 & 236 & 265 \\
 23 & \{1,0,0,23\} & 1 & 25 & -13 & 38 & 0 & 0 & 1 & 11 & 126 & 138 \\
 24 & \{1,0,0,24\} & 1 & 26 & -14 & 40 & 0 & 0 & 4 & 36 & 214 & 254 \\
 25 & \{1,0,0,25\} & 1 & 27 & -15 & 42 & 0 & 0 & 4 & 30 & 152 & 186 \\
 26 & \{1,0,0,26\} & 1 & 28 & -16 & 44 & 0 & 0 & 1 & 16 & 110 & 127 \\
 27 & \{1,0,0,27\} & 1 & 29 & -17 & 46 & 0 & 0 & 0 & 4 & 26 & 30 \\
 28 & \{1,0,0,28\} & 1 & 30 & -18 & 48 & 0 & 0 & 1 & 8 & 44 & 53 \\
 29 & \{1,0,0,29\} & 1 & 31 & -19 & 50 & 0 & 1 & 4 & 15 & 82 & 102 \\
 30 & \{1,0,0,30\} & 1 & 32 & -20 & 52 & 0 & 0 & 2 & 8 & 51 & 61 \\
 31 & \{1,0,0,31\} & 1 & 33 & -21 & 54 & 0 & 0 & 2 & 8 & 35 & 45 \\
 32 & \{1,0,0,32\} & 1 & 34 & -22 & 56 & 0 & 0 & 3 & 7 & 29 & 39 \\
 33 & \{1,0,0,33\} & 1 & 35 & -23 & 58 & 0 & 0 & 0 & 0 & 7 & 7 \\
 34 & \{1,0,0,34\} & 1 & 36 & -24 & 60 & 0 & 0 & 0 & 1 & 8 & 9 \\
 35 & \{1,0,0,35\} & 1 & 37 & -25 & 62 & 0 & 0 & 0 & 0 & 5 & 5 \\
 36 & \{1,0,0,36\} & 1 & 38 & -26 & 64 & 0 & 0 & 1 & 5 & 25 & 31 \\
 37 & \{1,0,0,37\} & 1 & 39 & -27 & 66 & 0 & 0 & 0 & 0 & 8 & 8 \\
 38 & \{1,0,0,38\} & 1 & 40 & -28 & 68 & 0 & 0 & 0 & 0 & 3 & 3 \\
 39 & \{1,0,0,39\} & 1 & 41 & -29 & 70 & 0 & 0 & 0 & 0 & 1 & 1 \\
 40 & \{1,0,0,40\} & 1 & 42 & -30 & 72 & 0 & 0 & 0 & 3 & 10 & 13 \\
 41 & \{1,0,0,41\} & 1 & 43 & -31 & 74 & 0 & 1 & 1 & 3 & 10 & 15 \\
 42 & \{1,0,0,42\} & 1 & 44 & -32 & 76 & 0 & 0 & 0 & 0 & 2 & 2 \\
 43 & \{1,0,0,43\} & 1 & 45 & -33 & 78 & 0 & 0 & 0 & 0 & 4 & 4 \\
 44 & \{1,0,0,44\} & 1 & 46 & -34 & 80 & 0 & 0 & 0 & 0 & 3 & 3 \\
 45 & \{1,0,0,48\} & 1 & 50 & -38 & 88 & 0 & 0 & 0 & 0 & 1 & 1 \\
 46 & \{1,0,0,49\} & 1 & 51 & -39 & 90 & 0 & 0 & 0 & 0 & 6 & 6 \\
 47 & \{1,0,0,50\} & 1 & 52 & -40 & 92 & 0 & 0 & 0 & 0 & 19 & 19 \\
 48 & \{1,0,0,54\} & 1 & 56 & -44 & 100 & 0 & 0 & 0 & 0 & 1 & 1 \\
 49 & \{1,0,0,55\} & 1 & 57 & -45 & 102 & 0 & 0 & 0 & 5 & 4 & 9 \\
 50 & \{1,0,0,57\} & 1 & 59 & -47 & 106 & 0 & 0 & 0 & 0 & 1 & 1 \\
 51 & \{1,0,0,60\} & 1 & 62 & -50 & 112 & 0 & 0 & 0 & 0 & 3 & 3 \\
 52 & \{1,0,0,66\} & 1 & 68 & -56 & 124 & 0 & 0 & 0 & 0 & 1 & 1 \\
 53 & \{1,0,0,71\} & 1 & 73 & -61 & 134 & 0 & 0 & 0 & 1 & 4 & 5 \\
 54 & \{1,0,0,72\} & 1 & 74 & -62 & 136 & 0 & 0 & 0 & 0 & 3 & 3 \\
 55 & \{1,0,0,78\} & 1 & 80 & -68 & 148 & 0 & 0 & 0 & 0 & 4 & 4 \\
 56 & \{1,0,0,81\} & 1 & 83 & -71 & 154 & 0 & 0 & 0 & 0 & 2 & 2 \\
 57 & \{1,0,0,84\} & 1 & 86 & -74 & 160 & 0 & 0 & 0 & 0 & 5 & 5 \\
 58 & \{1,0,0,87\} & 1 & 89 & -77 & 166 & 0 & 0 & 0 & 0 & 9 & 9 \\
 59 & \{1,0,0,89\} & 1 & 91 & -79 & 170 & 0 & 0 & 0 & 2 & 1 & 3 \\
 60 & \{1,0,0,90\} & 1 & 92 & -80 & 172 & 0 & 0 & 0 & 0 & 2 & 2 \\
 61 & \{1,0,0,96\} & 1 & 98 & -86 & 184 & 0 & 0 & 0 & 0 & 1 & 1 \\
 62 & \{1,0,0,98\} & 1 & 100 & -88 & 188 & 0 & 0 & 0 & 4 & 0 & 4 \\
 63 & \{1,0,0,109\} & 1 & 111 & -99 & 210 & 0 & 0 & 1 & 0 & 0 & 1 \\
&&&&&&&&&&& \\
\hline
&&&&&&&&&&& \\
 64 & \{1,0,1,19\} & 2 & 23 & 1 & 22 & 0 & 0 & 8 & 5 & 43 & 56 \\
{\bf 65} &{\bf  \{1,0,1,20\}} & {\bf 2} & {\bf 24} & {\bf 0} & {\bf 24} & {\bf 0} & {\bf 24} & {\bf 322} & {\bf 1879} & {\bf 10745} & {\bf 12970} \\
 66 & \{1,0,1,21\} & 2 & 25 & -1 & 26 & 0 & 8 & 124 & 965 & 6601 & 7698 \\
 67 & \{1,0,1,22\} & 2 & 26 & -2 & 28 & 0 & 4 & 53 & 391 & 2976 & 3424 \\
 68 & \{1,0,1,23\} & 2 & 27 & -3 & 30 & 0 & 0 & 20 & 213 & 1376 & 1609 \\
 69 & \{1,0,1,24\} & 2 & 28 & -4 & 32 & 0 & 4 & 25 & 125 & 927 & 1081 \\
 70 & \{1,0,1,25\} & 2 & 29 & -5 & 34 & 0 & 0 & 4 & 35 & 276 & 315 \\
 71 & \{1,0,1,26\} & 2 & 30 & -6 & 36 & 0 & 0 & 4 & 39 & 235 & 278 \\
 72 & \{1,0,1,27\} & 2 & 31 & -7 & 38 & 0 & 0 & 0 & 10 & 62 & 72 \\
 73 & \{1,0,1,28\} & 2 & 32 & -8 & 40 & 0 & 0 & 7 & 15 & 182 & 204 \\
 74 & \{1,0,1,29\} & 2 & 33 & -9 & 42 & 0 & 0 & 0 & 4 & 8 & 12 \\
 75 & \{1,0,1,30\} & 2 & 34 & -10 & 44 & 0 & 0 & 0 & 0 & 22 & 22 \\
 76 & \{1,0,1,31\} & 2 & 35 & -11 & 46 & 0 & 0 & 0 & 0 & 1 & 1 \\
 77 & \{1,0,1,32\} & 2 & 36 & -12 & 48 & 0 & 0 & 0 & 1 & 115 & 116 \\
 78 & \{1,0,1,36\} & 2 & 40 & -16 & 56 & 0 & 0 & 0 & 0 & 52 & 52 \\
 79 & \{1,0,1,38\} & 2 & 42 & -18 & 60 & 0 & 0 & 0 & 0 & 1 & 1 \\
 80 & \{1,0,2,26\} & 3 & 32 & 4 & 28 & 0 & 0 & 0 & 0 & 1 & 1 \\
 81 & \{1,0,2,27\} & 3 & 33 & 3 & 30 & 0 & 0 & 0 & 0 & 14 & 14 \\
 82 & \{1,0,2,28\} & 3 & 34 & 2 & 32 & 0 & 6 & 11 & 69 & 192 & 278 \\
 83 & \{1,0,2,29\} & 3 & 35 & 1 & 34 & 3 & 16 & 63 & 250 & 1136 & 1468 \\
 {\bf 84} & {\bf \{1,0,2,30\}} & {\bf 3} & {\bf 36} & {\bf 0} & {\bf 36} & {\bf 0} & {\bf 30} & {\bf 235} & {\bf 1145} & {\bf 5851} & {\bf 7261} \\
 85 & \{1,0,2,31\} & 3 & 37 & -1 & 38 & 0 & 0 & 7 & 86 & 799 & 892 \\
 86 & \{1,0,2,32\} & 3 & 38 & -2 & 40 & 0 & 0 & 4 & 48 & 412 & 464 \\
 87 & \{1,0,2,33\} & 3 & 39 & -3 & 42 & 0 & 0 & 4 & 24 & 241 & 269 \\
 88 & \{1,0,2,34\} & 3 & 40 & -4 & 44 & 0 & 0 & 2 & 17 & 104 & 123 \\
 89 & \{1,0,2,35\} & 3 & 41 & -5 & 46 & 0 & 0 & 0 & 4 & 52 & 56 \\
 90 & \{1,0,2,36\} & 3 & 42 & -6 & 48 & 0 & 0 & 2 & 3 & 43 & 48 \\
 91 & \{1,0,2,37\} & 3 & 43 & -7 & 50 & 0 & 0 & 0 & 2 & 5 & 7 \\
 92 & \{1,0,2,38\} & 3 & 44 & -8 & 52 & 0 & 0 & 0 & 0 & 8 & 8 \\
 93 & \{1,0,2,40\} & 3 & 46 & -10 & 56 & 0 & 0 & 0 & 0 & 1 & 1 \\
 94 & \{1,0,2,42\} & 3 & 48 & -12 & 60 & 0 & 0 & 0 & 0 & 1 & 1 \\
 95 & \{1,0,3,33\} & 4 & 41 & 7 & 34 & 0 & 0 & 0 & 3 & 1 & 4 \\
 96 & \{1,0,3,34\} & 4 & 42 & 6 & 36 & 0 & 0 & 0 & 2 & 5 & 7 \\
 97 & \{1,0,3,35\} & 4 & 43 & 5 & 38 & 0 & 0 & 1 & 8 & 27 & 36 \\
 98 & \{1,0,3,36\} & 4 & 44 & 4 & 40 & 0 & 0 & 4 & 57 & 117 & 178 \\
 99 & \{1,0,3,37\} & 4 & 45 & 3 & 42 & 4 & 10 & 45 & 166 & 809 & 1034 \\
 100 & \{1,0,3,38\} & 4 & 46 & 2 & 44 & 4 & 26 & 88 & 423 & 2063 & 2604 \\
 101 & \{1,0,3,39\} & 4 & 47 & 1 & 46 & 0 & 0 & 1 & 23 & 171 & 195 \\
 102 & \{1,0,3,40\} & 4 & 48 & 0 & 48 & 0 & 2 & 24 & 130 & 789 & 945 \\
 103 & \{1,0,3,41\} & 4 & 49 & -1 & 50 & 0 & 0 & 3 & 9 & 112 & 124 \\
 104 & \{1,0,3,42\} & 4 & 50 & -2 & 52 & 0 & 0 & 1 & 14 & 99 & 114 \\
 105 & \{1,0,3,43\} & 4 & 51 & -3 & 54 & 0 & 2 & 1 & 8 & 75 & 86 \\
 106 & \{1,0,3,44\} & 4 & 52 & -4 & 56 & 0 & 0 & 1 & 3 & 43 & 47 \\
 107 & \{1,0,3,45\} & 4 & 53 & -5 & 58 & 0 & 0 & 0 & 0 & 15 & 15 \\
 108 & \{1,0,3,46\} & 4 & 54 & -6 & 60 & 0 & 0 & 0 & 0 & 28 & 28 \\
 109 & \{1,0,3,48\} & 4 & 56 & -8 & 64 & 0 & 0 & 0 & 0 & 3 & 3 \\
 110 & \{1,0,3,49\} & 4 & 57 & -9 & 66 & 0 & 0 & 0 & 0 & 14 & 14 \\
 111 & \{1,0,3,52\} & 4 & 60 & -12 & 72 & 0 & 0 & 0 & 0 & 5 & 5 \\
 112 & \{1,0,4,42\} & 5 & 52 & 8 & 44 & 0 & 0 & 2 & 1 & 12 & 15 \\
 113 & \{1,0,4,43\} & 5 & 53 & 7 & 46 & 0 & 0 & 1 & 7 & 47 & 55 \\
 114 & \{1,0,4,44\} & 5 & 54 & 6 & 48 & 0 & 2 & 19 & 47 & 284 & 352 \\
 115 & \{1,0,4,45\} & 5 & 55 & 5 & 50 & 5 & 7 & 27 & 126 & 741 & 906 \\
 116 & \{1,0,4,46\} & 5 & 56 & 4 & 52 & 0 & 4 & 34 & 148 & 723 & 909 \\
 117 & \{1,0,4,47\} & 5 & 57 & 3 & 54 & 0 & 0 & 0 & 0 & 16 & 16 \\
 118 & \{1,0,4,48\} & 5 & 58 & 2 & 56 & 0 & 0 & 0 & 3 & 39 & 42 \\
 119 & \{1,0,4,49\} & 5 & 59 & 1 & 58 & 0 & 0 & 1 & 2 & 22 & 25 \\
 120 & \{1,0,4,50\} & 5 & 60 & 0 & 60 & 0 & 0 & 1 & 5 & 103 & 109 \\
 121 & \{1,0,4,51\} & 5 & 61 & -1 & 62 & 0 & 0 & 0 & 0 & 12 & 12 \\
 122 & \{1,0,4,52\} & 5 & 62 & -2 & 64 & 0 & 0 & 0 & 0 & 7 & 7 \\
 123 & \{1,0,4,53\} & 5 & 63 & -3 & 66 & 0 & 0 & 0 & 0 & 9 & 9 \\
 124 & \{1,0,4,54\} & 5 & 64 & -4 & 68 & 0 & 0 & 0 & 0 & 2 & 2 \\
 125 & \{1,0,4,55\} & 5 & 65 & -5 & 70 & 0 & 0 & 0 & 0 & 10 & 10 \\
 126 & \{1,0,4,56\} & 5 & 66 & -6 & 72 & 0 & 0 & 0 & 0 & 14 & 14 \\
 127 & \{1,0,4,59\} & 5 & 69 & -9 & 78 & 0 & 0 & 0 & 0 & 4 & 4 \\
 128 & \{1,0,5,49\} & 6 & 61 & 11 & 50 & 0 & 0 & 0 & 2 & 0 & 2 \\
 129 & \{1,0,5,50\} & 6 & 62 & 10 & 52 & 0 & 0 & 0 & 6 & 26 & 32 \\
 130 & \{1,0,5,51\} & 6 & 63 & 9 & 54 & 0 & 0 & 1 & 10 & 93 & 104 \\
 131 & \{1,0,5,52\} & 6 & 64 & 8 & 56 & 0 & 7 & 15 & 84 & 406 & 512 \\
 132 & \{1,0,5,53\} & 6 & 65 & 7 & 58 & 0 & 0 & 2 & 18 & 73 & 93 \\
 133 & \{1,0,5,54\} & 6 & 66 & 6 & 60 & 0 & 1 & 11 & 52 & 360 & 424 \\
 134 & \{1,0,5,55\} & 6 & 67 & 5 & 62 & 0 & 0 & 0 & 0 & 10 & 10 \\
 135 & \{1,0,5,56\} & 6 & 68 & 4 & 64 & 0 & 0 & 0 & 4 & 19 & 23 \\
 136 & \{1,0,5,60\} & 6 & 72 & 0 & 72 & 0 & 0 & 0 & 0 & 10 & 10 \\
 137 & \{1,0,5,61\} & 6 & 73 & -1 & 74 & 0 & 0 & 0 & 0 & 1 & 1 \\
 138 & \{1,0,5,62\} & 6 & 74 & -2 & 76 & 0 & 0 & 0 & 0 & 1 & 1 \\
 139 & \{1,0,5,63\} & 6 & 75 & -3 & 78 & 0 & 0 & 0 & 0 & 1 & 1 \\
 140 & \{1,0,6,56\} & 7 & 70 & 14 & 56 & 0 & 0 & 1 & 0 & 6 & 7 \\
 141 & \{1,0,6,57\} & 7 & 71 & 13 & 58 & 0 & 0 & 2 & 3 & 15 & 20 \\
 142 & \{1,0,6,58\} & 7 & 72 & 12 & 60 & 0 & 0 & 4 & 23 & 228 & 255 \\
 143 & \{1,0,6,59\} & 7 & 73 & 11 & 62 & 0 & 1 & 8 & 34 & 211 & 254 \\
 144 & \{1,0,6,60\} & 7 & 74 & 10 & 64 & 0 & 0 & 0 & 0 & 5 & 5 \\
 145 & \{1,0,6,61\} & 7 & 75 & 9 & 66 & 0 & 1 & 5 & 5 & 28 & 39 \\
 146 & \{1,0,6,62\} & 7 & 76 & 8 & 68 & 1 & 3 & 14 & 64 & 341 & 423 \\
 147 & \{1,0,6,63\} & 7 & 77 & 7 & 70 & 0 & 0 & 0 & 0 & 10 & 10 \\
 148 & \{1,0,6,64\} & 7 & 78 & 6 & 72 & 0 & 0 & 0 & 1 & 4 & 5 \\
 149 & \{1,0,6,65\} & 7 & 79 & 5 & 74 & 0 & 0 & 0 & 0 & 11 & 11 \\
 150 & \{1,0,6,66\} & 7 & 80 & 4 & 76 & 0 & 0 & 0 & 0 & 4 & 4 \\
 151 & \{1,0,6,67\} & 7 & 81 & 3 & 78 & 0 & 0 & 0 & 0 & 4 & 4 \\
 152 & \{1,0,6,68\} & 7 & 82 & 2 & 80 & 0 & 0 & 0 & 0 & 3 & 3 \\
 153 & \{1,0,7,62\} & 8 & 78 & 18 & 60 & 0 & 0 & 0 & 0 & 4 & 4 \\
 154 & \{1,0,7,64\} & 8 & 80 & 16 & 64 & 0 & 1 & 0 & 3 & 20 & 24 \\
 155 & \{1,0,7,65\} & 8 & 81 & 15 & 66 & 0 & 0 & 2 & 18 & 62 & 82 \\
 156 & \{1,0,7,66\} & 8 & 82 & 14 & 68 & 0 & 3 & 10 & 43 & 252 & 308 \\
 157 & \{1,0,7,67\} & 8 & 83 & 13 & 70 & 0 & 0 & 0 & 0 & 3 & 3 \\
 158 & \{1,0,7,68\} & 8 & 84 & 12 & 72 & 0 & 0 & 0 & 1 & 3 & 4 \\
 159 & \{1,0,7,69\} & 8 & 85 & 11 & 74 & 0 & 0 & 0 & 0 & 2 & 2 \\
 160 & \{1,0,7,70\} & 8 & 86 & 10 & 76 & 0 & 0 & 0 & 3 & 32 & 35 \\
 161 & \{1,0,7,72\} & 8 & 88 & 8 & 80 & 0 & 0 & 0 & 0 & 8 & 8 \\
 162 & \{1,0,7,73\} & 8 & 89 & 7 & 82 & 0 & 0 & 0 & 0 & 6 & 6 \\
 163 & \{1,0,8,69\} & 9 & 87 & 21 & 66 & 0 & 0 & 0 & 0 & 1 & 1 \\
 164 & \{1,0,8,70\} & 9 & 88 & 20 & 68 & 0 & 0 & 0 & 3 & 17 & 20 \\
 165 & \{1,0,8,71\} & 9 & 89 & 19 & 70 & 0 & 0 & 0 & 2 & 24 & 26 \\
 166 & \{1,0,8,72\} & 9 & 90 & 18 & 72 & 0 & 0 & 6 & 8 & 101 & 115 \\
 167 & \{1,0,8,73\} & 9 & 91 & 17 & 74 & 0 & 1 & 2 & 17 & 130 & 150 \\
 168 & \{1,0,8,74\} & 9 & 92 & 16 & 76 & 0 & 0 & 0 & 1 & 3 & 4 \\
 169 & \{1,0,8,76\} & 9 & 94 & 14 & 80 & 0 & 0 & 0 & 0 & 38 & 38 \\
 170 & \{1,0,8,78\} & 9 & 96 & 12 & 84 & 0 & 0 & 0 & 1 & 16 & 17 \\
 171 & \{1,0,9,69\} & 10 & 89 & 31 & 58 & 0 & 0 & 0 & 0 & 3 & 3 \\
 172 & \{1,0,9,76\} & 10 & 96 & 24 & 72 & 0 & 0 & 0 & 0 & 11 & 11 \\
 173 & \{1,0,9,77\} & 10 & 97 & 23 & 74 & 0 & 0 & 0 & 1 & 7 & 8 \\
 174 & \{1,0,9,78\} & 10 & 98 & 22 & 76 & 0 & 0 & 2 & 13 & 92 & 107 \\
 175 & \{1,0,9,79\} & 10 & 99 & 21 & 78 & 0 & 1 & 1 & 12 & 90 & 104 \\
 176 & \{1,0,9,80\} & 10 & 100 & 20 & 80 & 0 & 0 & 2 & 12 & 89 & 103 \\
 177 & \{1,0,9,83\} & 10 & 103 & 17 & 86 & 0 & 0 & 0 & 0 & 2 & 2 \\
 178 & \{1,0,9,84\} & 10 & 104 & 16 & 88 & 0 & 0 & 0 & 19 & 320 & 339 \\
 179 & \{1,0,9,85\} & 10 & 105 & 15 & 90 & 0 & 0 & 0 & 0 & 6 & 6 \\
 180 & \{1,0,9,86\} & 10 & 106 & 14 & 92 & 0 & 0 & 0 & 0 & 2 & 2 \\
 181 & \{1,0,10,80\} & 11 & 102 & 30 & 72 & 0 & 0 & 0 & 0 & 16 & 16 \\
 182 & \{1,0,10,81\} & 11 & 103 & 29 & 74 & 0 & 0 & 0 & 0 & 11 & 11 \\
 183 & \{1,0,10,82\} & 11 & 104 & 28 & 76 & 0 & 0 & 0 & 3 & 19 & 22 \\
 184 & \{1,0,10,83\} & 11 & 105 & 27 & 78 & 0 & 0 & 0 & 0 & 26 & 26 \\
 185 & \{1,0,10,84\} & 11 & 106 & 26 & 80 & 0 & 0 & 0 & 2 & 54 & 56 \\
 186 & \{1,0,10,85\} & 11 & 107 & 25 & 82 & 0 & 0 & 0 & 2 & 33 & 35 \\
 187 & \{1,0,10,86\} & 11 & 108 & 24 & 84 & 1 & 1 & 6 & 25 & 121 & 154 \\
 188 & \{1,0,10,87\} & 11 & 109 & 23 & 86 & 0 & 0 & 0 & 5 & 38 & 43 \\
 189 & \{1,0,10,88\} & 11 & 110 & 22 & 88 & 0 & 0 & 0 & 0 & 1 & 1 \\
 190 & \{1,0,10,90\} & 11 & 112 & 20 & 92 & 0 & 0 & 0 & 0 & 14 & 14 \\
 191 & \{1,0,10,91\} & 11 & 113 & 19 & 94 & 0 & 0 & 0 & 0 & 9 & 9 \\
 192 & \{1,0,10,92\} & 11 & 114 & 18 & 96 & 0 & 0 & 5 & 48 & 84 & 137 \\
 193 & \{1,0,10,94\} & 11 & 116 & 16 & 100 & 0 & 0 & 0 & 0 & 1 & 1 \\
 194 & \{1,0,10,96\} & 11 & 118 & 14 & 104 & 0 & 0 & 0 & 0 & 3 & 3 \\
 195 & \{1,0,10,98\} & 11 & 120 & 12 & 108 & 0 & 0 & 0 & 0 & 4 & 4 \\
 196 & \{1,0,11,88\} & 12 & 112 & 32 & 80 & 0 & 0 & 0 & 8 & 17 & 25 \\
 197 & \{1,0,11,89\} & 12 & 113 & 31 & 82 & 0 & 0 & 0 & 4 & 37 & 41 \\
 198 & \{1,0,11,90\} & 12 & 114 & 30 & 84 & 0 & 0 & 0 & 8 & 102 & 110 \\
 199 & \{1,0,11,91\} & 12 & 115 & 29 & 86 & 0 & 0 & 0 & 7 & 85 & 92 \\
 200 & \{1,0,11,92\} & 12 & 116 & 28 & 88 & 0 & 0 & 0 & 7 & 25 & 32 \\
 201 & \{1,0,11,93\} & 12 & 117 & 27 & 90 & 0 & 0 & 0 & 2 & 3 & 5 \\
 202 & \{1,0,11,94\} & 12 & 118 & 26 & 92 & 0 & 0 & 0 & 7 & 37 & 44 \\
 203 & \{1,0,11,96\} & 12 & 120 & 24 & 96 & 0 & 0 & 0 & 0 & 15 & 15 \\
 204 & \{1,0,11,97\} & 12 & 121 & 23 & 98 & 0 & 0 & 0 & 0 & 2 & 2 \\
 205 & \{1,0,11,98\} & 12 & 122 & 22 & 100 & 0 & 0 & 0 & 0 & 2 & 2 \\
 206 & \{1,0,11,100\} & 12 & 124 & 20 & 104 & 0 & 0 & 0 & 0 & 2 & 2 \\
 207 & \{1,0,12,89\} & 13 & 115 & 41 & 74 & 0 & 0 & 0 & 0 & 4 & 4 \\
 208 & \{1,0,12,90\} & 13 & 116 & 40 & 76 & 0 & 0 & 0 & 0 & 4 & 4 \\
 209 & \{1,0,12,94\} & 13 & 120 & 36 & 84 & 0 & 0 & 0 & 0 & 26 & 26 \\
 210 & \{1,0,12,95\} & 13 & 121 & 35 & 86 & 0 & 0 & 0 & 0 & 12 & 12 \\
 211 & \{1,0,12,96\} & 13 & 122 & 34 & 88 & 0 & 0 & 3 & 9 & 74 & 86 \\
 212 & \{1,0,12,97\} & 13 & 123 & 33 & 90 & 0 & 0 & 3 & 18 & 84 & 105 \\
 213 & \{1,0,12,98\} & 13 & 124 & 32 & 92 & 0 & 0 & 1 & 22 & 90 & 113 \\
 214 & \{1,0,12,101\} & 13 & 127 & 29 & 98 & 0 & 0 & 0 & 0 & 24 & 24 \\
 215 & \{1,0,12,102\} & 13 & 128 & 28 & 100 & 0 & 0 & 0 & 0 & 12 & 12 \\
 216 & \{1,0,12,103\} & 13 & 129 & 27 & 102 & 0 & 0 & 0 & 0 & 4 & 4 \\
 217 & \{1,0,12,105\} & 13 & 131 & 25 & 106 & 0 & 0 & 0 & 1 & 5 & 6 \\
 218 & \{1,0,12,106\} & 13 & 132 & 24 & 108 & 0 & 0 & 0 & 3 & 17 & 20 \\
 219 & \{1,0,13,93\} & 14 & 121 & 47 & 74 & 0 & 0 & 0 & 2 & 3 & 5 \\
 220 & \{1,0,13,94\} & 14 & 122 & 46 & 76 & 0 & 0 & 0 & 0 & 10 & 10 \\
 221 & \{1,0,13,95\} & 14 & 123 & 45 & 78 & 0 & 0 & 0 & 0 & 3 & 3 \\
 222 & \{1,0,13,96\} & 14 & 124 & 44 & 80 & 0 & 0 & 0 & 0 & 1 & 1 \\
 223 & \{1,0,13,98\} & 14 & 126 & 42 & 84 & 0 & 0 & 0 & 0 & 1 & 1 \\
 224 & \{1,0,13,100\} & 14 & 128 & 40 & 88 & 0 & 0 & 0 & 0 & 6 & 6 \\
 225 & \{1,0,13,102\} & 14 & 130 & 38 & 92 & 0 & 0 & 0 & 15 & 94 & 109 \\
 226 & \{1,0,13,103\} & 14 & 131 & 37 & 94 & 0 & 0 & 0 & 15 & 70 & 85 \\
 227 & \{1,0,13,104\} & 14 & 132 & 36 & 96 & 0 & 3 & 4 & 25 & 67 & 99 \\
 228 & \{1,0,13,105\} & 14 & 133 & 35 & 98 & 0 & 0 & 3 & 4 & 26 & 33 \\
 229 & \{1,0,13,108\} & 14 & 136 & 32 & 104 & 0 & 0 & 0 & 1 & 22 & 23 \\
 230 & \{1,0,13,112\} & 14 & 140 & 28 & 112 & 0 & 0 & 0 & 0 & 6 & 6 \\
 231 & \{1,0,14,102\} & 15 & 132 & 48 & 84 & 0 & 0 & 0 & 0 & 1 & 1 \\
 232 & \{1,0,14,104\} & 15 & 134 & 46 & 88 & 0 & 0 & 0 & 0 & 11 & 11 \\
 233 & \{1,0,14,105\} & 15 & 135 & 45 & 90 & 0 & 0 & 0 & 0 & 5 & 5 \\
 234 & \{1,0,14,106\} & 15 & 136 & 44 & 92 & 0 & 0 & 0 & 0 & 13 & 13 \\
 235 & \{1,0,14,107\} & 15 & 137 & 43 & 94 & 0 & 0 & 0 & 0 & 8 & 8 \\
 236 & \{1,0,14,108\} & 15 & 138 & 42 & 96 & 0 & 0 & 0 & 0 & 41 & 41 \\
 237 & \{1,0,14,109\} & 15 & 139 & 41 & 98 & 0 & 0 & 0 & 0 & 45 & 45 \\
 238 & \{1,0,14,110\} & 15 & 140 & 40 & 100 & 0 & 0 & 9 & 27 & 105 & 141 \\
 239 & \{1,0,14,111\} & 15 & 141 & 39 & 102 & 0 & 0 & 0 & 0 & 7 & 7 \\
 240 & \{1,0,14,112\} & 15 & 142 & 38 & 104 & 0 & 0 & 0 & 0 & 3 & 3 \\
 241 & \{1,0,14,115\} & 15 & 145 & 35 & 110 & 0 & 0 & 0 & 0 & 6 & 6 \\
 242 & \{1,0,14,120\} & 15 & 150 & 30 & 120 & 0 & 0 & 0 & 1 & 10 & 11 \\
 243 & \{1,0,15,107\} & 16 & 139 & 53 & 86 & 0 & 0 & 0 & 0 & 1 & 1 \\
 244 & \{1,0,15,108\} & 16 & 140 & 52 & 88 & 0 & 0 & 0 & 0 & 11 & 11 \\
 245 & \{1,0,15,109\} & 16 & 141 & 51 & 90 & 0 & 0 & 0 & 0 & 14 & 14 \\
 246 & \{1,0,15,110\} & 16 & 142 & 50 & 92 & 0 & 0 & 0 & 0 & 23 & 23 \\
 247 & \{1,0,15,111\} & 16 & 143 & 49 & 94 & 0 & 0 & 0 & 0 & 1 & 1 \\
 248 & \{1,0,15,112\} & 16 & 144 & 48 & 96 & 0 & 0 & 0 & 4 & 3 & 7 \\
 249 & \{1,0,15,113\} & 16 & 145 & 47 & 98 & 0 & 0 & 0 & 4 & 11 & 15 \\
 250 & \{1,0,15,114\} & 16 & 146 & 46 & 100 & 0 & 0 & 0 & 3 & 38 & 41 \\
 251 & \{1,0,15,115\} & 16 & 147 & 45 & 102 & 0 & 0 & 0 & 10 & 51 & 61 \\
 252 & \{1,0,15,116\} & 16 & 148 & 44 & 104 & 0 & 0 & 0 & 10 & 49 & 59 \\
 253 & \{1,0,15,122\} & 16 & 154 & 38 & 116 & 0 & 0 & 0 & 0 & 2 & 2 \\
 254 & \{1,0,15,124\} & 16 & 156 & 36 & 120 & 0 & 0 & 0 & 0 & 2 & 2 \\
 255 & \{1,0,15,126\} & 16 & 158 & 34 & 124 & 0 & 0 & 0 & 0 & 11 & 11 \\
 256 & \{1,0,15,128\} & 16 & 160 & 32 & 128 & 0 & 0 & 1 & 7 & 46 & 54 \\
 257 & \{1,0,15,129\} & 16 & 161 & 31 & 130 & 0 & 0 & 0 & 0 & 1 & 1 \\
 258 & \{1,0,15,133\} & 16 & 165 & 27 & 138 & 0 & 0 & 0 & 0 & 4 & 4 \\
 259 & \{1,0,15,134\} & 16 & 166 & 26 & 140 & 0 & 0 & 0 & 0 & 4 & 4 \\
 260 & \{1,0,15,140\} & 16 & 172 & 20 & 152 & 0 & 0 & 0 & 0 & 3 & 3 \\
 261 & \{1,0,16,114\} & 17 & 148 & 56 & 92 & 0 & 0 & 0 & 9 & 12 & 21 \\
 262 & \{1,0,16,115\} & 17 & 149 & 55 & 94 & 0 & 0 & 0 & 0 & 29 & 29 \\
 263 & \{1,0,16,116\} & 17 & 150 & 54 & 96 & 0 & 0 & 0 & 29 & 10 & 39 \\
 264 & \{1,0,16,117\} & 17 & 151 & 53 & 98 & 0 & 0 & 0 & 0 & 30 & 30 \\
 265 & \{1,0,16,118\} & 17 & 152 & 52 & 100 & 0 & 0 & 0 & 0 & 4 & 4 \\
 266 & \{1,0,16,119\} & 17 & 153 & 51 & 102 & 0 & 0 & 0 & 0 & 9 & 9 \\
 267 & \{1,0,16,120\} & 17 & 154 & 50 & 104 & 0 & 0 & 3 & 2 & 15 & 20 \\
 268 & \{1,0,16,121\} & 17 & 155 & 49 & 106 & 0 & 0 & 0 & 3 & 21 & 24 \\
 269 & \{1,0,16,122\} & 17 & 156 & 48 & 108 & 0 & 0 & 4 & 18 & 69 & 91 \\
 270 & \{1,0,16,123\} & 17 & 157 & 47 & 110 & 0 & 0 & 3 & 6 & 31 & 40 \\
 271 & \{1,0,16,124\} & 17 & 158 & 46 & 112 & 0 & 0 & 0 & 0 & 4 & 4 \\
 272 & \{1,0,16,127\} & 17 & 161 & 43 & 118 & 0 & 0 & 0 & 0 & 8 & 8 \\
 273 & \{1,0,16,134\} & 17 & 168 & 36 & 132 & 0 & 0 & 0 & 3 & 13 & 16 \\
 274 & \{1,0,16,142\} & 17 & 176 & 28 & 148 & 0 & 0 & 0 & 0 & 1 & 1 \\
 275 & \{1,0,17,117\} & 18 & 153 & 63 & 90 & 0 & 0 & 0 & 2 & 3 & 5 \\
 276 & \{1,0,17,118\} & 18 & 154 & 62 & 92 & 0 & 0 & 0 & 0 & 3 & 3 \\
 277 & \{1,0,17,119\} & 18 & 155 & 61 & 94 & 0 & 0 & 0 & 1 & 2 & 3 \\
 278 & \{1,0,17,120\} & 18 & 156 & 60 & 96 & 0 & 0 & 0 & 3 & 8 & 11 \\
 279 & \{1,0,17,121\} & 18 & 157 & 59 & 98 & 0 & 0 & 0 & 0 & 18 & 18 \\
 280 & \{1,0,17,122\} & 18 & 158 & 58 & 100 & 0 & 0 & 0 & 1 & 21 & 22 \\
 281 & \{1,0,17,123\} & 18 & 159 & 57 & 102 & 0 & 0 & 0 & 0 & 5 & 5 \\
 282 & \{1,0,17,124\} & 18 & 160 & 56 & 104 & 0 & 0 & 0 & 0 & 4 & 4 \\
 283 & \{1,0,17,125\} & 18 & 161 & 55 & 106 & 0 & 0 & 0 & 0 & 14 & 14 \\
 284 & \{1,0,17,126\} & 18 & 162 & 54 & 108 & 0 & 0 & 0 & 4 & 66 & 70 \\
 285 & \{1,0,17,127\} & 18 & 163 & 53 & 110 & 0 & 0 & 0 & 10 & 82 & 92 \\
 286 & \{1,0,17,128\} & 18 & 164 & 52 & 112 & 0 & 0 & 0 & 2 & 7 & 9 \\
 287 & \{1,0,17,129\} & 18 & 165 & 51 & 114 & 0 & 0 & 0 & 0 & 6 & 6 \\
 288 & \{1,0,17,130\} & 18 & 166 & 50 & 116 & 0 & 0 & 0 & 0 & 2 & 2 \\
 289 & \{1,0,17,132\} & 18 & 168 & 48 & 120 & 0 & 0 & 0 & 0 & 3 & 3 \\
 290 & \{1,0,17,139\} & 18 & 175 & 41 & 134 & 0 & 0 & 0 & 0 & 7 & 7 \\
 291 & \{1,0,17,140\} & 18 & 176 & 40 & 136 & 0 & 0 & 0 & 0 & 8 & 8 \\
 292 & \{1,0,18,127\} & 19 & 165 & 63 & 102 & 0 & 0 & 0 & 0 & 4 & 4 \\
 293 & \{1,0,18,128\} & 19 & 166 & 62 & 104 & 0 & 0 & 0 & 1 & 14 & 15 \\
 294 & \{1,0,18,129\} & 19 & 167 & 61 & 106 & 0 & 0 & 0 & 0 & 24 & 24 \\
 295 & \{1,0,18,130\} & 19 & 168 & 60 & 108 & 0 & 0 & 0 & 1 & 5 & 6 \\
 296 & \{1,0,18,131\} & 19 & 169 & 59 & 110 & 0 & 0 & 0 & 0 & 6 & 6 \\
 297 & \{1,0,18,132\} & 19 & 170 & 58 & 112 & 0 & 0 & 0 & 3 & 23 & 26 \\
 298 & \{1,0,18,133\} & 19 & 171 & 57 & 114 & 0 & 0 & 0 & 8 & 46 & 54 \\
 299 & \{1,0,18,134\} & 19 & 172 & 56 & 116 & 0 & 0 & 3 & 24 & 59 & 86 \\
 300 & \{1,0,18,135\} & 19 & 173 & 55 & 118 & 0 & 0 & 0 & 0 & 4 & 4 \\
 301 & \{1,0,18,137\} & 19 & 175 & 53 & 122 & 0 & 0 & 0 & 1 & 9 & 10 \\
 302 & \{1,0,18,148\} & 19 & 186 & 42 & 144 & 0 & 0 & 0 & 4 & 19 & 23 \\
 303 & \{1,0,18,150\} & 19 & 188 & 40 & 148 & 0 & 0 & 0 & 0 & 2 & 2 \\
 304 & \{1,0,18,154\} & 19 & 192 & 36 & 156 & 0 & 0 & 0 & 0 & 2 & 2 \\
 305 & \{1,0,19,132\} & 20 & 172 & 68 & 104 & 0 & 0 & 0 & 1 & 0 & 1 \\
 306 & \{1,0,19,133\} & 20 & 173 & 67 & 106 & 0 & 0 & 0 & 4 & 5 & 9 \\
 307 & \{1,0,19,134\} & 20 & 174 & 66 & 108 & 0 & 0 & 0 & 10 & 37 & 47 \\
 308 & \{1,0,19,135\} & 20 & 175 & 65 & 110 & 0 & 0 & 0 & 14 & 51 & 65 \\
 309 & \{1,0,19,136\} & 20 & 176 & 64 & 112 & 0 & 0 & 1 & 0 & 142 & 143 \\
 310 & \{1,0,19,138\} & 20 & 178 & 62 & 116 & 0 & 0 & 0 & 4 & 33 & 37 \\
 311 & \{1,0,19,139\} & 20 & 179 & 61 & 118 & 0 & 0 & 2 & 11 & 47 & 60 \\
 312 & \{1,0,19,140\} & 20 & 180 & 60 & 120 & 0 & 0 & 3 & 14 & 35 & 52 \\
 313 & \{1,0,19,141\} & 20 & 181 & 59 & 122 & 0 & 0 & 0 & 0 & 7 & 7 \\
 314 & \{1,0,19,152\} & 20 & 192 & 48 & 144 & 0 & 0 & 0 & 0 & 44 & 44 \\
 315 & \{1,0,19,153\} & 20 & 193 & 47 & 146 & 0 & 0 & 0 & 0 & 4 & 4 \\
 316 & \{1,0,19,154\} & 20 & 194 & 46 & 148 & 0 & 0 & 0 & 0 & 14 & 14 \\
 317 & \{1,0,20,138\} & 21 & 180 & 72 & 108 & 0 & 0 & 2 & 5 & 7 & 14 \\
 318 & \{1,0,20,139\} & 21 & 181 & 71 & 110 & 0 & 0 & 0 & 9 & 22 & 31 \\
 319 & \{1,0,20,140\} & 21 & 182 & 70 & 112 & 0 & 0 & 0 & 3 & 37 & 40 \\
 320 & \{1,0,20,141\} & 21 & 183 & 69 & 114 & 0 & 0 & 0 & 3 & 29 & 32 \\
 321 & \{1,0,20,142\} & 21 & 184 & 68 & 116 & 0 & 0 & 0 & 0 & 36 & 36 \\
 322 & \{1,0,20,143\} & 21 & 185 & 67 & 118 & 0 & 0 & 0 & 0 & 5 & 5 \\
 323 & \{1,0,20,144\} & 21 & 186 & 66 & 120 & 0 & 0 & 0 & 0 & 24 & 24 \\
 324 & \{1,0,20,145\} & 21 & 187 & 65 & 122 & 0 & 0 & 0 & 0 & 16 & 16 \\
 325 & \{1,0,20,146\} & 21 & 188 & 64 & 124 & 0 & 0 & 4 & 15 & 38 & 57 \\
 326 & \{1,0,20,147\} & 21 & 189 & 63 & 126 & 0 & 3 & 6 & 8 & 38 & 55 \\
 327 & \{1,0,20,148\} & 21 & 190 & 62 & 128 & 0 & 0 & 0 & 0 & 5 & 5 \\
 328 & \{1,0,20,151\} & 21 & 193 & 59 & 134 & 0 & 0 & 0 & 0 & 2 & 2 \\
 329 & \{1,0,20,161\} & 21 & 203 & 49 & 154 & 0 & 0 & 0 & 2 & 2 & 4 \\
 330 & \{1,0,20,162\} & 21 & 204 & 48 & 156 & 0 & 0 & 0 & 5 & 24 & 29 \\
 331 & \{1,0,20,165\} & 21 & 207 & 45 & 162 & 0 & 0 & 0 & 0 & 1 & 1 \\
 332 & \{1,0,20,166\} & 21 & 208 & 44 & 164 & 0 & 0 & 0 & 0 & 3 & 3 \\
 333 & \{1,0,21,145\} & 22 & 189 & 75 & 114 & 0 & 0 & 0 & 0 & 6 & 6 \\
 334 & \{1,0,21,146\} & 22 & 190 & 74 & 116 & 0 & 0 & 1 & 4 & 4 & 9 \\
 335 & \{1,0,21,147\} & 22 & 191 & 73 & 118 & 0 & 0 & 0 & 3 & 31 & 34 \\
 336 & \{1,0,21,148\} & 22 & 192 & 72 & 120 & 0 & 0 & 0 & 1 & 39 & 40 \\
 337 & \{1,0,21,150\} & 22 & 194 & 70 & 124 & 0 & 0 & 0 & 4 & 32 & 36 \\
 338 & \{1,0,21,151\} & 22 & 195 & 69 & 126 & 0 & 0 & 0 & 10 & 41 & 51 \\
 339 & \{1,0,21,152\} & 22 & 196 & 68 & 128 & 0 & 0 & 0 & 6 & 23 & 29 \\
 340 & \{1,0,21,154\} & 22 & 198 & 66 & 132 & 0 & 0 & 0 & 3 & 19 & 22 \\
 341 & \{1,0,21,157\} & 22 & 201 & 63 & 138 & 0 & 0 & 0 & 0 & 4 & 4 \\
 342 & \{1,0,21,168\} & 22 & 212 & 52 & 160 & 0 & 0 & 0 & 0 & 4 & 4 \\
 343 & \{1,0,21,170\} & 22 & 214 & 50 & 164 & 0 & 0 & 2 & 7 & 1 & 10 \\
 344 & \{1,0,21,176\} & 22 & 220 & 44 & 176 & 0 & 0 & 0 & 1 & 0 & 1 \\
 345 & \{1,0,22,151\} & 23 & 197 & 79 & 118 & 0 & 0 & 0 & 2 & 1 & 3 \\
 346 & \{1,0,22,152\} & 23 & 198 & 78 & 120 & 0 & 0 & 0 & 6 & 9 & 15 \\
 347 & \{1,0,22,153\} & 23 & 199 & 77 & 122 & 0 & 0 & 0 & 15 & 36 & 51 \\
 348 & \{1,0,22,154\} & 23 & 200 & 76 & 124 & 0 & 0 & 0 & 13 & 64 & 77 \\
 349 & \{1,0,22,156\} & 23 & 202 & 74 & 128 & 0 & 0 & 0 & 0 & 3 & 3 \\
 350 & \{1,0,22,157\} & 23 & 203 & 73 & 130 & 0 & 0 & 0 & 0 & 9 & 9 \\
 351 & \{1,0,22,158\} & 23 & 204 & 72 & 132 & 0 & 0 & 5 & 12 & 18 & 35 \\
 352 & \{1,0,22,159\} & 23 & 205 & 71 & 134 & 0 & 0 & 0 & 0 & 7 & 7 \\
 353 & \{1,0,22,160\} & 23 & 206 & 70 & 136 & 0 & 0 & 0 & 0 & 2 & 2 \\
 354 & \{1,0,22,174\} & 23 & 220 & 56 & 164 & 0 & 0 & 0 & 9 & 9 & 18 \\
 355 & \{1,0,22,176\} & 23 & 222 & 54 & 168 & 0 & 0 & 0 & 2 & 8 & 10 \\
 356 & \{1,0,22,178\} & 23 & 224 & 52 & 172 & 0 & 0 & 0 & 0 & 2 & 2 \\
 357 & \{1,0,23,157\} & 24 & 205 & 83 & 122 & 0 & 0 & 3 & 0 & 0 & 3 \\
 358 & \{1,0,23,158\} & 24 & 206 & 82 & 124 & 0 & 0 & 3 & 14 & 7 & 24 \\
 359 & \{1,0,23,159\} & 24 & 207 & 81 & 126 & 0 & 0 & 14 & 10 & 41 & 65 \\
 360 & \{1,0,23,160\} & 24 & 208 & 80 & 128 & 0 & 0 & 1 & 34 & 42 & 77 \\
 361 & \{1,0,23,162\} & 24 & 210 & 78 & 132 & 0 & 0 & 0 & 2 & 11 & 13 \\
 362 & \{1,0,23,163\} & 24 & 211 & 77 & 134 & 0 & 0 & 0 & 3 & 22 & 25 \\
 363 & \{1,0,23,164\} & 24 & 212 & 76 & 136 & 0 & 0 & 0 & 2 & 7 & 9 \\
 364 & \{1,0,23,166\} & 24 & 214 & 74 & 140 & 0 & 0 & 0 & 0 & 5 & 5 \\
 365 & \{1,0,23,168\} & 24 & 216 & 72 & 144 & 0 & 0 & 0 & 0 & 1 & 1 \\
 366 & \{1,0,24,162\} & 25 & 212 & 88 & 124 & 0 & 0 & 1 & 2 & 0 & 3 \\
 367 & \{1,0,24,163\} & 25 & 213 & 87 & 126 & 0 & 0 & 1 & 2 & 5 & 8 \\
 368 & \{1,0,24,164\} & 25 & 214 & 86 & 128 & 0 & 0 & 1 & 5 & 1 & 7 \\
 369 & \{1,0,24,165\} & 25 & 215 & 85 & 130 & 0 & 0 & 2 & 8 & 11 & 21 \\
 370 & \{1,0,24,166\} & 25 & 216 & 84 & 132 & 0 & 0 & 0 & 9 & 30 & 39 \\
 371 & \{1,0,24,169\} & 25 & 219 & 81 & 138 & 0 & 0 & 0 & 1 & 19 & 20 \\
 372 & \{1,0,24,170\} & 25 & 220 & 80 & 140 & 0 & 0 & 3 & 15 & 33 & 51 \\
 373 & \{1,0,24,175\} & 25 & 225 & 75 & 150 & 0 & 0 & 0 & 1 & 0 & 1 \\
 374 & \{1,0,24,190\} & 25 & 240 & 60 & 180 & 0 & 0 & 0 & 1 & 4 & 5 \\
 375 & \{1,0,25,171\} & 26 & 223 & 89 & 134 & 0 & 0 & 0 & 1 & 3 & 4 \\
 376 & \{1,0,25,172\} & 26 & 224 & 88 & 136 & 0 & 0 & 1 & 6 & 8 & 15 \\
 377 & \{1,0,25,174\} & 26 & 226 & 86 & 140 & 0 & 0 & 0 & 0 & 4 & 4 \\
 378 & \{1,0,25,175\} & 26 & 227 & 85 & 142 & 0 & 0 & 0 & 0 & 7 & 7 \\
 379 & \{1,0,25,176\} & 26 & 228 & 84 & 144 & 0 & 0 & 0 & 4 & 6 & 10 \\
 380 & \{1,0,25,177\} & 26 & 229 & 83 & 146 & 0 & 0 & 2 & 2 & 9 & 13 \\
 381 & \{1,0,25,196\} & 26 & 248 & 64 & 184 & 0 & 0 & 4 & 4 & 0 & 8 \\
 382 & \{1,0,26,177\} & 27 & 231 & 93 & 138 & 0 & 0 & 3 & 0 & 0 & 3 \\
 383 & \{1,0,26,178\} & 27 & 232 & 92 & 140 & 0 & 0 & 6 & 14 & 1 & 21 \\
 384 & \{1,0,26,181\} & 27 & 235 & 89 & 146 & 0 & 0 & 0 & 0 & 4 & 4 \\
 385 & \{1,0,26,182\} & 27 & 236 & 88 & 148 & 0 & 0 & 2 & 3 & 6 & 11 \\
 386 & \{1,0,26,184\} & 27 & 238 & 86 & 152 & 0 & 0 & 0 & 0 & 8 & 8 \\
 387 & \{1,0,26,204\} & 27 & 258 & 66 & 192 & 0 & 0 & 0 & 0 & 3 & 3 \\
 388 & \{1,0,27,182\} & 28 & 238 & 98 & 140 & 0 & 1 & 2 & 0 & 0 & 3 \\
 389 & \{1,0,27,183\} & 28 & 239 & 97 & 142 & 0 & 0 & 2 & 6 & 0 & 8 \\
 390 & \{1,0,27,184\} & 28 & 240 & 96 & 144 & 0 & 0 & 7 & 6 & 10 & 23 \\
 391 & \{1,0,27,186\} & 28 & 242 & 94 & 148 & 0 & 0 & 0 & 0 & 4 & 4 \\
 392 & \{1,0,27,187\} & 28 & 243 & 93 & 150 & 0 & 0 & 0 & 0 & 12 & 12 \\
 393 & \{1,0,27,188\} & 28 & 244 & 92 & 152 & 0 & 0 & 0 & 4 & 12 & 16 \\
 394 & \{1,0,27,191\} & 28 & 247 & 89 & 158 & 0 & 0 & 0 & 2 & 7 & 9 \\
 395 & \{1,0,27,193\} & 28 & 249 & 87 & 162 & 0 & 0 & 0 & 0 & 2 & 2 \\
 396 & \{1,0,28,189\} & 29 & 247 & 101 & 146 & 0 & 1 & 0 & 0 & 0 & 1 \\
 397 & \{1,0,28,190\} & 29 & 248 & 100 & 148 & 0 & 0 & 2 & 2 & 0 & 4 \\
 398 & \{1,0,28,194\} & 29 & 252 & 96 & 156 & 0 & 0 & 0 & 3 & 6 & 9 \\
 399 & \{1,0,28,195\} & 29 & 253 & 95 & 158 & 0 & 0 & 0 & 0 & 3 & 3 \\
 400 & \{1,0,28,218\} & 29 & 276 & 72 & 204 & 0 & 1 & 1 & 0 & 0 & 2 \\
 401 & \{1,0,29,196\} & 30 & 256 & 104 & 152 & 0 & 0 & 2 & 1 & 0 & 3 \\
 402 & \{1,0,29,199\} & 30 & 259 & 101 & 158 & 0 & 0 & 0 & 2 & 5 & 7 \\
 403 & \{1,0,29,200\} & 30 & 260 & 100 & 160 & 0 & 0 & 0 & 4 & 12 & 16 \\
 404 & \{1,0,29,202\} & 30 & 262 & 98 & 164 & 0 & 0 & 0 & 0 & 2 & 2 \\
 405 & \{1,0,29,204\} & 30 & 264 & 96 & 168 & 0 & 0 & 0 & 0 & 4 & 4 \\
 406 & \{1,0,30,201\} & 31 & 263 & 109 & 154 & 0 & 1 & 0 & 0 & 0 & 1 \\
 407 & \{1,0,30,202\} & 31 & 264 & 108 & 156 & 0 & 5 & 2 & 0 & 1 & 8 \\
 408 & \{1,0,30,205\} & 31 & 267 & 105 & 162 & 0 & 0 & 0 & 0 & 11 & 11 \\
 409 & \{1,0,30,206\} & 31 & 268 & 104 & 164 & 0 & 0 & 1 & 3 & 4 & 8 \\
 410 & \{1,0,31,206\} & 32 & 270 & 114 & 156 & 0 & 1 & 0 & 0 & 0 & 1 \\
 411 & \{1,0,31,207\} & 32 & 271 & 113 & 158 & 0 & 0 & 2 & 0 & 0 & 2 \\
 412 & \{1,0,31,208\} & 32 & 272 & 112 & 160 & 0 & 1 & 2 & 2 & 0 & 5 \\
 413 & \{1,0,31,212\} & 32 & 276 & 108 & 168 & 0 & 0 & 0 & 3 & 5 & 8 \\
 414 & \{1,0,31,213\} & 32 & 277 & 107 & 170 & 0 & 0 & 2 & 2 & 3 & 7 \\
 415 & \{1,0,32,216\} & 33 & 282 & 114 & 168 & 0 & 0 & 0 & 0 & 2 & 2 \\
 416 & \{1,0,32,217\} & 33 & 283 & 113 & 170 & 0 & 0 & 0 & 0 & 9 & 9 \\
 417 & \{1,0,32,218\} & 33 & 284 & 112 & 172 & 0 & 0 & 0 & 0 & 3 & 3 \\
 418 & \{1,0,32,220\} & 33 & 286 & 110 & 176 & 0 & 0 & 0 & 0 & 1 & 1 \\
 419 & \{1,0,33,224\} & 34 & 292 & 116 & 176 & 0 & 0 & 0 & 1 & 3 & 4 \\
 420 & \{1,0,34,225\} & 35 & 295 & 125 & 170 & 1 & 0 & 0 & 0 & 0 & 1 \\
 421 & \{1,0,34,226\} & 35 & 296 & 124 & 172 & 0 & 1 & 0 & 0 & 0 & 1 \\
 422 & \{1,0,34,230\} & 35 & 300 & 120 & 180 & 0 & 0 & 0 & 7 & 8 & 15 \\
 423 & \{1,0,34,234\} & 35 & 304 & 116 & 188 & 0 & 0 & 0 & 0 & 13 & 13 \\
 424 & \{1,0,34,235\} & 35 & 305 & 115 & 190 & 0 & 0 & 0 & 0 & 2 & 2 \\
 425 & \{1,0,35,232\} & 36 & 304 & 128 & 176 & 1 & 0 & 0 & 0 & 0 & 1 \\
 426 & \{1,0,35,234\} & 36 & 306 & 126 & 180 & 0 & 0 & 0 & 0 & 2 & 2 \\
 427 & \{1,0,35,235\} & 36 & 307 & 125 & 182 & 0 & 0 & 0 & 0 & 2 & 2 \\
 428 & \{1,0,35,236\} & 36 & 308 & 124 & 184 & 0 & 0 & 0 & 0 & 2 & 2 \\
 429 & \{1,0,35,240\} & 36 & 312 & 120 & 192 & 0 & 0 & 0 & 0 & 6 & 6 \\
 430 & \{1,0,36,242\} & 37 & 316 & 128 & 188 & 0 & 0 & 0 & 1 & 1 & 2 \\
 431 & \{1,0,38,253\} & 39 & 331 & 137 & 194 & 0 & 0 & 0 & 0 & 4 & 4 \\
 432 & \{1,0,38,255\} & 39 & 333 & 135 & 198 & 0 & 0 & 2 & 1 & 13 & 16 \\
 433 & \{1,0,39,259\} & 40 & 339 & 141 & 198 & 0 & 0 & 0 & 1 & 4 & 5 \\
 434 & \{1,0,39,260\} & 40 & 340 & 140 & 200 & 0 & 0 & 0 & 3 & 8 & 11 \\
 435 & \{1,0,41,271\} & 42 & 355 & 149 & 206 & 0 & 0 & 0 & 0 & 11 & 11 \\
 436 & \{1,0,41,276\} & 42 & 360 & 144 & 216 & 0 & 0 & 0 & 0 & 9 & 9 \\
 437 & \{1,0,42,278\} & 43 & 364 & 152 & 212 & 0 & 0 & 0 & 1 & 7 & 8 \\
 438 & \{1,0,42,283\} & 43 & 369 & 147 & 222 & 0 & 0 & 0 & 3 & 2 & 5 \\
 439 & \{1,0,44,288\} & 45 & 378 & 162 & 216 & 0 & 0 & 0 & 0 & 51 & 51 \\
 440 & \{1,0,44,289\} & 45 & 379 & 161 & 218 & 0 & 0 & 0 & 0 & 8 & 8 \\
 441 & \{1,0,44,290\} & 45 & 380 & 160 & 220 & 0 & 0 & 0 & 0 & 14 & 14 \\
 442 & \{1,0,45,296\} & 46 & 388 & 164 & 224 & 0 & 0 & 0 & 3 & 3 & 6 \\
 443 & \{1,0,46,303\} & 47 & 397 & 167 & 230 & 0 & 0 & 1 & 4 & 0 & 5 \\
 444 & \{1,0,47,306\} & 48 & 402 & 174 & 228 & 0 & 0 & 0 & 0 & 4 & 4 \\
 445 & \{1,0,47,307\} & 48 & 403 & 173 & 230 & 0 & 0 & 0 & 0 & 14 & 14 \\
 446 & \{1,0,47,312\} & 48 & 408 & 168 & 240 & 0 & 0 & 0 & 0 & 4 & 4 \\
 447 & \{1,0,48,314\} & 49 & 412 & 176 & 236 & 0 & 0 & 0 & 4 & 5 & 9 \\
 448 & \{1,0,49,320\} & 50 & 420 & 180 & 240 & 0 & 0 & 0 & 0 & 2 & 2 \\
 449 & \{1,0,50,325\} & 51 & 427 & 185 & 242 & 0 & 0 & 0 & 0 & 4 & 4 \\
 450 & \{1,0,51,331\} & 52 & 435 & 189 & 246 & 0 & 0 & 0 & 11 & 0 & 11 \\
 451 & \{1,0,51,332\} & 52 & 436 & 188 & 248 & 0 & 0 & 0 & 5 & 5 & 10 \\
 452 & \{1,0,53,348\} & 54 & 456 & 192 & 264 & 0 & 0 & 0 & 0 & 2 & 2 \\
 453 & \{1,0,54,350\} & 55 & 460 & 200 & 260 & 0 & 0 & 0 & 2 & 0 & 2 \\
 454 & \{1,0,55,357\} & 56 & 469 & 203 & 266 & 0 & 0 & 2 & 0 & 0 & 2 \\
 455 & \{1,0,57,368\} & 58 & 484 & 212 & 272 & 0 & 0 & 0 & 1 & 0 & 1 \\
 456 & \{1,0,58,374\} & 59 & 492 & 216 & 276 & 0 & 0 & 4 & 0 & 0 & 4 \\
 457 & \{1,0,59,384\} & 60 & 504 & 216 & 288 & 0 & 0 & 0 & 0 & 3 & 3 \\
 458 & \{1,0,65,417\} & 66 & 549 & 243 & 306 & 0 & 1 & 0 & 0 & 0 & 1 \\
&&&&&&&&&&& \\
\hline
&&&&&&&&&&& \\
{\bf  459} & {\bf\{1,1,0,2\} }& {\bf 0} & {\bf 0} & {\bf 0} & {\bf 0} & {\bf 0} & {\bf 0} & {\bf 17} & {\bf 214} & {\bf 1940} & {\bf 2171} \\
 460 & \{1,1,0,3\} & 0 & 1 & -1 & 2 & 0 & 0 & 0 & 1 & 66 & 67 \\
 461 & \{1,1,0,4\} & 0 & 2 & -2 & 4 & 0 & 0 & 0 & 8 & 177 & 185 \\
 462 & \{1,1,0,5\} & 0 & 3 & -3 & 6 & 0 & 0 & 0 & 5 & 153 & 158 \\
 463 & \{1,1,0,6\} & 0 & 4 & -4 & 8 & 0 & 0 & 0 & 4 & 55 & 59 \\
 464 & \{1,1,0,8\} & 0 & 6 & -6 & 12 & 0 & 0 & 0 & 0 & 42 & 42 \\
 465 & \{1,1,0,10\} & 0 & 8 & -8 & 16 & 0 & 0 & 0 & 1 & 3 & 4 \\
 466 & \{1,1,0,11\} & 0 & 9 & -9 & 18 & 0 & 0 & 0 & 0 & 1 & 1 \\
 467 & \{1,2,0,2\} & -1 & -4 & -8 & 4 & 0 & 1 & 16 & 145 & 925 & 1087 \\
 468 & \{1,2,0,3\} & -1 & -3 & -9 & 6 & 0 & 0 & 0 & 2 & 43 & 45 \\
 469 & \{1,2,0,4\} & -1 & -2 & -10 & 8 & 0 & 0 & 1 & 19 & 136 & 156 \\
 470 & \{1,2,0,5\} & -1 & -1 & -11 & 10 & 0 & 0 & 0 & 1 & 20 & 21 \\
 471 & \{1,2,0,6\} & -1 & 0 & -12 & 12 & 0 & 0 & 0 & 8 & 48 & 56 \\
 472 & \{1,2,0,7\} & -1 & 1 & -13 & 14 & 0 & 0 & 0 & 0 & 6 & 6 \\
 473 & \{1,2,0,8\} & -1 & 2 & -14 & 16 & 0 & 0 & 0 & 0 & 13 & 13 \\
 474 & \{1,2,0,10\} & -1 & 4 & -16 & 20 & 0 & 0 & 0 & 0 & 4 & 4 \\
 475 & \{1,2,0,12\} & -1 & 6 & -18 & 24 & 0 & 0 & 0 & 0 & 1 & 1 \\
 476 & \{1,2,0,14\} & -1 & 8 & -20 & 28 & 0 & 0 & 0 & 0 & 1 & 1 \\
 477 & \{1,3,0,2\} & -2 & -8 & -16 & 8 & 0 & 1 & 17 & 127 & 773 & 918 \\
 478 & \{1,3,0,3\} & -2 & -7 & -17 & 10 & 0 & 0 & 0 & 0 & 17 & 17 \\
 479 & \{1,3,0,4\} & -2 & -6 & -18 & 12 & 0 & 0 & 0 & 2 & 25 & 27 \\
 480 & \{1,3,0,5\} & -2 & -5 & -19 & 14 & 0 & 0 & 0 & 1 & 15 & 16 \\
 481 & \{1,3,0,6\} & -2 & -4 & -20 & 16 & 0 & 0 & 1 & 17 & 149 & 167 \\
 482 & \{1,3,0,7\} & -2 & -3 & -21 & 18 & 0 & 0 & 0 & 0 & 2 & 2 \\
 483 & \{1,3,0,8\} & -2 & -2 & -22 & 20 & 0 & 0 & 0 & 0 & 7 & 7 \\
 484 & \{1,3,0,10\} & -2 & 0 & -24 & 24 & 0 & 0 & 0 & 6 & 55 & 61 \\
 485 & \{1,3,0,14\} & -2 & 4 & -28 & 32 & 0 & 0 & 0 & 0 & 13 & 13 \\
 486 & \{1,4,0,2\} & -3 & -12 & -24 & 12 & 0 & 1 & 11 & 26 & 275 & 313 \\
 487 & \{1,4,0,3\} & -3 & -11 & -25 & 14 & 0 & 0 & 0 & 0 & 1 & 1 \\
 488 & \{1,4,0,4\} & -3 & -10 & -26 & 16 & 0 & 0 & 0 & 0 & 1 & 1 \\
 489 & \{1,4,0,5\} & -3 & -9 & -27 & 18 & 0 & 0 & 0 & 1 & 30 & 31 \\
 490 & \{1,4,0,6\} & -3 & -8 & -28 & 20 & 0 & 0 & 0 & 0 & 2 & 2 \\
 491 & \{1,4,0,7\} & -3 & -7 & -29 & 22 & 0 & 0 & 0 & 0 & 2 & 2 \\
 492 & \{1,4,0,8\} & -3 & -6 & -30 & 24 & 0 & 0 & 1 & 1 & 20 & 22 \\
 493 & \{1,4,0,11\} & -3 & -3 & -33 & 30 & 0 & 0 & 0 & 0 & 2 & 2 \\
 494 & \{1,4,0,12\} & -3 & -2 & -34 & 32 & 0 & 0 & 0 & 0 & 1 & 1 \\
 495 & \{1,4,0,14\} & -3 & 0 & -36 & 36 & 0 & 0 & 0 & 0 & 3 & 3 \\
 496 & \{1,5,0,2\} & -4 & -16 & -32 & 16 & 0 & 0 & 0 & 12 & 75 & 87 \\
 497 & \{1,5,0,4\} & -4 & -14 & -34 & 20 & 0 & 0 & 0 & 0 & 4 & 4 \\
 498 & \{1,6,0,2\} & -5 & -20 & -40 & 20 & 0 & 0 & 8 & 35 & 273 & 316 \\
 499 & \{1,6,0,3\} & -5 & -19 & -41 & 22 & 0 & 0 & 0 & 0 & 6 & 6 \\
 500 & \{1,6,0,4\} & -5 & -18 & -42 & 24 & 0 & 0 & 0 & 0 & 1 & 1 \\
 501 & \{1,6,0,5\} & -5 & -17 & -43 & 26 & 0 & 0 & 0 & 0 & 3 & 3 \\
 502 & \{1,6,0,7\} & -5 & -15 & -45 & 30 & 0 & 0 & 0 & 0 & 33 & 33 \\
 503 & \{1,6,0,8\} & -5 & -14 & -46 & 32 & 0 & 0 & 0 & 0 & 1 & 1 \\
 504 & \{1,6,0,9\} & -5 & -13 & -47 & 34 & 0 & 0 & 0 & 0 & 3 & 3 \\
 505 & \{1,6,0,12\} & -5 & -10 & -50 & 40 & 0 & 0 & 0 & 0 & 14 & 14 \\
 506 & \{1,6,0,16\} & -5 & -6 & -54 & 48 & 0 & 0 & 0 & 0 & 1 & 1 \\
 507 & \{1,7,0,2\} & -6 & -24 & -48 & 24 & 0 & 0 & 0 & 6 & 41 & 47 \\
 508 & \{1,7,0,8\} & -6 & -18 & -54 & 36 & 0 & 0 & 0 & 0 & 3 & 3 \\
 509 & \{1,7,0,14\} & -6 & -12 & -60 & 48 & 0 & 0 & 0 & 0 & 1 & 1 \\
 510 & \{1,8,0,2\} & -7 & -28 & -56 & 28 & 0 & 0 & 1 & 3 & 40 & 44 \\
 511 & \{1,8,0,7\} & -7 & -23 & -61 & 38 & 0 & 0 & 0 & 0 & 2 & 2 \\
 512 & \{1,8,0,12\} & -7 & -18 & -66 & 48 & 0 & 0 & 0 & 0 & 1 & 1 \\
 513 & \{1,9,0,2\} & -8 & -32 & -64 & 32 & 0 & 0 & 1 & 23 & 160 & 184 \\
 514 & \{1,9,0,4\} & -8 & -30 & -66 & 36 & 0 & 0 & 0 & 0 & 5 & 5 \\
 515 & \{1,9,0,6\} & -8 & -28 & -68 & 40 & 0 & 0 & 0 & 0 & 17 & 17 \\
 516 & \{1,9,0,8\} & -8 & -26 & -70 & 44 & 0 & 0 & 0 & 0 & 5 & 5 \\
 517 & \{1,9,0,10\} & -8 & -24 & -72 & 48 & 0 & 0 & 0 & 0 & 11 & 11 \\
 518 & \{1,9,0,14\} & -8 & -20 & -76 & 56 & 0 & 0 & 0 & 0 & 2 & 2 \\
 519 & \{1,9,0,18\} & -8 & -16 & -80 & 64 & 0 & 0 & 0 & 0 & 2 & 2 \\
 520 & \{1,10,0,2\} & -9 & -36 & -72 & 36 & 0 & 0 & 3 & 26 & 76 & 105 \\
 521 & \{1,10,0,5\} & -9 & -33 & -75 & 42 & 0 & 0 & 0 & 0 & 2 & 2 \\
 522 & \{1,10,0,7\} & -9 & -31 & -77 & 46 & 0 & 0 & 0 & 0 & 2 & 2 \\
 523 & \{1,10,0,8\} & -9 & -30 & -78 & 48 & 0 & 0 & 0 & 4 & 13 & 17 \\
 524 & \{1,10,0,12\} & -9 & -26 & -82 & 56 & 0 & 0 & 0 & 0 & 1 & 1 \\
 525 & \{1,10,0,14\} & -9 & -24 & -84 & 60 & 0 & 0 & 0 & 2 & 5 & 7 \\
 526 & \{1,11,0,2\} & -10 & -40 & -80 & 40 & 0 & 0 & 0 & 3 & 14 & 17 \\
 527 & \{1,11,0,4\} & -10 & -38 & -82 & 44 & 0 & 0 & 0 & 0 & 1 & 1 \\
 528 & \{1,11,0,6\} & -10 & -36 & -84 & 48 & 0 & 0 & 0 & 0 & 1 & 1 \\
 529 & \{1,12,0,2\} & -11 & -44 & -88 & 44 & 0 & 0 & 3 & 13 & 56 & 72 \\
 530 & \{1,12,0,3\} & -11 & -43 & -89 & 46 & 0 & 0 & 0 & 0 & 4 & 4 \\
 531 & \{1,12,0,4\} & -11 & -42 & -90 & 48 & 0 & 0 & 0 & 0 & 2 & 2 \\
 532 & \{1,12,0,5\} & -11 & -41 & -91 & 50 & 0 & 0 & 0 & 0 & 2 & 2 \\
 533 & \{1,12,0,7\} & -11 & -39 & -93 & 54 & 0 & 0 & 0 & 0 & 3 & 3 \\
 534 & \{1,12,0,8\} & -11 & -38 & -94 & 56 & 0 & 0 & 0 & 0 & 1 & 1 \\
 535 & \{1,12,0,12\} & -11 & -34 & -98 & 64 & 0 & 0 & 0 & 0 & 1 & 1 \\
 536 & \{1,13,0,2\} & -12 & -48 & -96 & 48 & 0 & 0 & 0 & 4 & 8 & 12 \\
 537 & \{1,14,0,2\} & -13 & -52 & -104 & 52 & 0 & 0 & 1 & 2 & 12 & 15 \\
 538 & \{1,15,0,2\} & -14 & -56 & -112 & 56 & 0 & 1 & 4 & 11 & 42 & 58 \\
 539 & \{1,15,0,4\} & -14 & -54 & -114 & 60 & 0 & 0 & 0 & 0 & 2 & 2 \\
 540 & \{1,15,0,6\} & -14 & -52 & -116 & 64 & 0 & 0 & 0 & 0 & 4 & 4 \\
 541 & \{1,15,0,9\} & -14 & -49 & -119 & 70 & 0 & 0 & 0 & 0 & 2 & 2 \\
 542 & \{1,15,0,10\} & -14 & -48 & -120 & 72 & 0 & 0 & 0 & 0 & 2 & 2 \\
 543 & \{1,15,0,16\} & -14 & -42 & -126 & 84 & 0 & 0 & 0 & 0 & 1 & 1 \\
 544 & \{1,16,0,2\} & -15 & -60 & -120 & 60 & 0 & 0 & 3 & 2 & 8 & 13 \\
 545 & \{1,17,0,2\} & -16 & -64 & -128 & 64 & 0 & 0 & 0 & 0 & 5 & 5 \\
 546 & \{1,18,0,2\} & -17 & -68 & -136 & 68 & 0 & 0 & 1 & 2 & 7 & 10 \\
 547 & \{1,18,0,3\} & -17 & -67 & -137 & 70 & 0 & 0 & 0 & 0 & 1 & 1 \\
 548 & \{1,18,0,4\} & -17 & -66 & -138 & 72 & 0 & 0 & 0 & 0 & 1 & 1 \\
 549 & \{1,20,0,2\} & -19 & -76 & -152 & 76 & 0 & 0 & 0 & 3 & 6 & 9 \\
 550 & \{1,21,0,2\} & -20 & -80 & -160 & 80 & 0 & 1 & 1 & 2 & 6 & 10 \\
 551 & \{1,22,0,2\} & -21 & -84 & -168 & 84 & 0 & 0 & 0 & 0 & 3 & 3 \\
 552 & \{1,24,0,2\} & -23 & -92 & -184 & 92 & 0 & 0 & 0 & 0 & 1 & 1 \\
 553 & \{1,25,0,2\} & -24 & -96 & -192 & 96 & 0 & 0 & 0 & 0 & 19 & 19 \\
 554 & \{1,27,0,2\} & -26 & -104 & -208 & 104 & 0 & 0 & 0 & 0 & 1 & 1 \\
 555 & \{1,28,0,2\} & -27 & -108 & -216 & 108 & 0 & 0 & 0 & 5 & 1 & 6 \\
 556 & \{1,30,0,2\} & -29 & -116 & -232 & 116 & 0 & 0 & 0 & 0 & 3 & 3 \\
 557 & \{1,33,0,2\} & -32 & -128 & -256 & 128 & 0 & 0 & 0 & 0 & 1 & 1 \\
 558 & \{1,36,0,2\} & -35 & -140 & -280 & 140 & 0 & 0 & 0 & 1 & 7 & 8 \\
 559 & \{1,39,0,2\} & -38 & -152 & -304 & 152 & 0 & 0 & 0 & 0 & 4 & 4 \\
 560 & \{1,42,0,2\} & -41 & -164 & -328 & 164 & 0 & 0 & 0 & 0 & 5 & 5 \\
 561 & \{1,43,0,2\} & -42 & -168 & -336 & 168 & 0 & 0 & 0 & 0 & 9 & 9 \\
 562 & \{1,45,0,2\} & -44 & -176 & -352 & 176 & 0 & 0 & 0 & 2 & 2 & 4 \\
 563 & \{1,48,0,2\} & -47 & -188 & -376 & 188 & 0 & 0 & 0 & 0 & 1 & 1 \\
 564 & \{1,49,0,2\} & -48 & -192 & -384 & 192 & 0 & 0 & 0 & 4 & 0 & 4 \\
 565 & \{1,55,0,2\} & -54 & -216 & -432 & 216 & 0 & 0 & 1 & 0 & 0 & 1 \\
&&&&&&&&&&& \\
\hline
\end{longtable}
\end{center}


\bibliographystyle{JHEP}
\bibliography{reference}


\end{document}